\documentclass[trackchanges,twocolumn,twocolappendix]{aastex701}
\newcommand{\ixpe}{\textit{IXPE}}
\newcommand{\chandra}{\textit{Chandra}}
\newcommand{\xmm}{XMM-\textit{Newton}}
\newcommand{\swift}{\textit{Swift}}

\newcommand{\nustar}{\textit{NuSTAR}}
\newcommand{\skirt}{\texttt{X-skirtor}}
\newcommand{\rxt}{\texttt{RXTorusD}}
\newcommand{\ux}{\texttt{UXCLUMPY}}

\newcommand{\nh}{N$_{\rm H}$}
\newcommand{\nhlos}{N$_{\rm H,LoS}$}

\def\code#1{\texttt{#1}}

\defcitealias{pizzetti2022}{I}
\defcitealias{torres23}{II}
\defcitealias{pizzetti2025}{III}
\defcitealias{sengupta2025}{IV}
\defcitealias{torres2025}{V}
\usepackage{natbib,twoopt}
\usepackage{makecell}
\usepackage{colortbl,xcolor,color}

\usepackage{comment}
\usepackage{amsmath}
\usepackage{algorithm}
\usepackage{amsfonts}
\usepackage{mathrsfs}
\usepackage{amssymb}
\usepackage{graphicx} 
\let\tablenum\relax
\usepackage{siunitx}
\usepackage[table]{xcolor}
\usepackage{subcaption}
\usepackage{longtable}
\definecolor{redcell}{RGB}{224, 102, 102}
\definecolor{greencell}{RGB}{144, 238, 144}
\definecolor{whitecell}{RGB}{211, 211, 211}
\newcommand{\graytableline}{%
\arrayrulecolor{gray!25}\hline
\arrayrulecolor{black}
}
\usepackage{txfonts}

\begin{document}

\title{The hidden variability of the torus in local Active Galactic Nuclei: \\ 20 years of \chandra, \xmm, and \nustar\ observations}

\author[orcid=0000-0002-9719-8740,sname='Gianolli']{Vittoria Elvezia Gianolli}
\affiliation{Department of Physics and Astronomy, Clemson University, Kinard Lab of Physics, Clemson, SC 29634, USA}
\affiliation{INAF-Osservatorio Astronomico di Brera, Via Brera 28, 20121 Milano, Italy}
\email[show]{vgianol@clemson.edu}

\author[orcid=0000-0003-3638-8943]{Núria Torres-Albà}
\affiliation{Department of  Physics, University of Virginia, Charlottesville, VA 22904, USA}
\altaffiliation{GECO fellow}
\affiliation{Department of Physics and Astronomy, Clemson University, Kinard Lab of Physics, Clemson, SC 29634, USA}
\email[noshow]{nuria@virginia.edu}

\author[0000-0002-6584-1703]{Stefano Marchesi}
\affiliation{Dipartimento di Fisica e Astronomia (DIFA) Augusto Righi, Università di Bologna, via Gobetti 93/2, 40129 Bologna, Italy}
\affiliation{INAF-Osservatorio di Astrofisica e Scienza dello Spazio (OAS), via Gobetti 93/3, 40129 Bologna, Italy}
\affiliation{Department of Physics and Astronomy, Clemson University, Kinard Lab of Physics, Clemson, SC 29634, USA}
\email[noshow]{stefano.marchesi@inaf.it}

\author[0000-0001-5544-0749]{Marco Ajello}
\affiliation{Department of Physics and Astronomy, Clemson University, Kinard Lab of Physics, Clemson, SC 29634, USA}
\email[show]{majello@g.clemson.edu}

\author[0000-0001-5231-2645]{Claudio Ricci}
\affiliation{Department of Astronomy, University of Geneva, ch. d’Ecogia 16, 1290, Versoix, Switzerland}
\affiliation{Instituto de Estudios Astrof\'isicos, Facultad de Ingenier\'ia y Ciencias, Universidad Diego Portales, Av. Ej\'ercito Libertador 441, Santiago, Chile}
\email[noshow]{claudio.ricci@unige.ch}

\author[0000-0003-2287-0325]{Isaiah Cox}
\affiliation{Department of Physics and Astronomy, Clemson University, Kinard Lab of Physics, Clemson, SC 29634, USA}
\email[noshow]{isc@clemson.edu}

\author[0000-0003-2754-9258]{Massimo Gaspari}
\affiliation{Department of Physics, Informatics and Mathematics, University of Modena and Reggio Emilia, 41125 Modena, Italy}
\email[noshow]{massimo.gaspari@unimore.it}

\author[]{Dhrubojyoti Sengupta}
\affiliation{CNRS, CEA, AIM, Université Paris-Saclay, Université Paris Cité, 91191, Gif-sur-Yvette, France}
\email[noshow]{phydhrubo@gmail.com}

\author[0000-0001-6564-0517]{Ross Silver}
\affiliation{NASA Goddard Space Flight Center, Greenbelt, MD 20771, USA}
\affiliation{Southeastern Universities Research Association, Washington, DC 20005, USA}
\affiliation{School of Physics and Astronomy, University of Minnesota, Minneapolis, MN 55455, USA}
\email[noshow]{ross.m.silver@nasa.gov}

\author[0000-0002-7825-1526]{Indrani Pal}
\affiliation{Department of Physics and Astronomy, Clemson University, Kinard Lab of Physics, Clemson, SC 29634, USA}
\email[noshow]{ipal@clemson.edu}

\author[0000-0001-6412-2312]{Andrealuna Pizzetti}
\affiliation{European Southern Observatory, Alonso de Còrdova 3107, Casilla 19, Santiago, 19001, Chile}
\email[noshow]{Andrealuna.Pizzetti@eso.org}

\author[0000-0002-7791-3671]{Xiurui Zhao}
\affiliation{Department of Astronomy, University of Illinois at Urbana - Champaign, Urbana, IL 61801, USA}
\affiliation{Cahill Center for Astrophysics, California Institute of Technology, 1216 East California Boulevard, Pasadena, CA 91125, USA}
\email[noshow]{xiurui.zhao.work@gmail.com}

\begin{abstract}
X-ray absorption variability in active galactic nuclei (AGN) provides key constraints on the structure and dynamics of the circumnuclear obscuring medium, the so-called torus. A fraction of nearby AGN, however, have been classified as non-variable in line-of-sight (LoS) column density based on limited temporal coverage. We present the first systematic study of a sample of 11 local (z $\leq$ 0.1) obscured (N$_{\rm H} \geq 10^{22}$ cm$^{-2}$) AGN, initially classified as non-variable. The sample is selected from the \textit{Swift}–BAT 100-month catalog, and comprises 60 observations from \textit{Chandra}, XMM-\textit{Newton}, and \textit{NuSTAR}, spanning timescales from days to nearly two decades.
We simultaneously model all available spectra for each source adopting physically motivated torus models: \skirt, \rxt, and \ux. This approach allows us to derive the global properties of the obscurer while tracking possible epoch-to-epoch variations in the LoS column density and intrinsic X-ray emission. We find that the original non-variable classification (based on only two X-ray observations) is frequently not robust: clear \nhlos\ variability is detected in half of the sample, whereas 7 out of 10 AGN require intrinsic flux variability, with the rest showing flux--\nhlos\ degeneracies.
We also find that the probability of identifying absorption variability increases with the number of observations, and the largest column density changes preferentially occur on long timescales, consistent with absorption by extended, structured clouds on torus scales. These findings support a clumpy and dynamic obscuring medium as a common feature of nearby AGN and highlight the importance of long-term X-ray monitoring for accurately characterizing AGN obscuration.

\end{abstract}

\section{Introduction}
\label{sec:intro}
Active Galactic Nuclei (AGN) display a remarkable diversity in luminosity (with bolometric luminosities ranging from 10$^{41}$ to 10$^{48}$ erg s$^{-1}$), spectral properties, and classification. Despite this apparent variety, they are all powered by the same fundamental mechanism: the accretion of matter onto a supermassive black hole (SMBH; M$_{\rm BH}\sim 10^{6}-10^{10}\rm M_{\rm{\odot}}$) located at the gravitational center of $\sim$10\% of known galaxies \citep{soltan1982,richstone1998}. According to the Unified Model \citep{antonucci93,urry95}, the central engine (i.e., a SMBH surrounded by an accretion disk) emits primarily in the optical and ultraviolet, while a hot corona close to the inner disk up-scatters photons to X-rays via inverse Compton processes \citep{haardt91,haardt93,fabian2015,ricci2017}. Meanwhile, the observed differences among AGN arise mainly from geometrical and orientation effects caused by an optically thick, dusty, molecular torus surrounding the nucleus \citep[for a review see][]{hickox2018}.
This molecular torus, extending from $\sim$1 pc up to few times 100 pc from the SMBH \citep[e.g.,][]{jaffe04,burtscher13}, absorbs ultraviolet and optical radiation from the accretion disk and thermally re-emits it in the infrared, producing a characteristic peak around tens of microns \citep{krolik2000}. Its orientation relative to the observer determines the observed AGN type. When the line-of-sight (LoS) does not intercept the torus, both the broad-line region (BLR) and the narrow-line region (NLR) are visible, yielding a Type 1 optical spectrum; when the LoS intercepts the torus, the BLR is hidden and only narrow lines remain, resulting in a Type 2 AGN. In X-rays, the same geometry leads to different levels of absorption by material characterized by a hydrogen column density (\nhlos): sources with \nhlos\ $<$ 10$^{22}$ cm$^{-2}$ are unobscured, those with $10^{22}$ $\leq$ \nhlos\ $<$ 1.5 $\times$ 10$^{24}$ cm$^{-2}$ are Compton-thin (CTH), and those exceeding \nhlos\ $\geq$ 1.5 $\times$ 10$^{24}$ cm$^{-2}$ are Compton-thick (CTK) AGN \citep[e.g.,][]{matt2002,comastri2004,caccianiga07,dellaceca08}. In the nearby Universe, the majority of AGN are obscured, with approximately 70\% exhibiting column densities $\geq$ 10$^{22}$ cm$^{-2}$. Of these, $\sim$20–50\% are predicted to be CTK-AGN \citep[e.g.,][]{ghisellini1994,Ajello2009,ueda2014,ananna2019}, although observations recover them to varying degrees of completeness \citep[e.g.,][]{burlon2011,ricci2015,Kammoun2020,torres2021,Sengupta2023,Boorman2024}. The optical and X-ray classifications therefore generally represent two complementary manifestations of the same geometry, governed by the viewing angle and the column density of the obscuring material.

The physical nature and geometry of the torus remain one of the most debated aspects of AGN unification. Early radiative-transfer models described it as a smooth, continuous distribution of dust and molecular gas, in which the temperature declines with distance from the SMBH \citep{pier1992}. However, interferometric and spectral studies have shown that this description cannot reproduce the observed infrared emission and variability. Instead, a clumpy torus, composed of discrete, optically thick clouds embedded in a more diffuse medium, provides a better match to both theoretical expectations and observations \citep[e.g.,][]{nenkova08,netzer2015,ramos2017,honig2019}. In this scenario, the obscuration depends probabilistically on the number and distribution of clumps along the line of sight, while the dust temperature and composition may vary even at fixed radius. Hydrodynamical simulations of chaotic cold accretion (CCA) and radiation-driven turbulence \citep{wada2012,gaspari2020} also produce irregular, dynamic structures consistent with such clumpiness. In the CCA scenario, cold clouds and filaments condense out of a turbulently stirred hot medium, fragment, and circulate through inflow and fountain-like motions while feedback maintains global quasi-equilibrium. In this framework, LoS column density variability naturally traces both compact inner structures (short timescales) and the evolution or rearrangement of larger cloud complexes at larger radii (long timescales), linking observed absorption changes directly to the granularity and dynamics of the circumnuclear medium \citep[e.g.,][]{gaspari2013,gaspari2017,gaspari2020}.

A major step forward in constraining the geometry of this obscuring medium came from X-ray polarimetry. The launch of the Imaging X-ray Polarimetry Explorer \citep[\ixpe;][]{weisskopf22} in 2021 enabled, for the first time, direct measurements of polarization in heavily obscured AGN. Observations of the Compton-thick sources Circinus galaxy and NGC~1068 revealed polarization degrees and angles consistent with scattering in a geometrically thick equatorial structure with a half-opening angle of about 45$^\circ$–55$^\circ$ from the vertical axis of the system, in line with the Unification Model \citep{ursini2023,marin2024,marin24}. These measurements provided a further direct confirmation of the torus geometry, demonstrating that the equatorial obscurer is indeed extended and likely inhomogeneous. Polarimetry thus complements spectroscopy: while the former reveals the global scattering and orientation of the system, the latter traces its dynamic and inhomogeneous behavior through temporal variations in absorption.

X-ray absorption variability offers one of the most direct probes of the obscuring material clumpiness. Changes in the \nhlos\ over timescales ranging from hours to years are interpreted as the transits of individual clouds across the observer’s line of sight. For example, NGC~1365 presents rapid transitions between Compton-thin and thick states over days, providing the first strong evidence for BLR-scale obscuring clouds \citep[e.g.,][]{risaliti2002,risaliti2007,risaliti2009,risaliti09}; NGC~4151, showing complex, multi-epoch absorption changes \citep[e.g.,][]{puccetti07,Beuchert2017}; and NGC~7582, where recurrent \nh\ variations revealed complex absorption: short-timescale variability likely arising from BLR clouds, together with longer-term changes driven by a patchy, Compton-thick torus that both obscures the nucleus and contributes significant reflection \citep[e.g.,][]{bianchi2009,rivers2015}. 
Long-term variability has also been detected in NGC~4945 \citep{yaqoob2012} and NGC~2992 \citep{shu2010}, while even the archetypal Compton-thick AGN NGC~1068 has shown partial-covering changes over multi-year timescales \citep{marinucci2016}. These represent only a few examples among the many that have established circumnuclear obscuration as a dynamic, multi-phase phenomenon operating across multiple spatial scales, and that have laid the groundwork for large-scale statistical studies.

Over the past decade, considerable effort has gone into systematically investigating the \nhlos\ variability in nearby AGN using data from \chandra, XMM-{\it Newton}, \swift-XRT, and \nustar\ \citep[e.g.,][]{risaliti2002, Markowitz2014,Laha2020,Esparza2021,zhao2021,marchesi2022,Tanimoto2022}. The Clemson-INAF Compton thick AGN (CI-CTAGN) project extends this effort by assembling a large, homogeneous sample of obscured AGN to constrain both line of sight and global torus properties through simultaneous, multi-epoch spectral modeling with physically motivated torus models such as \texttt{MYTorus} \citep{murphy2009}, \texttt{borus02} \citep{bolokovic2018}, \texttt{UXCLUMPY} \citep{buchner2019}, and \texttt{XCLUMPY} \citep{tanimoto2019}. These studies collectively show that approximately one-third to one-half of local CTH-AGN exhibit significant \nhlos\ variability, while the rest appear consistent with being non-variable within observational uncertainties \citep[][see Sect.~\ref{sec:data} for more details]{torres23,pizzetti2025}.

However, AGN with no detected \nhlos\ variability challenge our understanding of AGN obscuration and the clumpy torus paradigm. A true lack of column density variability (or only yearly-scale changes) would require clouds much larger, denser, and more uniform than typically inferred \citep[e.g.,][]{nenkova08,Laha2020}, suggesting a smoother or more extended torus, possibly dominated by stable circumnuclear dust lanes \citep[consistent with interferometric evidence for extended, stable dust distributions on parsec to hundred-parsec scales in nearby AGN; e.g.,][]{jaffe2004,packman2005,honig2010,tristam2014,mexcua2016} or galactic-scale absorbers \citep[such as misaligned or warped disks and kpc-scale gas concentrations; e.g.,][]{malkan1998,goulding2012,buchner17}.
Alternatively, this apparent stability may arise from observational biases, such as limited temporal coverage or sensitivity \citep{bianchi2012,hickox2018,mazzolari2024,cox2025}. 
Disentangling these scenarios would require continuous X-ray monitoring over days to years to resolve the full structure of the obscuring medium in individual AGN, an observational effort of considerable scale. However, decades of archival X-ray observations offer a powerful resource for investigating \nhlos\ variability and constraining the geometry and distribution of the absorbing material.
Within this framework, this paper inaugurates a series dedicated to the study of Compton-thin and Compton-thick AGN initially classified as non-variable \citep[][see Sect.~\ref{sec:data} for the selection criteria]{zhao2021}. We conduct a systematic study of a sample of 11 such sources, comprising 60 X-ray observations from \chandra, \xmm, and \nustar, with three to nine observations per AGN with separations ranging from 1 day (i.e., consecutive observations) to $\sim$20 years. The paper is organized as follows: Sect.~\ref{sec:data} details the sample selection and data reduction procedures; Sect.~\ref{method} reports the methodology adopted during the analysis; Sect.~\ref{model_comp} provides a comparison between the three torus models; Sect.~\ref{sec:res} presents the X-ray (CTH- or CTK-AGN) and variability classifications for the sample; Sect.~\ref{evol_nh_with_t} discusses the time evolution of epoch-to-epoch changes in \nhlos; Sect.~\ref{stat} reports the statistical analysis of the torus and AGN parameters derived from the best-fit models; Sect.~\ref{mcg} shows the results of the spectral analysis for MCG-03-34-064; and Sect.~\ref{sec:conclusions} summarizes the main findings.

In this paper, we use as cosmological constants H$_{\rm 0}$ = 70 km s$^{-1}$ Mpc$^{-1}$, $\Omega \Lambda _{\rm 0}$ = 0.73, and $\Omega _{\rm M}$ = 0.27. Errors are quoted at the 90\% confidence level (c.l.). We consider a probability of 0.01 (roughly corresponding to $\sim$2.6$\sigma$ for a Gaussian distribution) as a statistically significant threshold for the null hypothesis probability (NHP; i.e., Log(NHP) $<$ -2.00).
\begin{figure*}
\centering
    \includegraphics[width=0.8\textwidth, , trim=0 10 0 0,clip]{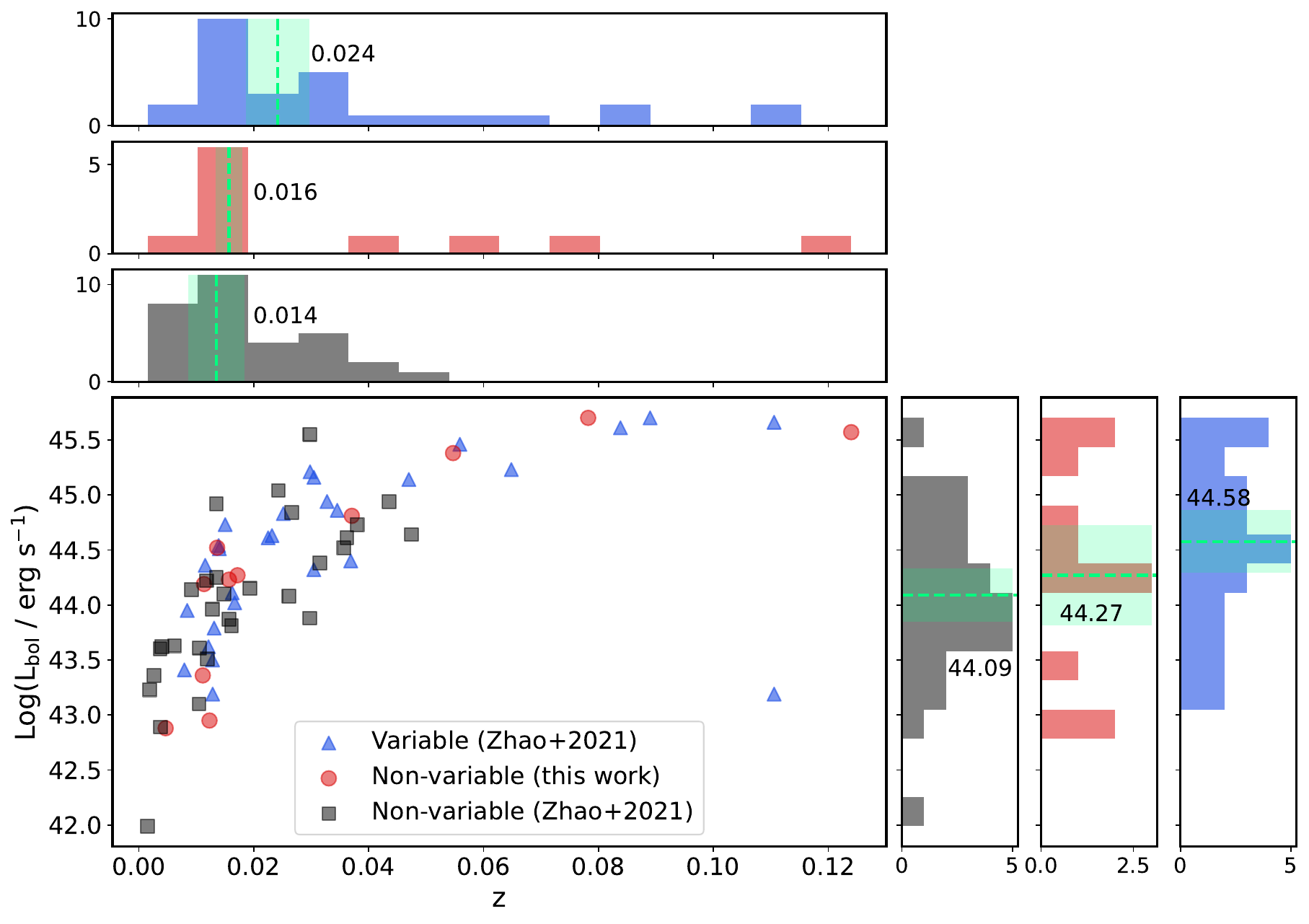}
    \caption[]
    {{Samples comparison: bolometric luminosity versus redshift. The variable sample (in blue triangles; \citetalias{pizzetti2022}-\citetalias{torres2025}), the sub-sample of non-variable AGN studied in this paper (in red circles), and the non-variable sources to be analyzed in upcoming papers (in gray squares) are shown. For each sample, the distributions are displayed above the scatter plot, and the green dashed lines with shaded bands indicate the median values and the corresponding median absolute deviations, ${\rm MAD}={\rm median}(|x_i-{\rm median}(x)|)$.
    The outlier AGN of the variable sample is NGC~6300, analyzed in \citetalias{sengupta2025}.
    The KS-tests show no statistically significant differences between the bolometric luminosity or redshift distributions of the sub-samples, supporting a direct comparison between the variable and non-variable sources.}}
    \label{fig:z_bol}
\end{figure*}

\section{Data}
\label{sec:data}
\subsection{Sample selection}

The present paper is part of the CI-CTAGN\footnote{\url{https://science.clemson.edu/ctagn/}} project and within it, of the \nh\ variability project that aims at deriving and studying the torus properties of a well-defined sample of obscured AGN, and their variability (or lack thereof) through multi-epoch X-ray observations. The original sample was selected by \citet{zhao2021}, who performed a broadband X-ray spectral analysis of 93 Compton-thin AGN in the nearby Universe (z $\leq$ 0.1), all with archival \nustar\ coverage. 
The sources were drawn from the Swift-BAT \citep[Burst Alert Telescope;][]{Barthelmy2005} 100-month catalog\footnote{\url{http://bat.ifc.inaf.it/100m_bat_catalog/100m_bat_catalog_v0.0.htm}}. For each selected AGN, \citet{zhao2021} jointly fitted one \nustar\ observation, to account for the hard X-ray (3 -- 79 keV) coverage, together with a single soft (0.3 -- 12 keV) X-ray observation (from either \xmm, \chandra, or \swift-XRT), to constrain both the torus geometry parameters and the line-of-sight column density. The presence of \nustar\ data is critical for obscured AGN, where soft X-rays are attenuated, as it enables disentangling reflection from LoS absorption and provides robust constraints on the properties of the obscuring torus \citep{civano2015,marchesi2018}.
In their preliminary study, \citet{zhao2021} identified 31 AGN as variable in intrinsic flux or \nhlos\ (``variable sample'', hereafter), and 62 as non-variable (``non-variable sample'', hereafter).
To be noted that for 22 out of 93 (i.e., 23.7\%) AGN, \citet{zhao2021} considered \nustar\ and soft X-ray simultaneous observations, introducing a possible bias towards a non-variable classification for these sources.
In papers \textsc{I -- V} \citep[][respectively]{pizzetti2022,torres23,pizzetti2025,sengupta2025,torres2025}, 28 out of 31 variable AGN have been the subject of detailed follow-up investigations. 
Given the large number of available observations, papers \citetalias{pizzetti2022,sengupta2025,torres2025} were dedicated to single-source studies of the variable AGN NGC~7582, NGC~6300, and Mrk~477, respectively.
Meanwhile, in papers \citetalias{torres23} and \citetalias{pizzetti2025}, sub-sets of 12 (total of 53 observations) and 13 (total of 52 observations) variable sources have been studied.
Across the combined sample of 28 variable AGN, 39\% (11/28) shows clear \nhlos\ variability to adequately reproduce the data, and also require flux changes. The remaining 61\% (17/28) is consistent with flux variability alone.
Considering both geometrical and intrinsic properties, we find no intrinsic differences between the Variable, Non-variable, and Undetermined\footnote{We report a complete description of the criteria used to classify a source as ``Undetermined'' in Sect.~\ref{var_class}.} groups. Nearly half the sources (13/28) require a Compton-thick reflector in \ux\ fits. Importantly, obscuration variability is found to be more common on longer timescales: increasing from $\sim$20\% at $\Delta$t $<$ 10 days to 60-70\% at $\Delta$t $>$ 5 yr, with Mrk~477 providing a clear example \citepalias{torres2025}.

Here, we study the complementary non-variable sample to i) assess the robustness of this classification in light of the all the available multi-epoch data, and ii) to characterize their torus properties. From the non-variable sample, we selected a sub-set designed to be statistically comparable in size and number of observations to the variable sub-groups studied in \citetalias{torres23} and \citetalias{pizzetti2025}. This approach allows us to perform an analysis analogous to that carried out on the variable sample, enabling a direct comparison between the two populations. 
Specifically, our sub-sample contains 11 sources, for a total of 60 observations. The remaining non-variable sources will be analyzed in future dedicated studies. Table~\ref{tab:1_properties} lists names, redshifts, classification in the optical and radio bands, bolometric luminosities \citep[taken from the BASS sample;][]{koss2022}, and X-ray (2 - 10 keV) fluxes (derived from our spectral analysis, see Sect~\ref{sec:res}) of the 11 AGN. In Fig.~\ref{fig:z_bol}, we show the bolometric luminosities as a function of redshift, as well as the parameters distributions, for the variable and non-variable samples. 
We show only 42 AGN for the non-variable sample in Fig.~\ref{fig:z_bol}, given that this is the fraction of sources with more than 2 available X-ray observation and thus, the AGN for which a complete broad-band multi-observations analysis can be carried out. 
We performed the two-sample Kolmogorov–Smirnov (KS) test to compare the bolometric luminosity and redshift distributions of the samples, and find no statistically significant differences between any pair of samples in either parameter.

\begin{deluxetable*}{ccccccccc}[h]
\tablewidth{0pt}
\tablecaption{AGN properties.\label{tab:1_properties}}
\tablehead{
\colhead{\scriptsize (1)} &
\colhead{\scriptsize (2)} &
\colhead{\scriptsize (3)} &
\colhead{\scriptsize (4)} &
\colhead{\scriptsize (5)} &
\colhead{\scriptsize (6)} &
\colhead{\scriptsize (7)} &
\colhead{\scriptsize (8)} &
\colhead{\scriptsize (9)} \\
\colhead{Source} &
\colhead{$z$} &
\colhead{Optical} &
\colhead{X-ray} &
\colhead{Radio} &
\colhead{$\log(M_{\rm BH})$} &
\colhead{$\log(L_{\rm bol})$} &
\colhead{$\log(\lambda_{\rm Edd}$)}& 
\colhead{$F_{\rm 2-10 keV}$} \\
&
&
&
&
&
\colhead{[$M_\odot$]} &
\colhead{[erg/s]} & &
\colhead{[erg/s/cm$^{2}$]}
}
\startdata
ESO~464-G016 & 0.037 & Sy2 & CTH & RQ & 8.40 & 44.81 & -1.76 & $0.46\pm0.12$ \\
IC~5063 & 0.011 & Sy2 & CTH & RL & 8.24 & 44.19 & -2.22 & $7.27^{+0.10}_{-0.25}$ \\
LEDA~511869 & 0.078 & Sy2 & CTH & RQ & 8.66 & 45.70 & -1.14 & $1.19^{+0.10}_{-0.15}$ \\
MCG-03-34-064 & 0.017 & Sy1.8 & CTH & RQ & 8.37 & 44.27 & -2.28 & $2.27^{+0.18}_{-0.20}$ \\
Mrk~18 & 0.011 & Sy2 & CTH & RL & 7.72 & 43.36 & -2.53 & $0.70^{+0.11}_{-0.14}$ \\
Mrk~1498 & 0.055 & Sy1.9 & CTH & RQ & 7.19 & 45.38 & 0.02 & $7.23^{+0.05}_{-0.20}$ \\
NGC~1194 & 0.014 & Sy1.9 & CTK & RL & 7.83 & 44.52 & -1.49 & $1.13^{+0.23}_{-0.11}$ \\
NGC~2655 & 0.005 & Sy2 & CTH & RQ & 8.34 & 42.88 & -3.63 & $0.65^{+0.62}_{-0.19}$ \\
NGC~4785 & 0.013 & Sy2 & CTH & RQ & 8.00$^{1}$ & 42.95 & -3.15 & $1.15^{+0.11}_{-0.10}$ \\
UGC~03752 & 0.013 & Sy2 & CTH/CTK & RQ & 7.65 & 44.23 & -1.59 & $0.43^{+0.01}_{-0.39}$ \\
WISE~J144850 & 0.124 & Sy1.2 & CTH & RQ & 8.14 & 45.57 & -0.74 & $4.48^{+0.11}_{-0.36}$ \\
\enddata
\tablecomments{
\footnotesize {\it Columns:} 1) source name; 2) redshift; 3) optical classification; 4) radio classification: radio-quiet (RQ) or radio-loud (RL); 5) X-ray classification: Compton-thin (CTH) or Compton-thick (CTK); 6) black hole mass; 7) bolometric luminosity; 8) logarithm of the Eddington ratio; (9) X-ray flux in the 2--10 keV energy range from the best-fit with the \ux in units of 10$^{-12}$ erg/s/cm$^{2}$. 
The reported mass and bolometric luminosity are taken from the BASS sample \citep{koss2022}. In particular, they derive L$_{\rm bol}$ from the intrinsic 14-150 keV luminosity using a bolometric correction of 8, while 
M$_{\rm BH}$ is derived from broad-line virial estimators when available, or from stellar velocity dispersion measurements for 
obscured sources. For NGC~4785, we derived L$_{\rm bol}$ from the intrinsic, absorption-corrected 2-10 keV luminosity by applying the bolometric correction of \citet{duras2020} with a bolometric correction of 15.4.
We note that these estimates are highly uncertain for Seyfert~2 galaxies and should therefore be taken with caution. The Eddington ratios are taken from \citep{koss2022}, except for NGC~4785 for which we derive it.
The reported flux refers to the first \chandra\ observation, or XMM observation if no \chandra\ data are available. For the other observations, the flux can be derived by multiplying for the cross-normalization constant registered in the source best-fit Tables~\ref{tab:spec_1}--\ref{tab:spec_11}. 
{\bf $^{1}$}: see note on the source in Appendix~\ref{source_note}.}
\end{deluxetable*}

\subsection{Data reduction}

During the analysis of the sample presented in this paper, we use all the available X-ray observations from \chandra \footnote{This paper employs a list of \chandra\ datasets, obtained by the \chandra\ X-ray Observatory,
contained in the Chandra Data Collection~\href{https://doi.org/10.25574/cdc.467}{DOI: 10.25574/cdc.467}.}, XMM-{\it Newton} \footnote{\url{https://www.cosmos.esa.int/web/xmm-newton/xsa}}, and \nustar \footnote{\url{https://heasarc.gsfc.nasa.gov/docs/nustar/nustar_archive.html}}. 

For XMM-{\it Newton}, we use spectra from the EPIC pn \citep{struder01}. Data reduction was performed with SAS v22.1, applying the standard filtering procedure to remove background flares. 
All {\it NuSTAR} \citep{harrison13} observations include both coaligned X-ray telescopes, Focal Plane Module A (FPMA) and B (FPMB). Cleaned and calibrated event files were produced using the \texttt{Nupipeline} task together with the calibration database (CALDB 20250122). For both XMM and \nustar, source and background extraction radii were determined iteratively to maximize the signal-to-noise (S/N) ratio in the 2 -- 10 keV band, following the procedure described in \cite{piconcelli04}.
\chandra\ \citep{Weisskopf00} spectra are reduced with CIAO v4.17 and calibration files (CALDB 4.11.6), following standard procedures. To extract the source (background) data a 5\arcsec\ circle (annulus with internal radius of 6\arcsec\ and outer of 15\arcsec) was adopted. Each region was visually inspected to avoid contamination from nearby sources.
XMM, \nustar, and \chandra\ have been rebinned considering \citet{kaastra2016} optimal binning scheme. In particular, for \nustar\ spectra, we adopted an iterative procedure to ensure the required S/N and same bin width in both FPMA/B. We adopted the C-statistic as fitting statistic. However, for 5 AGN (LEDA~511869, Mrk~18, NGC~4785, NGC~1194, and MCG-03-34-064), the \chandra\ spectra are rebinned to $\geq$15 photon counts per bin and, in these cases, the $\chi^2$ statistic is used. Tables~\ref{tab:spec_1}-\ref{tab:spec_11} report the statistic adopted for each AGN. Table~\ref{tab:1_observations} lists the observation ID, date, and effective exposure time after the filtering process for each observation.

\section{Methodology}\label{method}

To perform the broad-band X-ray spectral analysis, we follow the methodology described in \citetalias{torres23} and \citetalias{pizzetti2025} papers, using \textsc{XSPEC} version 12.14.1 \citep{Arnaud1996} to fit simultaneously the 0.5 -- 10.0 keV XMM-{\it Newton}, 0.5 -- 8.0 keV \chandra, and 3.0 -- 79.0 keV \nustar\ spectra.
In the following sections, we describe the different model components adopted during our fit.

\subsection{Model components}\label{model_components}

\noindent{\bf Galactic neutral absorption:} \texttt{phabs} component in \textsc{XSPEC} with column densities appropriate for the sky coordinates of each source from \citet{kalberla2005}, representing absorption due to gas in the Milky Way (order of $10^{20}$ cm$^{-2}$). The value was fixed during the fit.

\noindent{\bf Soft X-ray model:} below $\sim$2 keV, the soft X-ray spectrum is affected by the host galaxy emission and is characterized by emission lines \citep[e.g.][]{Schurch2004}, likely produced by photoionizated gas in the NLR, as commonly found in obscured AGN \citep{bianchi06,Guainazzi2007,Bianchi2019}, or by star-formation processes \citep[e.g.,][]{ranalli2003,mineo2012}. We thus include a collisionally ionized diffuse gas component \citep[\texttt{apec};][]{smith2001}. UGC~3752 requires two \texttt{apec} components to describe the multiphase thermal gas with the hotter component closer to the nucleus and more obscured \citep[see][]{torres2018}. Furthermore, where significant residuals ($\geq 3\sigma$) remain, at the energies of known emission lines, we model them phenomenologically by adding Gaussian components. In Appendix~\ref{source_note}, we report the number of added Gaussian components and their energies.

\noindent{\bf Intrinsic flux variability:} to account for differences between observations, we apply a multiplicative factor, $C_{\rm AGN}$, to the intrinsic power-law emission. 

\noindent{\bf Torus model:} it describes the primary and reprocessed AGN emissions, including reflection by obscuring Compton-thick material.  
The models applied in this paper are described in Sects.~\ref{sec:uxc}, \ref{sec:rxt}, and \ref{sec:skirt}. For MCG-03-34-064, Mrk 1498, and NGC 1194, we find that the adopted torus models do not adequately reproduce the iron K band, leaving significant residuals. To account for this, we include a Gaussian line and report its best-fit parameters in Appendix~\ref{source_note}.

\noindent{\bf Scattered component:} 
representing the fraction of the intrinsic AGN power-law that passes through the torus without interacting or interacting elastically, a few percent or less, and primarily arising from non-variable scattered emission and recombination from the NLR, which decreases with increasing column density \citep[e.g.,][]{bianchi06,McKaig2023,ricci2017}.

In \textsc{XSPEC}, it is parameterized as \texttt{F$_s$*cutoffpl}, where \texttt{F$_s$} is a multiplicative constant accounting for the scattered fraction and \texttt{cutoffpl} is the primary power-law component.
During the fit, we assume that the photon index ($\Gamma$) and torus geometry parameters (see Sects.~\ref{sec:uxc}, \ref{sec:rxt}, and \ref{sec:skirt} for details) do not vary on timescales up to $\sim$20 years. Thus, they are linked across epochs to reduce uncertainties \citep[see][]{marchesi2022}.
While continuum flux may vary between epochs, the corresponding changes in Eddington ratio ($\lambda_{\rm Edd}$) are small enough that they are not expected to significantly affect the intrinsic continuum shape or $\Gamma$.
This first order approximation is not valid for MCG-03-34-064 and thus, we fit each observation separately (see Sect.~\ref{mcg} for the detailed analysis).
In all fits, the high-energy cutoﬀ is fixed to 300 keV.

In summary, the model can be written in \textsc{XSPEC} as:

\code{phabs * (Soft Model + C$_{\rm AGN}$ * (Torus Model + F$_s$ * cutoffpl)}.

Best-fit results and spectra are presented in Sect.~\ref{sec:res} and Appendix~\ref{bestfit}. In Appendix~\ref{source_note}, we report and discuss whether modifications are applied to the standard fitting methodology.

\subsection{Torus model: \ux}\label{sec:uxc}

\ux\footnote{\url{https://github.com/JohannesBuchner/xars/blob/master/doc/uxclumpy.rst}} \citep{buchner2019} is a clumpy torus model designed to reproduce the observed distribution of column densities and the frequency of cloud eclipsing events in AGN. Unlike smooth-density torus models, \ux\ accounts for the clumpy nature of the obscurer by assuming clouds distributed on circular Keplerian orbits with random orientations. Their angular dispersion is parameterized by $\sigma_{\rm tor}$, which regulates how widely the clouds spread around the equatorial plane.

A key feature of \ux\ is the inclusion of an optional inner Compton-thick reflector, characterized by the parameter \texttt{CTKcover}, which represents the covering factor of this material. Given its nature, in the best-fit tables (Tables~\ref{tab:spec_1} to \ref{tab:spec_11}), this parameter will be presented together with the covering factors as C$_{\rm f}$. However, due to its distinct physical origin, it cannot be directly compared with the covering factors derived from the other torus models. 
This inner component is motivated by the need to reproduce strong reflection features in AGN spectra and can be interpreted as part of the dust-free broad-line region or the inner torus wall.
Together, $\sigma_{\rm tor}$, \texttt{CTKcover}, and the inclination angle $\theta$ allow the model to probe a wide range of line-of-sight hydrogen column densities 20 $\leq$ Log(\nhlos\ / cm$^{-2}$) $\leq$ 26. 

The two tables of \ux\ treat the direct and torus reflection (first table below), and scattering emissions (second table below) self-consistently. The latter represents emission that escapes the torus after undergoing at least one interaction with the medium.
In \textsc{XSPEC}, the model can be implemented as:

\code{Torus Model = \{uxclumpy.fits\} + \\F$_s$ * \{uxclumpy\_scattered.fits\}}.

\subsection{Torus model: \rxt}\label{sec:rxt}

\rxt\footnote{\url{https://www.astro.unige.ch/reflex/xspec-models}} \citep{paltani2017,ricci2023} is the first X-ray spectral model that reproduces the emission of AGN considering a dusty gas. In particular, \rxt\ incorporates a wide range of physical processes associated with dust, including scattering, near-edge X-ray absorption fine structure, and self-shielding, as well as additional effects such as Rayleigh scattering and scattering on molecular gas.

\rxt\ assumes a toroidal homogeneous geometry, allowing a variable covering factor defined as the ratio of the minor to major axes (C$_{\rm f}$ = r/R) that can vary between 0.01 and 1. Meanwhile, the inclination, i.e., viewing angle of the observer, spans 0$^\circ$-90$^\circ$ (with $\theta$ = 0$^\circ$ being a face-on AGN). It distinguishes between line-of-sight (\nhlos) and equatorial (N$_{\rm H,eq}$) column densities, where \nhlos\ affects the transmitted continuum and N$_{\rm H,eq}$ determines the intensity of the reprocessed  emission, which self-consistently includes reflection and fluorescence.

In this work, we adopt the tables computed for solar metallicity and a dust fraction of 1, thus assuming that all iron atoms are in dust form. 
In \textsc{XSPEC}, the model can be implemented as follows with the first table accounting for the reprocessed components (i.e., reflection and fluorescence) and the second for absorption:

\code{Torus Model = \{RXTorus\_rprc.mod\} + \\ \{RXTorus\_cont.mods\}*cutoffpl}.

\subsection{Torus model: \skirt}\label{sec:skirt}

\skirt\footnote{\url{https://github.com/BertVdM/xskirtor}} (Vander Meulen et al., subm.) is the first AGN torus model calculated using the X-ray Monte Carlo radiative transfer code \texttt{SKIRT} \citep{vandermeulen2023}. It simulates reflection and obscuration by cold atomic gas in a toroidal geometry, with a uniform wedge shape defined by the covering factor (C$_{\rm f}$ = 0.25-0.85) and equatorial hydrogen column density (22 $\leq$ Log(N$_{\rm H,eq}$ / cm$^{-2}$) $\leq$ 25), observed at an inclination $\theta$ (between 0$^\circ$ and 90$^\circ$). What sets \skirt\ apart is its advanced treatment of X-ray physics and high spectral resolution. In particular, it accounts for bound-electron scattering by incorporating the momentum distribution of electrons bound to neutral atoms. Covering a broad energy range (0.2–200 keV) and providing a detailed description of reprocessed emission, \skirt\ is among the most advanced AGN torus models currently available. The model treats both transmitted and reprocessed emission using separate tables for the direct continuum (second table below) and the reprocessed component (first table below), in \textsc{XSPEC}:

\code{Torus Model = \{xskirtor\_smooth\_ccd\_rpc.mod\} + \{xskirtor\_smooth\_ccd\_ext.mods\}*cutoffpl}. \\

In Table~\ref{tab:torus_mod}, we report a summary of the parameters fitted during the analysis of the three torus models.

\begin{table}[h]
\centering
\caption{Summary of the torus-model parameters.\label{tab:torus_mod}}
\label{tab:torus_model_parameters}
\begin{tabular}{llll}
\hline
\hline
Geometry & Parameter & Meaning \\
\hline

&\texttt{\ux}& \\
Clumpy
& $\Gamma$
& X-ray photon index \\

& \nhlos
& Line-of-sight column density \\

& $\theta$
& Viewing angle \\

& $\sigma_{\rm tor}$
& Angular width of the clouds distribution \\

& \texttt{CTKcover}
& Covering factor inner CTK reflector \\
\graytableline
&\texttt{\rxt}& \\

& $\Gamma$
& X-ray photon index \\
Smooth 
& \nhlos
& Line-of-sight column density \\
toroidal
& $\rm N_{H,eq}$
& Equatorial column density \\
circle
& $\theta$
& Viewing angle \\

& $r/R$
& Inner-to-outer radius ratio \\
\graytableline
&\texttt{\skirt}& \\

& $\Gamma$
& X-ray photon index \\
Smooth 
& \nhlos
& Line-of-sight column density \\
toroidal 
& $\rm N_{H,eq}$
& Equatorial column density \\
wedge
& $\theta$
& Viewing angle \\

& $C_{\rm f}$
& Covering factor \\
\hline
\end{tabular}\tablecomments{
\footnotesize For a detailed discussion of the geometries and assumptions of the torus models adopted here, we refer to the corresponding papers. \ux: \citet{Nenkova2008,buchner2019}, \rxt: \citet{paltani2017,ricci2023}, and \skirt: \citet{vandermeulen2023}; Vander Meulen et al., subm.
}
\end{table}

\subsection{Variability classification} \label{var_class}

Sources in this work were originally classified as non-variable, based on only two observations. To assess whether incorporating an additional 2 to 7 observations (depending on the AGN) alters this classification, we derive the \nhlos\ for each available observation and assess whether it changes significantly across epochs. As a first step, we check whether the \nhlos\ values from different observations are consistent within their  uncertainties. For example, if the value obtained in the first observation lies within the uncertainty range of the second, and both of these are also consistent with the third, then the source is indeed non-variable. If, instead, at least one \nhlos\ measurement falls outside the uncertainties of the others, variability is suggested. 
Subsequently, to statistically evaluate variability, we apply the two independent methods defined in \citetalias{torres23}.

\begin{itemize}
    \item \textbf{Reduced statistic method} \\
    In this approach, the reduced $\chi^{2}$ of the best-fit model (which allows \nhlos\ and intrinsic flux variability) is compared to constrained models under three assumptions: 1) no variability at all, 2) only $N_{\mathrm{H,los}}$ variability, 3) only intrinsic flux variability. A \emph{tension parameter} ({\it T}) is then defined as:
    \[
    T = \frac{|1 - \chi^2_{\mathrm{red}}|}{\sigma}, 
    \]
    where $\sigma$ = $\frac{1}{\sqrt{N}}$ and $N$ is the number of degrees of freedom. A source is classified as ``Variable'' if the best-fit model including $N_{\mathrm{H,los}}$ variability has $T < 3$, while the fit assuming non-variability yields $T > 5$. If both models yield $T < 3$, the source is considered ``Non-variable'', since $N_{\mathrm{H,los}}$ variability is not required. When the torus models disagree or classifications are inconsistent, the source is labeled ``Undetermined''. In cases where the C-statistic is applied, we consider reduced statistic instead of $\chi^2_{\mathrm{red}}$. 
    We note that the interpretation is less straightforward since the distribution does not strictly follow a Gaussian function.
    This test therefore provides a way to evaluate whether the best-fit model is statistically satisfactory and whether an equally good fit could be obtained without invoking $N_{\rm H,los}$ variability. In practice, this helps in robustly assessing the number of variable sources.
    \medskip
    \item \textbf{p-value method} \\
    This method tests the null hypothesis that no $N_{\mathrm{H,los}}$ variability exists across observations of the same source. By comparing the best-fit $N_{\mathrm{H,los}}$ values\footnote{We note that the $N_{\mathrm{H,los}}$ values adopted here are taken from the best-fit model accounting for both column density and intrinsic flux variability, and thus are independent of the assumptions made in the tension test.} at different epochs with their average, we can define:
    \[
    \chi^{2} = \sum_{i=1}^{n} \frac{\big(N_{\mathrm{H,los},i} - \langle N_{\mathrm{H,los}} \rangle \big)^{2}}{\delta(N_{\mathrm{H,los},i})^{2}}
    \]
    where $N_{\mathrm{H,los},i}$ is the line-of-sight column density at epoch $i$, $\langle N_{\mathrm{H,los}} \rangle$ is the mean value across epochs, and $\delta(N_{\mathrm{H,los},i})$ is the measurement uncertainty (using asymmetric errors when appropriate). The $\chi^{2}$ is then converted into a \emph{p-value}. If the p-value is $\leq 0.01$ for all the three torus models, the AGN is classified as ``Variable''. If the p-value is above $0.01$ for all models, it is considered ``Non-variable''. Disagreement among the models (some above, some below the threshold) leads to an ``Undetermined'' classification.
\end{itemize}

In Appendix~\ref{source_note}, we report the classification of the sources across different wavelength bands (if the information is available in the literature), and we discuss the adopted classification for Mrk~1498 and IC~5063.
The {\it tension} and {\it p-value} results for each AGN are reported at the end of Tables~\ref{tab:spec_1} -- \ref{tab:spec_11}.

\section{Torus model comparison}\label{model_comp}
\begin{figure*}[]
\centering
\includegraphics[width=0.9\textwidth, trim=260 0 300 10, clip]{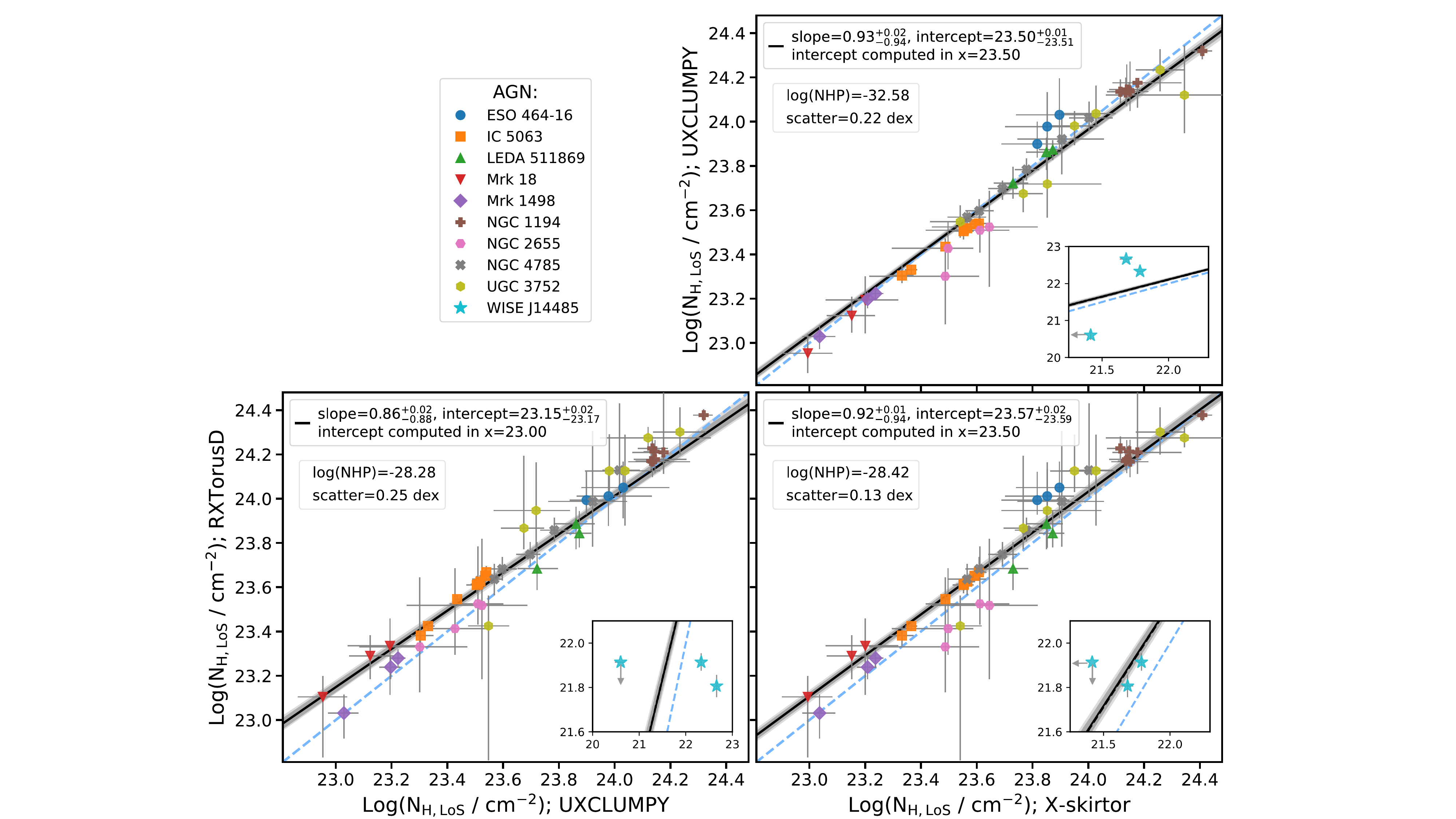}
    \caption[]
    {{\nhlos\ comparison: \nhlos\ best-fit values derived from \rxt, \ux, and \skirt\ models for our AGN sub-sample. Insets highlight the low-density regime, populated by WISE~J144850, whose \nustar\ observation shows upper limits (gray arrows) in the \rxt\ and \skirt\ models. The dashed light blue lines correspond to y = x. The solid black lines represent the best-fitting linear correlation and the dark and light gray shadowed areas indicate the 68\% and 90\% confidence bands, respectively. In the legend, we report the best-fit coefficients, log NHP, and the intrinsic scatters for the correlations. The different models generally provide consistent \nhlos\ estimates for most observations, with the largest deviations occurring in sources with peculiar obscuration behavior (UGC~3752) or in the low-obscuration regime (WISE~J144850).}}
    \label{fig:3mod_comp}
\end{figure*}
In this section, we compare the results from the three different models. Since \rxt\ and \skirt\ are newly developed compared to \ux, a direct comparison is important to assess their mutual consistency and reliability. 
To provide a preliminary assessment of how model estimates vary across different observations of the same source and to highlight potential discrepancies between models, Fig.~\ref{fig:3mod_comp} presents a comparison of the \nhlos\ values derived from the three models for each observation in our AGN sample.
Each panel compares a pair of models, displaying all sources and epochs together with their corresponding uncertainties. From the scatter plots, the best-fit values show a strong, statistically significant correlation and overall good agreement among the models. 
From a visual inspection, \rxt\ appears to yield slightly higher \nhlos\ at low column densities, while \ux\ and \skirt\ lie closer to the 1:1 line. This difference could potentially be related to the inclusion of dust in the former torus model, which is not accounted for in the other two. Statistically, however, the differences between all three models are below 1$\sigma$, confirming consistency with the 1:1 relation.
From the visual inspection of Fig.~\ref{fig:3mod_comp}, we find that, for most sources, the different observations cluster within a similar range of \nhlos\ values. The sole exception is UGC~3752, consistent with our classification of this source as a candidate changing-look/obscuration AGN (see Sect.~\ref{ctk_cth_class}).
We quantified these correlations through the Spearman test, whose rank and p-value evaluate the strength and significance of monotonic trends between parameters. To properly account for asymmetric uncertainties, we adopted the perturbation method described in \citet{gianolli2024}, which also allows estimating the linear regressions and corresponding uncertainty between model pairs. 
We report in Fig.~\ref{fig:3mod_comp} the linear regression best-fit parameters, Log(NHP), and intrinsic scatter for each of the model pair.

To further assess the consistency of the models within each variability class and to obtain an overall view of their behavior, we compute the pairwise normalized differences:

\[
Z = \frac{|m_1 - m_2|}{\sqrt{\sigma_1^2 + \sigma_2^2}},
\]

where m$_i$ and $\sigma_i$ are the median and uncertainty estimated. We define two model estimates as inconsistent when Z $\geq$ 2.6, which corresponds approximately to a probability threshold of 0.01 for a Gaussian distribution, consistent with the significance criterion adopted throughout this work.
Concerning the median values of the \nhlos\ for each source and model, we find that eight of the ten AGN show mutual agreement among all models, suggesting that the choice of torus model has little impact on the derived obscuration parameters.
Only IC~5063 and WISE~J14485 display significant inter-model discrepancies, indicating that their derived \nhlos\ values depend more strongly on the assumed torus geometry. In particular, IC~5063 median column densities differ by up to $\sim$4$\sigma$ (maximum Z = 4.11). The value derived by adopting \rxt\ is significantly higher than both alternative models, and the strongest discrepancy occurs between \ux\ and \rxt, with a median difference of 8.53. Interestingly, if we consider other spectral parameters (i.e., the photon index and normalization, and thus the AGN luminosity), \ux\ appears most discrepant relative to \skirt\ and \rxt. This suggests that IC~5063 is a complex source, possibly affected by multiple factors \citep[e.g., spectral degeneracies or partial jet contribution; see][for a discussion on the jet emission]{travescio2021}, and that different combinations of model parameters can reproduce the observed spectrum, leading to the observed spread in \nhlos. 
Interestingly, from Fig.~\ref{fig:3mod_comp}, Mrk~18 appears to deviate from the 1:1 line to a similar extent as IC~5063, suggesting a comparable discrepancy. However, the Z-test does not flag Mrk~18 as significantly discrepant. This is likely due to the larger uncertainties in the \nhlos\ measurements for Mrk~18, which effectively encompass the 1:1 relation and reduce the significance of any apparent offset.
Even more significant model-to-model discrepancies are present in WISE~J14485, with Z = 14.86. 
Here, \ux\ yields a significantly larger median \nhlos. 
However, we note that during the fitting process, only \ux\ was able to constrain the \nhlos\ for the third observation (taken with \nustar), whereas the other two models yielded only upper limits. This AGN was observed in a weakly absorbed state, with inferred obscuration ranging from 0.43 to 5 $\times$ 10$^{22}$ cm$^{-2}$ (when constrained). Hence, this outcome likely reflects the challenge of modeling sources in a low-obscuration regime with \skirt\ and \rxt\ torus models.

In addition, we applied the same analysis to the other key spectral and geometrical parameters constrained by the three torus models (photon index $\Gamma$, fraction of scattered continuum F$_\mathrm{s}$, and torus covering factor C$_\mathrm{f}$ (considering only \rxt\ and \skirt\ results, see Sect.~\ref{sec:uxc}), inclination angle $\theta$, and equatorial column density N$_{\rm H,eq}$). For $\Gamma$, we find overall good agreement among the models for most sources, with significant discrepancies emerging only for IC~5063 for which a higher value is inferred with \ux, suggesting that differences in the treatment of the transmitted continuum and scattering components can affect the spectral slope.
A stronger inter-model tension appears for the covering factor. Here, half of the sample displays statistically significant discrepancies. The inclination angle is often only weakly constrained. NGC~1194 shows the higher significant difference, with \rxt\ preferring lower inclination angles than the the other two models. These mismatches likely reflect the fundamentally different geometric assumptions encoded in each torus framework: \ux\ prescribes a clumpy medium, whereas \skirt\ and \rxt\ impose smooth or quasi-smooth angular density distributions. As a result, the same spectra can be reproduced with very different combinations of covering factor, cloud distribution, and equatorial opacity, leading to large model-dependent shifts.

IC~5063 and WISE~J144850 also show strong model-to-model inconsistencies in the scattered-fraction values. Additionally, a significant difference is also present in NGC~4785, for which \ux\ results to be higher than \skirt\ and \rxt\ values, which instead are in agreement.
Differences in the equatorial column density are seen in LEDA~511869, Mrk~18, NGC~1194, NGC~2655, and UGC~3752. In the cases of LEDA~511869, NGC~2655, and UGC~3752 the difference is between 0.5 and 1.5. While for Mrk~18 and NGC~1194, the difference is small, less than 4\% and 12\%, respectively.

Overall, while the line-of-sight absorption appears robust against torus-model choice for the majority of sources, several secondary parameters are significantly model dependent. This highlights the importance of interpreting such inferred torus properties in the context of the specific geometrical assumptions of each model.

\section{X-ray and variability classifications}\label{sec:res}

In the following sections, we discuss the results obtained for 10 of the 11 AGN of the sub-sample. Given that MCG‑03‑34‑064 cannot be fit under our assumptions (constant photon index and torus geometry) and exhibits significant residuals in the soft X-rays and Fe K band, it requires a separate fitting procedure and is thus excluded from the statistical and physical analysis below. A complete discussion of the fitting procedure and results for MCG-03-34-064 is provided separately in Sect.~\ref{mcg}.

\subsection{Compton-thin and -thick classification}\label{ctk_cth_class}

Considering the results of our fitting procedure, reported in Tables~\ref{tab:spec_1}-\ref{tab:spec_11}, we classify 9 out of 10 AGN as Compton-thin, while NGC 1194 is confirmed as Compton-thick \citep[in agreement with previous studies; e.g.,][]{kuo2011,severgnini2012,marchesi2018,Turner2020}. We adopt a conservative classification based on overall behavior across epochs and models, rather than on isolated upper limits. In particular, UGC~3752 is identified as a candidate changing-obscuration AGN, with the line-of-sight column density exceeding the Compton-thick threshold (\nhlos $>$ 1.5 $\times$ 10$^{24}$ cm$^{-2}$) in multiple epochs and models (see Table~\ref{tab:spec_10}).
ESO~464-G016 and NGC~4785 remain classified as Compton-thin as their \nhlos\ show upper limits in the CTK regime only in a single observation and model (see Tables~\ref{tab:spec_1} and \ref{tab:spec_9}, respectively). 
NGC~1194 shows evidence for possible transitions between Compton-thin and Compton-thick states on timescales of roughly 1–2 years, although the uncertainties remain consistent with a persistently CTK classification in some models (see Table~\ref{tab:spec_7} and Fig.~\ref{fig:1194_plots} panel b). NGC~1194 is currently being monitored with \nustar+XMM (proposal 11159, PI: S. Marchesi) to study weeks-to-months \nhlos\ variability, probing the obscuring medium structure and AGN feeding/feedback processes. We report the X-ray classification in Table~\ref{tab:1_properties}. Further details on the individual sources of the sample are provided in Appendix ~\ref{source_note}.

\begin{deluxetable*}{c||c||c|c|c|c|c|c|c||c|c||c}
\tablecaption{Classification summary. \label{tab:clas_summary}}
\tablehead{
\colhead{\scriptsize (1)} & \colhead{\scriptsize (2)} & \colhead{\scriptsize (3)} & \multicolumn{2}{c}{\scriptsize (4)} &
\multicolumn{2}{c}{\scriptsize (5)} & \multicolumn{2}{c}{\scriptsize (6)} & \multicolumn{2}{c}{\scriptsize (7)} & \colhead{\scriptsize (8)} \\
\colhead{Source} & \colhead{\# obs} & \colhead{1st order} & 
\multicolumn{2}{c}{\skirt} & \multicolumn{2}{c}{\rxt} & \multicolumn{2}{c}{\ux} & 
\multicolumn{2}{c}{Comparison} & \colhead{Is N\textsubscript{H,los} variable?} \\
\colhead{} & \colhead{} & \colhead{} & \colhead{T} & \colhead{p-value} & \colhead{T} & \colhead{p-value} & \colhead{T} & \colhead{p-value} & \colhead{T} & \colhead{p-value} & \colhead{}
}
\startdata
ESO~464-G016 & 3 & NV & U & NV & U & NV & NV & NV & \cellcolor{whitecell}? & \cellcolor{redcell} & \cellcolor{whitecell} ? \\
LEDA 511869 & 3 & NV & NV & NV & NV & NV & NV & NV & \cellcolor{redcell} & \cellcolor{redcell} & \cellcolor{redcell} \\
Mrk~1498 & 3 & V & V & V & V{\bf *} & V & V{\bf *} & V & \cellcolor{greencell} & \cellcolor{greencell} & \cellcolor{greencell} \\
WISE~J144850.99-400845.6 & 3 & U{\bf *} & U & V & U & V & U & V & \cellcolor{whitecell}? & \cellcolor{greencell} & \cellcolor{whitecell} ? \\
Mrk~18 & 4 & NV & NV & NV & NV & NV & U & NV & \cellcolor{redcell} & \cellcolor{redcell} & \cellcolor{redcell} \\
NGC~2655 & 6 & NV & NV & NV & NV & NV & NV & NV & \cellcolor{redcell} & \cellcolor{redcell} & \cellcolor{redcell} \\
NGC~4785 & 7 & V & V & V & V & V & V & V & \cellcolor{greencell} & \cellcolor{greencell} & \cellcolor{greencell} \\
UGC~3752 & 7 & V & V & V & V & V & V & V & \cellcolor{greencell} & \cellcolor{greencell} & \cellcolor{greencell} \\
IC~5063 & 8 & V & V & V & U{\bf *} & V & U{\bf *} & V & \cellcolor{greencell} & \cellcolor{greencell} & \cellcolor{greencell} \\
NGC~1194 & 8 & V & V & V & V & V & V & V & \cellcolor{greencell} & \cellcolor{greencell} & \cellcolor{greencell} \\
MCG-03-34-064 {\bf +} & 9 & V & / & / & / & / & / & / & / & / & / \\
\enddata
\tablecomments{{\it Columns:} 
1) Number of observations; 
2) First-order variability classification, based on whether \nhlos\ values from different observations overlap within uncertainties; 
3--6) classifications from the two statistical tests, reported separately for each of the three adopted models; 
7) variability summary across models for each test; 
8) final classification combining the results of the two tests (T and p-value) reported in column 7. 
Gray with a question mark (or ``U''), red (or ``NV''), and green (or ``V'') indicate ``Undetermined'', ``Non-variable'', and ``Variable'', respectively.\\
{\bf *} see Appendix~\ref{source_note} with notes on each AGN. \\
{\bf +} refer to Sect.~\ref{mcg} for the classification of the source.}
\end{deluxetable*}
\subsection{\nhlos\ and intrinsic flux variability classification}\label{var_classification_res}

To assess the variability classification of each AGN (see Sect.~\ref{var_class}), we first compare the \nhlos\ values from individual observations to evaluate whether they are consistent within their uncertainties. The results are reported in Table~\ref{tab:clas_summary}, column 3. For ESO~464-G016, LEDA~511869, Mrk~18, and NGC~2655 (4 out of 10 AGN, $\sim$40\%) the best-fit \nhlos\ values from different epochs lie within uncertainties, indicating that no significant variability can be inferred. In contrast, for Mrk~1498, NGC~4785, UGC~3752, IC~5063, and NGC~1194 (5 out of 10 AGN, $\sim$50\%), significant variability in \nhlos\ on timescales of days, months, and/or years is observed. Thus, identifying these AGN as good candidates for variable sources. WISE~J144850.99-400845.6 (hereafter ``WISE~J144850'') shows an intermediate behavior, as only upper limits on \nhlos\ could be derived from the \nustar\ observation when adopting \skirt\ and \rxt\ models.

We then statistically classify each AGN as \nhlos Variable or Non-variable using the two tests described in Sect.~\ref{var_class}. For each source, the {\it T} and {\it p-value} tests are applied to the best-fit results from the three spectral models, \skirt, \rxt, and \ux, yielding six determinations in total.
 The results are reported in Table~\ref{tab:clas_summary}, columns 4--6. For each test independently, if at least two of the three model-based determinations agree on Variable or Non-variable (V/NV), that classification is assigned to the corresponding comparison column (column 7); otherwise, the result is marked as Undetermined (U). The final classification combines the two test-comparison results: if both agree on V or NV, the AGN is assigned that class; if they disagree or at least one is U, the final class is Undetermined (column 8).
This procedure yields 2 out of 10 Undetermined AGN (ESO~464-G016 and WISE~J144850), 3 out of 10 Non-variable AGN in \nhlos (LEDA~511869, Mrk~18, and NGC~2655), and 5 out of 10 Variable AGN in \nhlos (Mrk~1498, NGC~4785, UGC~3752, IC~5063, and NGC~1194). 
Overall, for 9/10 sources, the statistical tests confirm the first-order classification. The only discrepancy is ESO~464-G016, classified as NV by the {\it p-value} test and U by the {\it tension} test, leading to a final Undetermined classification (see Table~\ref{tab:clas_summary}).
Finally, most sources in the sample exhibit intrinsic flux variability. When neither \nhlos\ nor intrinsic flux variability is imposed, 7 out of 10 AGN show significantly worse best-fits (see the ``No variability'' tests in Appendix~\ref{bestfit}). However, for ESO~464-G016, LEDA~511869, and NGC~4785, the tensions remain comparable across hypotheses or reveal degeneracies between flux and \nhlos\ variability, preventing a firm assessment (see Appendix~\ref{source_note} for more details).

These results highlight a key aspect of our study: the initial classification, based only on two observations, one in the soft X-ray band and one in the hard X-ray band, is insufficient to properly assess variability.
In addition, we note that, out of the 22 AGN for which \citet{zhao2021} considered simultaneous observations (see Sect.~\ref{sec:data}), our sample includes three: NGC~4785, UGC~3752, and MCG-03-34-064. While we refer to Sect.~\ref{mcg} for MCG-03-34-064, our analysis reclassifies NGC~4785 and UGC~3752 as Variable sources. This reflects the larger multi-epoch dataset analyzed here and highlights the need to account for spectral variability, both in \nhlos\ and intrinsic flux.

Consistent with this interpretation, the detection of \nhlos\ variability appears to depend strongly on observational sampling. In our sample, all targets with at least seven observations are found to be \nhlos\ variable, as shown in Table~\ref{tab:clas_summary}. The only exception is Mrk~1498, which shows \nhlos\ variability despite having only three observations. However, this threshold may differ from source to source and may depend on whether the variability occurs on short or long timescales.
\begin{figure}
\centering
    \includegraphics[width=0.47\textwidth, trim=6 10 0 0, clip]{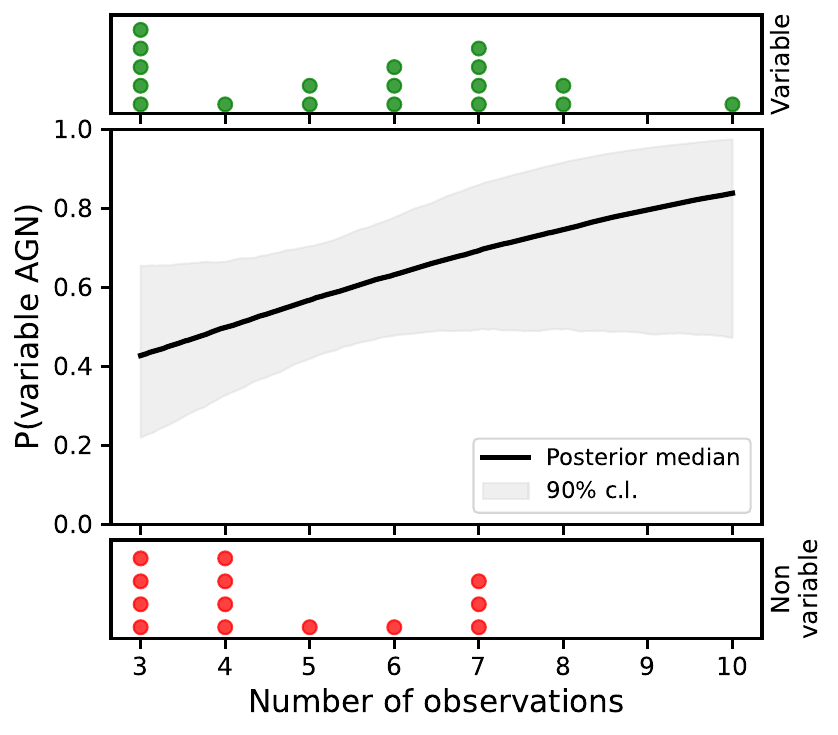}
    \caption[]
    {{Probability of detecting \nhlos\ variability as a function of the number of available X-ray observations, considering the 10 AGN analyzed here and the sources analyzed by \citetalias{pizzetti2022}-\citetalias{torres2025}. Green and red points show, respectively, AGN classified as Variable and Non-variable. The black solid line represents the posterior median of a Bayesian logistic regression fitted to the binary variability outcomes, and the gray shaded regions show the corresponding 90\% confidence bands. The posterior median probability increases with $N_{\rm obs}$, with average slopes of $0.06$ and $0.05$ per observation over the $N_{\rm obs}=3$--5 and $N_{\rm obs}=5$--10 ranges, respectively. The increase is visually most apparent around $N_{\rm obs}\sim5$--6, indicating that denser multi-epoch X-ray coverage improves the detectability of obscuration changes.}}    
    \label{fig:bayesian}
\end{figure}
To statistically test whether and how the probability of detecting \nhlos\ variability depends on the number of available X-ray observations, we model variability detection as a binary outcome using a Bayesian logistic regression \citep{McElreath2020}. For each AGN, variability is therefore treated as a simple yes/no outcome, i.e., either detected or not detected, and we infer the probability of detecting variability for a given number of observations, $P(\mathrm{variable}\,|\,N_{\rm obs})$.
Because this probability must lie between 0 and 1, we describe its dependence on $N_{\rm obs}$ using a logistic function. This provides a continuous empirical description of how the likelihood of detecting variability changes with observational sampling, rather than imposing an arbitrary threshold in $N_{\rm obs}$. We adopt a Bayesian framework to account for the small sample size and to infer a posterior distribution of plausible probability curves (instead of a single best-fitting curve).
We use a prior that does not favor any particular probability, so that the inferred relation is driven primarily by the observed distribution of Variable and Non-variable sources.
We emphasize that this analysis probes the detectability of variability given a finite number of observations, rather than the intrinsic fraction of variable AGN.
Since variability is encoded as detected (Variable source) or not detected (Non-variable source), Undetermined sources (i.e., sources with ambiguous classifications) are excluded from this analysis. Given the limited size of our sub-sample, we apply the Bayesian analysis to the expanded sample, combining the AGN analyzed in this work with the pre-selected variable sample (\citetalias{pizzetti2022}-\citetalias{torres2025}). 
The posterior median relation for the expanded sample is shown in Fig.~\ref{fig:bayesian}. AGN with more observations are increasingly dominated by Variable sources, whereas sources with fewer observations include a mixed population of Variable and Non-variable AGN. This increase is visually most apparent around $N_{\rm obs}\sim5$--6. We note that, although the posterior median trend is positive with the probability for a positive slope of $P({\rm slope}>0)=0.91$), the uncertainty on the slope remains large. 
Therefore, while the data suggest an increasing trend, this result should be regarded as tentative and may change as more AGN are analyzed.

A similar trend is reported by \citet{cox2025}, who performed a systematic search for \nhlos\ variability in nearby AGN using the \chandra\ archive, further supporting the evidence that expanded multi-epoch coverage enhances our ability to detect obscuration changes. This consistency suggests that the dependence on observational sampling reflects a genuine behavior of multi-epoch X-ray AGN data, rather than a feature specific to our classification method. Quantifying this dependence therefore provides an empirical benchmark for variability studies and clumpy-obscurer models.

\section{$\Delta$N$_{\rm H,LoS}$ time evolution}\label{evol_nh_with_t}
\begin{figure*}
\centering
\includegraphics[width=\textwidth, trim=0 10 0 0, clip]{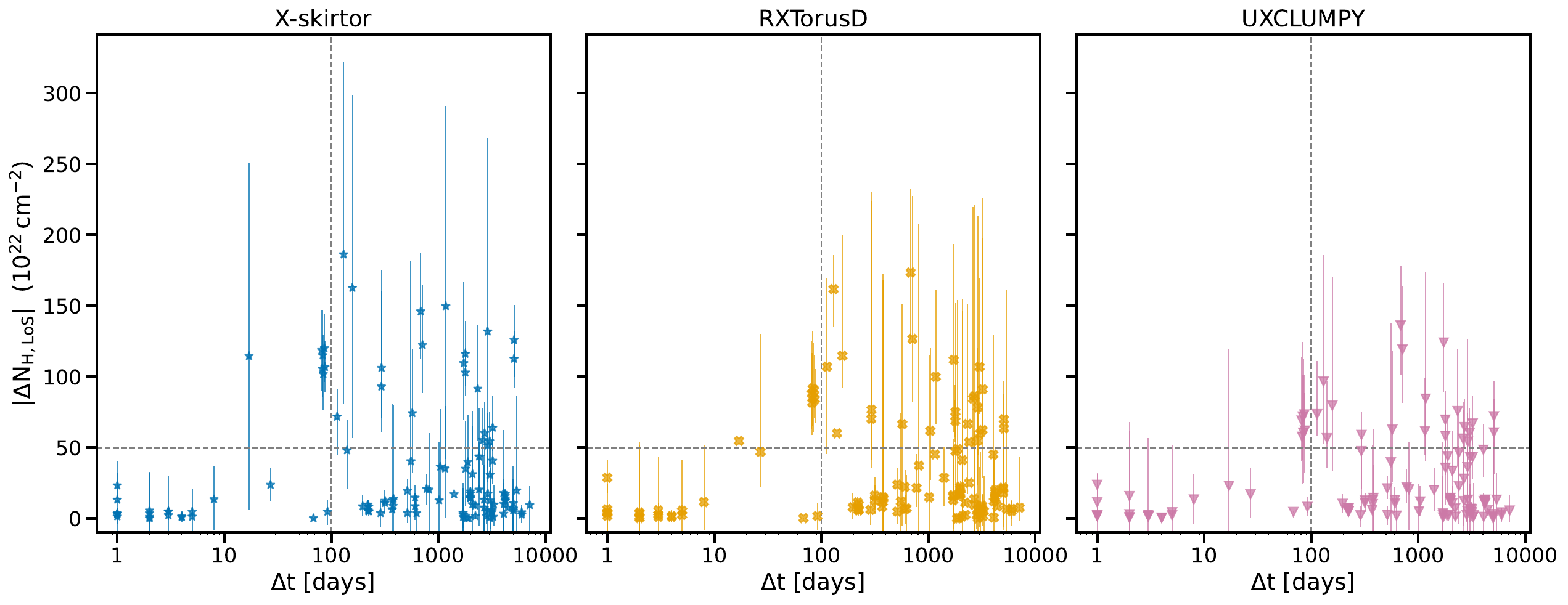}
    \caption[]
    {{Distribution of $\Delta$\nhlos\ as a function of time interval ($\Delta$t) for the sample.  The amplitude of \nhlos\ variability is timescale-dependent: small variations in $\Delta$N$_{\rm H,LoS}$ are observed at all timescales, whereas large changes ($\gtrsim$ 5 $\times$ 10$^{23}$ cm$^{-2}$), occur predominantly at long $\Delta$t ($>$100 days). The dashed lines show the thresholds adopted for changes in the column density (horizontal line) and time intervals (vertical line), see \citetalias{torres23}.
    The blue, yellow, and pink sub-panels correspond to \skirt, \rxt, and \ux, respectively. For all three models, the KS-tests between the short- and long-term distributions confirm a significant difference, with Log(NHP) = -2.97 for \skirt; Log(NHP) = -4.39 for \rxt; and Log(NHP) = -2.06 for \ux.}}
    \label{fig:large_nh_large_t}
\end{figure*}
In this section, we discuss the derived \nhlos\ and the epoch-to-epoch changes in column density ($\Delta$N$_{\rm H,LoS}$) as a function of time. We computed the epoch-to-epoch changes in column density as $\Delta$N$_{\rm H,LoS}$ = N$_{\rm H,LoS}^{i+1}$ - N$_{\rm H,LoS}^{i}$, and propagated the asymmetric uncertainties via quadrature of the appropriate adjacent errors.  
An important aspect of interpreting absorption variability is the characteristic timescale of the change. In a Keplerian potential, short-timescale (i.e., hours to weeks) variations are expected to originate from compact, rapidly moving clouds located at radii comparable to, or smaller than, the BLR, whereas variability on longer timescales (i.e., months to years) is associated with material located at larger radii, such as torus clumps \citep[e.g.,][]{risaliti2002,Markowitz2014}. In general, large $\Delta$N$_{\rm H,LoS}$ events imply denser or more extended structures, while small changes can indicate partially covering clouds or more distant material \citep[see][for a review]{netzer15}.

A useful way to visualize this behavior is to plot the variation in column density as a function of the timescale normalized by the black hole mass, $\Delta$t/M$_{\rm BH}$. Given that dynamical timescales scale with the mass, this normalization allows different AGN to be compared.
For each of the ten AGN in our sample, Appendix~\ref{bestfit} shows the time evolution of \nhlos\ and the plots with $\Delta$\nhlos\ versus $\Delta$t/M$_{\rm BH}$ (middle and lower panels, respectively, of Figs.~\ref{fig:eso_plots}–\ref{fig:wise_plots}). 
To provide a qualitative check on the possible physical scale of the absorber, we perform a simple cloud-crossing test under the assumptions of Keplerian motion and coherent obscuring structures. For each pair of consecutive epochs, we compared the observed variability timescale with the characteristic timescales expected for BLR- and torus-scale absorbers. The results suggest that the observed variability is not associated with a single characteristic scale (i.e., BLR-scale or torus-scale material). However, this interpretation relies on simplifying assumptions that may not hold in a clumpy, turbulent, or multi-phase circumnuclear medium. In particular, the absorber may not consist of long-lived individual clouds crossing the line of sight, and non-Keplerian motions such as turbulence, inflows, or outflows may also contribute \citep[e.g.,][]{Wittor2020}. Moreover, given that the AGN are not continuously monitored, our conclusions are necessarily limited by the available sampling. We therefore interpret the inferred locations only as indicative and as a qualitative consistency check indicating that both compact inner absorbers and larger-scale circumnuclear material could contribute to the observed variability.

In \citetalias{torres23} and \citetalias{pizzetti2025}, the authors report that large changes in LoS column density (i.e., $\Delta$N$_{\rm H,LoS}$ $\gtrsim$ 5 $\times$ 10$^{23}$ cm$^{-2}$) are preferentially associated with longer time intervals, while smaller variations occur across all timescales. To investigate this, we consider the computed $\Delta$N$_{\rm H,LoS}$ (between all epochs for each source and model) as a function of the difference in time for our sample. The results are shown in Fig.~\ref{fig:large_nh_large_t} (left panel for \skirt, middle panel for \rxt, and right panel for \ux), where the time intervals are divided into short ($\Delta$t $<$ 100 days) and long ($\Delta$t $>$ 100 days) epochs.
We observe that small changes in \nhlos\ are present at all timescales, whereas large variations occur only over longer time separations also for our sample. We run KS tests between the short- and long-term distributions, confirming significant differences between them (\skirt: Log(NHP) = -2.97; \rxt: Log(NHP) = -4.39; \ux: Log(NHP) = -2.06).
Depending on the size and distance of the absorbing clouds, this result could have different interpretations. For extended clouds, it may reflect a density profile, with density increasing toward the cloud centers, as also inferred by \citetalias{torres23}. Alternatively, if the clouds are roughly uniform, short-term observations would sample the same cloud, while longer intervals between observations could map a different cloud, resulting in a markedly different column density and, consequently, a larger $\Delta$\nhlos. A third interpretation, motivated by hydrodynamical simulations, suggests that the dominant structures are extended cloud or filament bundles shaped by turbulence, collisions, and shear \citep[e.g.,][]{gaspari2017,gaspari2020}. A further possibility is that the line of sight intersects multiple clouds spanning a range of scales and velocities. In this picture, slow column density variations from a large-scale structure (e.g., the torus) are modulated by rapid fluctuations from smaller, faster inner clouds. Alignment of several components would temporarily increase the \nhlos, producing short-timescale variability superimposed on longer-term changes.
We note, however, that our sample is biased towards longer timescales as those in \citetalias{torres23} and \citetalias{pizzetti2025}. In particular, although short intervals ($\Delta$t $<$ 100 days) are present in several sources (NGC~4785, UGC~3752, IC~5063, NGC~1194, Mrk~18, NGC~2655, ESO~464-G016, and WISE~J144850), the majority of intervals (77\%) are long ($\Delta$t $>$ 100 days).
Adopting a different methodology, i.e., a population-based approach to identify \nhlos\ variability, \citet{cox2025} independently reinforce our results and those of \citetalias{torres23,pizzetti2025}. In particular, the authors assess variability through a $\chi^2$ statistic computed between spectral pairs as a function of time separation, rather than by directly measuring $\Delta$N$_{\rm H,LoS}$. By dividing the sample into short and long timescales (adopting the same $\Delta t$ threshold as used here), they find that the fraction of pairs flagged as variable increases from $\sim$10\% at short timescales to $\sim$24\% at long timescales. Although this method does not allow the variability amplitude to be directly quantified, since the $\chi^2$ statistic depends on both data quality and intrinsic variability, the conclusion is consistent: large \nhlos\ variability is more prevalent over longer timescales.

\subsection{Torus model comparison: $\Delta$N$_{\rm H,LoS}$ values}\label{model_comp1}

In this section, we expand the torus model comparison by focusing on temporal consistency within each model and agreement across models.
To assess the former, we consider the interval
$[\Delta$N$_{\rm H,LoS}^i$ - $\sigma^-_{\Delta,i}$, $\Delta$N$_{\rm H,LoS}^i$ + $\sigma^+_{\Delta,i}]$.  If consecutive $\Delta$N$_{\rm H,LoS}$ intervals do not overlap, this indicates a significant change in column density within a single model. To test agreement between models, we computed the difference
$\delta_i$ = $\Delta$N$_{\rm H,LoS}^{\rm Mod \, A}$ - $\Delta$N$_{\rm H,LoS}^{\rm Mod \, B}$, propagating the associated asymmetric uncertainty. If the resulting interval contains zero, the two models are considered consistent, meaning that any observed difference could plausibly be explained by measurement uncertainties.

We find that all the Non-variable AGN (i.e., LEDA~511869, Mrk~18, and NGC~2655) along with the Undetermined ESO~464-G016, show no between-model discrepancies, although LEDA~511869 exhibits consecutive-delta non-overlaps with \skirt. We refrain from discussing WISE~J144850, as only \ux\ provides well-constrained \nhlos\ values; thus, any observed discrepancy found would reflect differences in constraining power rather than genuine model conflict. 
By applying the same criteria to the Variable sub-sample, we instead obtain the following results:

\begin{itemize}
    \item IC~5063, the within-model comparisons reveal that the first two intervals (obs 1-2 vs 2-3, and 2-3 vs 3-4) as well as the last (obs 6-7 vs 7-8) do not overlap for all three models. The remaining consecutive intervals do. Between-model comparisons confirm agreement across nearly all intervals, except for the final interval when comparing \rxt\ and \ux\, which shows a disagreement; 
    \item Mrk~1498 shows clear non-overlap between individual observations within the same model. However, when considering the full \nhlos\ ranges (including uncertainties) across all models, the $\Delta N_{\rm H,LoS}$ values remain consistent; 
    \item NGC~1194 exhibits the clearest model-dependent tension. In particular, \skirt\ predicts a substantially larger $\Delta N_{\rm H,LoS}$ for obs 3-4 compared to those derived from \rxt\ and \ux. Overall, this source exhibits the largest number of non-overlapping consecutive intervals among all sources, highlighting strong short-term variability. Between-model comparisons indicate agreement for most intervals; however, clear disagreements appear also for obs 4-5 when comparing \skirt\ with \ux;
    \item In NGC~4785, the first interval (obs 1–2) does not overlap in any model, indicating a significant change. In contrast, subsequent intervals mostly overlap within each model, except for \ux\ where obs 3–4 vs 4–5 show no overlap, suggesting that variations in column density between later observations are generally consistent within the uncertainties of each model;
    \item UGC~3752 fifth interval (obs 5-6) shows no overlap for any of the three models, while the other intervals overlap within their respective uncertainties.
\end{itemize}

\begin{figure*}
\centering
    \includegraphics[width=\textwidth, trim=0 10 5 0, clip]{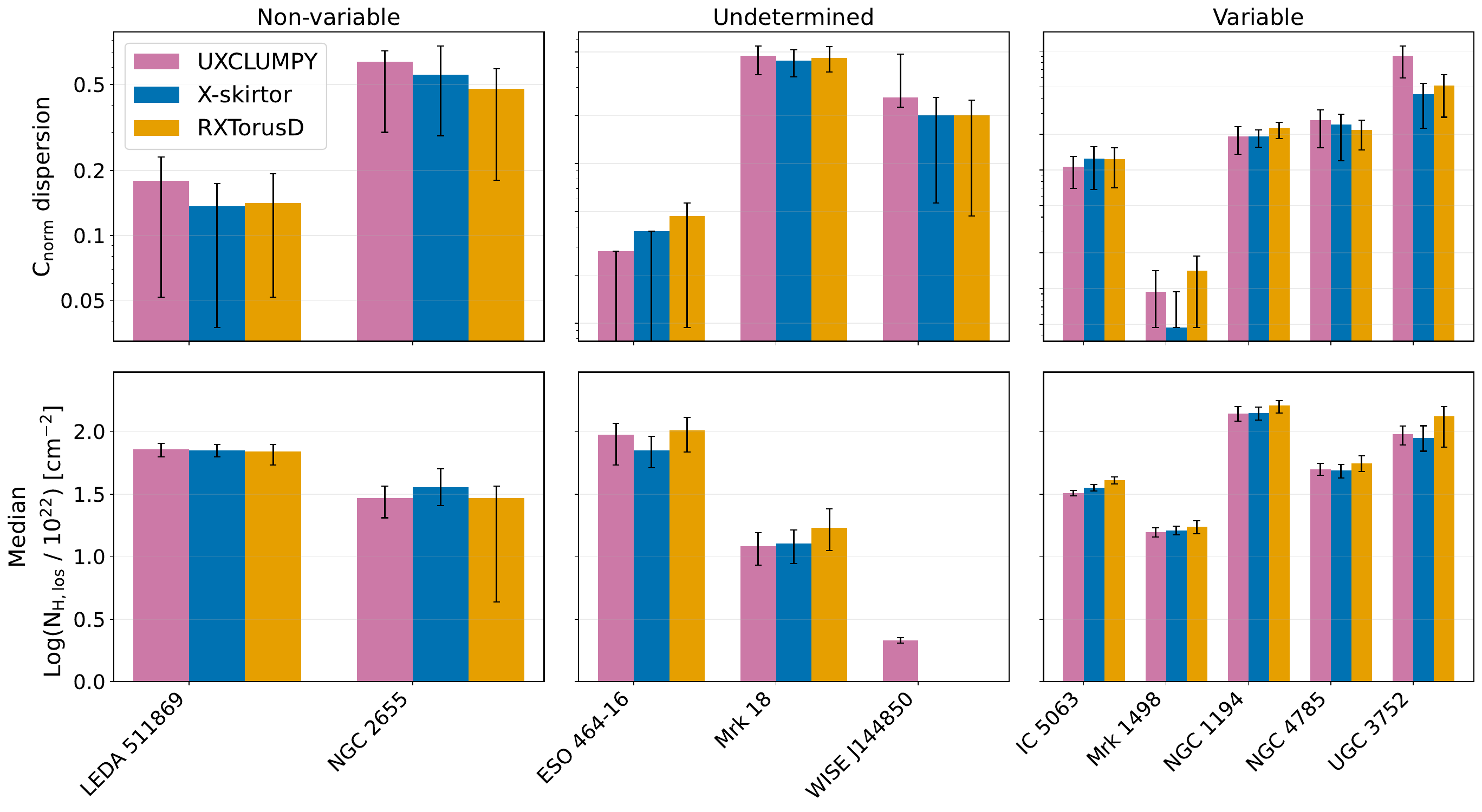}
    \caption[]{Distributions of the dispersion for the cross-normalization constants and the logarithm of the median \nhlos\ values, grouped by variability class. 
    Results for the three adopted torus models are shown as: \ux\ in pink, \skirt\ in blue, and \rxt\ in yellow, including asymmetric uncertainties at 90\% c.l..
    {\it Left panels} present the non-variable, {\it middle panels} the undetermined, and {\it right panels} the variable AGN samples.
    The middle panel of the lower row show only the best-fit result for \ux\ for WISE~J14485, as the other two models provide only upper limits.
    Within the limitations of the current small and heterogeneous sample, the two distributions do not exhibit significant differences among variability classes.}
    \label{fig:distr_3models}
\end{figure*}
Regarding the temporal evolution of $\Delta N_{\mathrm{H,LoS}}$ between models, we find that although the absolute values may differ, the overall trend is largely consistent. This evolution as a function of $\Delta t / M_{\rm BH}$ is shown in the lower panels of Figs.~\ref{fig:eso_plots}–\ref{fig:UGC_plots}.

Our analysis shows that between-model differences are rare, occurring in only $\sim$4\% of all intervals, and primarily involving \skirt\ and \ux. Discrepancies are most evident in AGN with strong short-term variability (e.g., NGC~1194), while Non-variable and Undetermined sources generally show agreement across all models. We note, however, that systematic shifts between models could introduce shifts in the \nhlos\ values, so overlapping ranges across all models do not necessarily imply true non-variability. In summary, while the choice of model affects the amplitude of inferred column density variations, the classification of AGN as Variable or Non-variable remains robust, indicating that the observed trends largely reflect intrinsic source behavior rather than systematic biases. 

\section{Statistical properties}\label{stat}

Considering the best-fit results of the 10 AGN in our sample, we analyze the statistical properties of the model parameters by examining their distributions and potential correlations.

\subsection{Parameter distributions}\label{distributions}
\begin{figure*}
\centering
\includegraphics[width=0.8\textwidth, trim=10 10 10 10, clip]{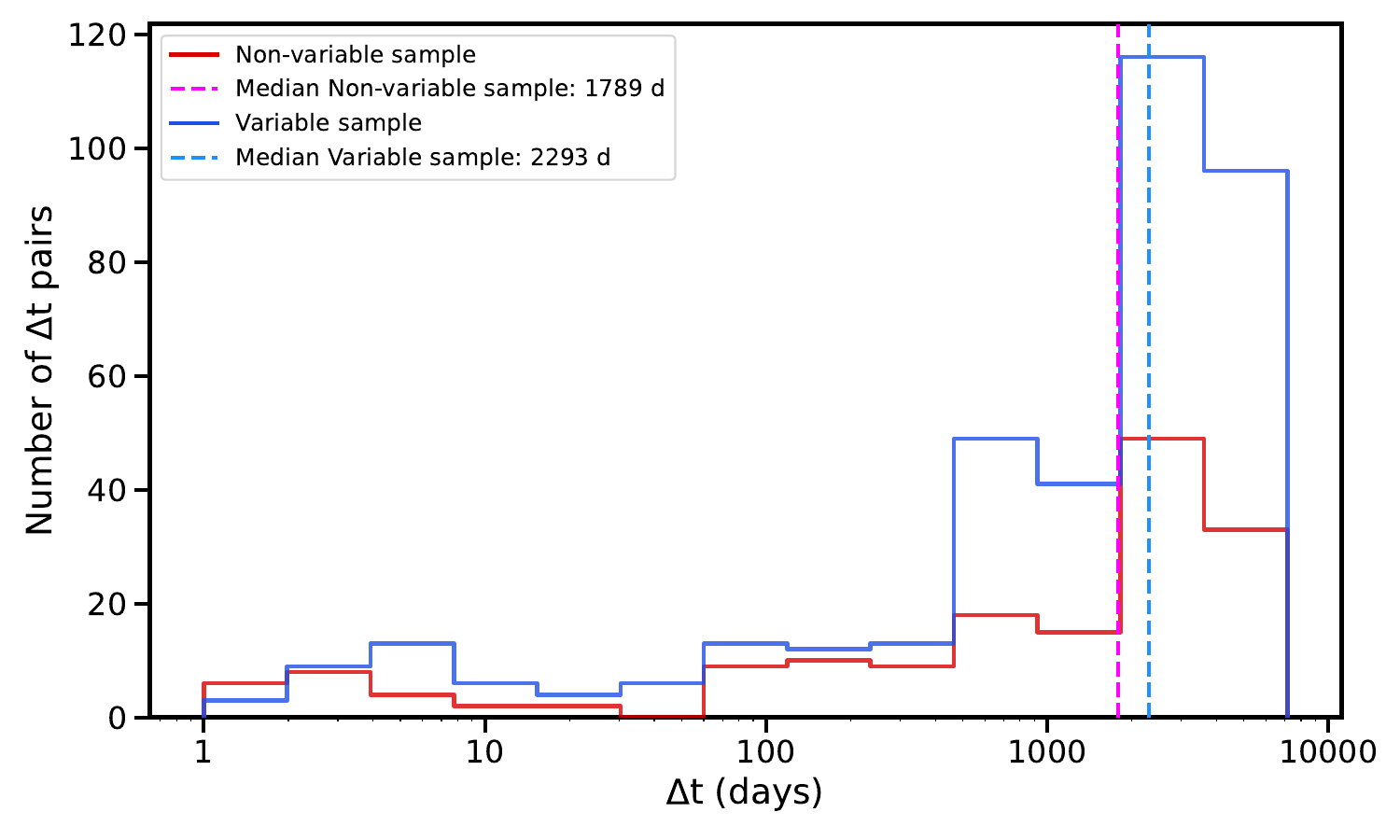}
    \caption[]
    {{Comparison of $\Delta$t distributions between pre-selected non-variable and variable samples. Histograms show the number of $\Delta$t pairs in log-scaled bins for the Non-variable sample in red, and the variable sample in blue. The dashed vertical lines indicate the median $\Delta$t for each sample (magenta for non-variable, light blue for variable). The KS test performed to quantify whether the two distributions are statistically different yields to a KS statistic of 0.12 and Log(NHP) = -1.22, indicating no statistically significant difference between the samples, suggesting that differences in variability classification are not primarily driven by different time-interval coverage.}}
    \label{fig:delta_t_distr}
\end{figure*}
In the following, we investigate whether the torus (covering factor C$_\mathrm{f}$ for \rxt\ and \skirt, inclination angle $\theta$, equatorial column density N$_{\rm H,eq}$, vertical extent of the cloud population $\sigma_{\rm tor}$, and line-of-sight column density \nhlos) and AGN continuum properties (photon index $\Gamma$, fraction of scattered continuum F$_\mathrm{s}$, and cross-normalization constant) differ among AGN with different variability classifications (i.e., Variable, Non-variable, and Undetermined). The distributions of the parameters are shown in Figs.~\ref{fig:distr_3models} and \ref{fig:distr_3models_bestfit}. 
For each source and model, we consider the median value across observations for \nhlos\ to provide a robust estimate. Group medians are then computed separately for the three classification sub-samples.
Given that the cross-normalization constant represents a relative scaling with respect to the first \chandra\ observation of each AGN, its median value depends on the arbitrary choice of reference epoch. We therefore characterize the cross-normalization using its dispersion, which is independent of the normalization choice and captures the relative spread of the constants across observations.
We stress that the present sample is small and heterogeneous, and therefore any comparison among variability classes should be interpreted with caution. Moreover, as discussed above, the probability of identifying variability depends on the number and temporal sampling of the available observations. Consequently, some sources currently classified as Non-variable may be reclassified once additional observations become available.

To test for possible differences among the three variability classes, we adopt the Kruskal-Wallis statistical test \citep{kruskal1952}, a non-parametric method suitable for small samples that may not follow a Gaussian distributions. To further explore pairwise differences, we apply the KS-test to evaluate whether the two distributions differ in shape. We report the derived medians and p-values for the two tests in Appendix~\ref{distribution_test}. 
Across all parameters, Kruskal-Wallis tests yield non-significant differences (p $>$ 0.2, i.e., Log(NHP) = -0.7, in all cases, the values are reported in Table~\ref{tab:p-values_ktests}). 
The KS tests generally present non-significant p-values (see Table~\ref{tab:p-values_ktests}), indicating no statistically significant differences in the distributions of the parameters between the variability classes across the models. These findings suggest that in the current sample and for the majority of parameters, the distributions of torus and continuum properties do not strongly depend on AGN variability classification. 

Although no significant differences are found, we note that the photon index shows a decrease from Non-variable to Variable sources across all models, implying marginally harder spectra in the latter class of AGN. 
In the literature, a well-established trend exists between the X-ray photon index and the Eddington ratio, with higher values typically associated with softer spectra (larger $\Gamma$) and lower $\lambda_{\rm Edd}$ corresponding to harder spectra \citep[smaller $\Gamma$; e.g.,][]{brandt1998,laor2000,shemmer2008, ris2009,brightman2013}. Considering the Eddington ratios of the sources in our sample\footnote{The Eddington ratios are reported in Table~\ref{tab:1_properties}. Values are taken from the BASS sample \citep{koss2022}, except for NGC~4785, for which we derive $\lambda_{\rm Edd}$ using the bolometric luminosity we estimated.}, we find that Non-variable AGN have a higher median $\lambda_{\rm Edd}$ (0.036) than Variable AGN (0.026). This is consistent with the literature trend, suggesting that the harder spectra observed in Variable AGN can be interpreted as a consequence of their lower accretion rates and, correspondingly, reduced radiative cooling of the X-ray emitting corona. 
A more robust statistical assessment will be possible once the full sample of the project has been analyzed.

In Fig.~\ref{fig:delta_t_distr}, we show the distribution of the pre-selected non-variable and variable samples in terms of the number of $\Delta$t pairs. The pre-selected non-variable sample exhibits a larger number of observations at short time intervals compared to the variable sample. This trend is observed in the sub-sample analyzed in this work (see Sect.~\ref{evol_nh_with_t}). However, the KS-test performed on the two distributions confirms that no statistically significant difference is present, Log(NHP) = -1.22.

\subsection{Parameter correlations}\label{corr}

We performed a Spearman test (accounting for asymmetric errors) to investigate possible correlation between torus and AGN parameters derived from the best-fit results of each of the three models independently (see Appendixes~\ref{tab:spec_1}-\ref{tab:spec_11} for the best-fits). 
For \nhlos\ and the cross-normalization constants, we adopt the median value across observations.
No significant correlations emerged. However, we note a marginally significant negative correlation between C$_\mathrm{f}$ and $\theta$ best-fits from \skirt\ (Log(NHP) = -1.89\footnote{The derived probability is above our adopted significance threshold, but it is below the conventional p = 0.05 (Log(NHP) = -1.30) level.}. A complete correlation analysis including all sources (both initially classified as variable and non-variable) will be presented in a forthcoming paper, once the non-variable sub-sample has been fully analyzed. This will ensure a comprehensive sample with sufficient AGN in each variability class, allowing us to investigate not only correlations across different models but also within variability classes themselves.

\section{MCG-03-34-064}\label{mcg}

In the following sections, we present the X-ray spectral analysis of MCG-03-34-064, a Compton-thin Seyfert 1.8 \citep{veron2006} AGN, candidate to host a dual system \citep{trindale2024}.

\subsection{Spectral modeling}
To optimize the signal-to-noise ratio and minimize the impact of the instrumental background at high energies, different energy bands are selected for each instrument: 0.5-10 keV for the XMM-{\it Newton}, 0.8-8 keV for \chandra, and 3.5-28 keV for \nustar\ data.
For this AGN, our first-order approximation (i.e., photon index and torus geometry parameters do not vary over the timescales of the available observations, see Sect.~\ref{model_components}) is not valid. Indeed, for all three models applied, we obtain poor fits, with $\chi^2$/d.o.f. $\sim$ 3, and clear residuals in the soft band ($<$2 keV) and around the iron K band (6-8 keV). Even after introducing Gaussian lines to account for significant emission and absorption features, the fits remain statistically unacceptable, with $\chi^2$/d.o.f. $\sim$ 1.7. 
We note that the complexities in the iron K band have been previously reported, including evidence for line broadening and a strong reflection component \citep{miniutti2007}. This suggests that additional physical processes may be at play and thus motivates our modeling approach.

To carry out a more detailed analysis, we focus on the computationally fastest of the three models, \skirt.
As a first test, we allow the photon index $\Gamma$ to vary freely for each observation. This results in an improvement of $\Delta\chi^2 = 40$ for eight additional free parameters. $\Gamma$ is generally clustered around $\sim$2.3–2.4 for most XMM and \chandra\ spectra, indicating a broadly consistent intrinsic continuum. We report the contour plot of the photon index versus the normalization of the primary continuum in Fig.~\ref{fig:cp_mgc}, upper panel. We note that while the photon indices of the XMM spectra are well constrained at 3$\sigma$ (i.e., 99.87\%), those of the \chandra\ observations are less constrained. In particular, Chandra\_2 (2021-03-29) and Chandra\_3 (2023-04-19) observations present only a lower limit on $\Gamma$, whereas Chandra\_4 (2023-04-20) and Chandra\_5 (2023-04-21) are constrained only at the 90\% c.l., and 2$\sigma$ (95.45\%) c.l., respectively. Overall, most observations favor a soft spectral shape; however, a significantly harder slope, with $\Gamma$ = 1.78 $\pm$ 0.11, is observed in Chandra\_1 (2006-07-31). This hard spectrum suggests either a variation in the intrinsic continuum or the effect of increased parameter degeneracy with absorption, coupled with reduced signal-to-noise ratio. In heavily absorbed spectra, a degeneracy between $\Gamma$ and \nhlos\ is expected: increasing column densities suppress soft photons, which can be compensated in the fit by a flatter $\Gamma$, and vice versa. In Chandra\_1, the combination of a moderately high \nhlos\ and relatively large uncertainties allows the fit to prefer a flatter continuum. We note that other spectra have similar column densities, but do not require such a hard photon index when the data quality is higher or the constraints are tighter. 

As a second step, we analyze each spectrum individually. The XMM\_3 and \nustar\ observations are simultaneous and are therefore fitted jointly. We also note that the reduced chi-squared values for the Chandra\_2, 3, 4, and 5 observations are the closest to unity in the combined fit. In addition, the photon index is poorly constrained, the remaining parameters from the individual fits are consistent within uncertainties, and the Chandra\_3, 4, and 5 observations are separated by only one day. For these reasons, we link these observations, allowing only the cross-calibration constant and \nhlos\ to vary, and refer to this combined dataset as ``Ch\_(2+3+4+5)''. The results of the spectral fitting are presented in Table~\ref{tab:spec_4} and the spectra are shown in Fig.~\ref{fig:mcg_spectra}.

\subsection{Variability across epochs}
The thermal plasma temperature remains consistent across all observations, with kT $\sim$ 0.85 - 0.90 keV. No evidence for variability is found, suggesting an extended and stable plasma component, likely associated with circumnuclear or host-galaxy-scale emission rather than the central engine (see contour plot in Fig.~\ref{fig:cp_mgc}, middle panel). In contrast, the photon index shows significant changes, with soft $\Gamma$ during the XMM\_1, XMM\_2, and \chandra\ observations and a harder continuum during the simultaneous XMM\_3+\nustar\ epoch, indicating intrinsic spectral variability. 
The covering factor of the neutral reflector is well constrained only for the XMM\_3+\nustar\ observation, and lower limits are found for the remaining observations.
Both the scattered fraction and continuum normalization vary between epochs. The XMM\_3+\nustar\ observation shows an enhanced scattered component and a lower normalization, consistent with the observed flux decrease. The equatorial column density is generally poorly constrained, particularly in the \chandra\ observations, preventing firm conclusions regarding its variability. The line-of-sight column density exhibits clear variability, both among XMM observations and among the closely spaced \chandra\ epochs, indicating changes in the absorbing material on timescales as short as one day. We report the contour plot between the X-ray photon index and the \nhlos\ in Fig.~\ref{fig:cp_mgc}, lower panel. The time evolution of \nhlos\ and $\Delta$\nhlos\ versus $\Delta$t/M$_{\rm BH}$ for MCG-03-34-064 are shown in Fig.~\ref{fig:mcg_plots}.
Applying the {\it p-value} method (see Sect.~\ref{var_class}) to the best-fit \nhlos\ values yields p-value = 1.49 $\times$ 10$^{-3}$, confirming the variability classification for this AGN.
Finally, the observed 2–10 keV flux varies reaching a factor of $\sim$1.6 between the brightest (XMM\_2) and faintest (XMM\_3+\nustar) epochs, consistent with the combined effects of intrinsic continuum variability and variable absorption.

\subsection{Physical interpretation}

We note that the obtained results are potentially influenced by the nature of MCG-03-34-064 as a candidate dual AGN with a projected separation of $\sim$100 pc \citep[$\sim$0.3\arcsec;][]{trindale2024}, corresponding to a system in the transition between the dual and binary phases. At this separation, the angular resolution of \xmm\ and \nustar\ is insufficient to disentangle the two nuclei, and the \chandra\ data reduction adopted here was not specifically optimized to isolate their individual contributions, as such an analysis is beyond the scope of this work, which may affect the modeling of the Fe K band. Within this context, the observed variability of the line-of-sight column density on timescales as short as one day suggests that at least part of the absorbing material is located at relatively small radii, possibly within the broad-line region or in clumpy structures at the inner edge of the dusty torus. In a dual AGN system transitioning towards a bound binary, the circumnuclear gas distribution is expected to be dynamically complex, and rapid changes in \nhlos\ could arise from multiple absorbing structures at different distances, whose relative contribution along the LoS varies with time.
However, we stress that comparable levels of rapid absorption variability are also commonly observed in isolated AGN, where clumpy absorbers alone can reproduce similar phenomenology. 
Therefore, while the dual AGN nature of MCG-03-34-064 provides a plausible framework for interpreting part of the observed spectral complexity, it is not required by the present data, and further investigation of this scenario is beyond the scope of this work. We thus conservatively classify the source as X-ray variable, with the caveat that the observed variability may be driven either by changes in the absorbing material along the LoS or by the unresolved dual nature of the system.

\section{Conclusions}
\label{sec:conclusions}

In this work, we carried out the first systematic multi-epoch X-ray study of a sub-sample of AGN previously classified as non-variable (in both line-of-sight column density and intrinsic flux) by \citet{zhao2021} based on the analysis of only two X-ray observations per source.
Our primary goals are to reassess the robustness of this classification by adopting the full archival X-ray coverage available to date, and to characterize the geometry and physical properties of the obscuring material using physically motivated torus models. We select and analyze a sample of 11 local ($z \leq 0.1$) AGN selected from the \swift--BAT 100-month catalog, with a total of 60 observations from \chandra, XMM-{\it Newton}, and \nustar, spanning timescales from days to nearly two decades (see Sect.~\ref{sec:data}). The AGN properties and information on the available observations are reported in Tables~\ref{tab:1_properties} and \ref{tab:1_observations}. By simultaneously fitting all available spectra for each source with three torus models (\skirt, \rxt, and \ux, see Sect.~\ref{model_components} for details), we constrained global torus parameters while allowing the line-of-sight column density, \nhlos, and the cross-normalization constant, C$_{\rm inst.,num}$ accounting for the intrinsic flux variability, to vary between epochs.
In Appendix~\ref{source_note}, we provide a description of each source, including its multiwavelength classifications and any modifications adopted to the standard fitting procedure.
The LoS column density variability was assessed through both direct comparison of \nhlos\ measurements and two complementary statistical tests (Sect.~\ref{var_class}).

Our main results can be summarized as follows:

\begin{enumerate}
\item \textbf{Comparison of torus models.} 
    Even though the three adopted torus models (\ux, \rxt, and \skirt) rely on different geometrical and physical assumptions, they provide consistent estimates of \nhlos\ (see Fig.~\ref{fig:3mod_comp}) and lead to robust variability classifications (Sect.~\ref{model_comp}). Eight out of ten sources show mutual agreement in their median values (Fig.~\ref{fig:distr_3models}; Appendix~\ref{bestfit}).
    Significant discrepancies are limited to IC~5063 and WISE~J14485, likely reflecting increased model sensitivity to complex geometries or low-obscuration regimes. For example, both \rxt\ and \skirt\ are unable to constrain the column density in the low-obscuration states of WISE~J14485. \ux\ can provide constrained values even for \nhlos\ $< 10^{21}$ cm$^{-2}$. The constants accounting for intrinsic flux variability, are also generally in agreement between the models (Sect.~\ref{model_comp}). 
    Concerning the remaining AGN and torus parameters (i.e., photon index, scattered fraction, covering factor for \rxt\ and \skirt, equatorial column density, and inclination angle) exhibit stronger inter-model discrepancies, reflecting the intrinsic differences in the torus geometry and cloud distributions assumed by each model (Sect.~\ref{model_comp}). 
    Moreover, we note that the $\Delta$\nhlos\ derived for each pair of observations differs in absolute values across the three models (see Sect.~\ref{model_comp1}), but the evolution is consistent as seen from the $\Delta$\nhlos\ versus $\Delta$t / M$_{\rm BH}$ plots in the lower panels of Figs.~\ref{fig:eso_plots}-~\ref{fig:UGC_plots}. 
    
    \item \textbf{Compton-thin and Compton-thick classification.}  
    The analysis confirms that the majority of  our sample (9 out of 11) are Compton-thin AGN (see Table~\ref{tab:1_properties}). We identify UGC~3752 as a candidate transitioning source and confirm the Compton-thick nature of NGC~1194 (Sect.~\ref{evol_nh_with_t}). The latter shows possible transitions between CTH and CTK states on timescales of $\sim$1-2 years (see Table~\ref{tab:spec_7} and Fig.~\ref{fig:1194_plots}), providing strong evidence for a clumpy and dynamic obscuring medium.
    
    \vspace{0.05cm}
    \item \textbf{Reassessment of the non-variable classification.}  
    When fitting more than two observations, we find that the original ``non-variable'' classification is often not robust. Out of the ten AGN for which a homogeneous analysis was possible (excluding MCG-03-34-064; Sect.~\ref{mcg}), five sources (50\%) show clear $N_{\mathrm{H,LoS}}$ variability, three (30\%) remain consistent with being non-variable, and two (20\%) are classified as undetermined due to model-dependent results (see Sect.~\ref{var_classification_res} and Tables \ref{tab:spec_1}-\ref{tab:spec_11} for the best-fit results). A classification summary for all the analyzed AGN is reported in Table~\ref{tab:clas_summary}. 
    For comparison, follow-up studies of the pre-selected variable sample find clear \nhlos\ variability in 39\% (11/28) of AGN, while the rest show either no column-density changes or Undetermined behavior (see \citetalias{pizzetti2022}-\citetalias{torres2025} papers).

    We note that 7 out of 10 AGN in our sample require intrinsic flux variability to reproduce the data (see Sect.~\ref{var_classification_res} and Tables \ref{tab:spec_1}-\ref{tab:spec_11}). In the remaining three cases, the tensions are comparable across hypotheses or reveal degeneracies between intrinsic flux and \nhlos\ variability, preventing a firm determination of the physical driver.

    \vspace{0.05cm}
    \item \textbf{Dependence of variability detection on observational cadence.}  
    By applying a Bayesian logistic regression to the combined sample, which includes the AGN analyzed in this work together with the previously studied variable sources from the CI-CTAGN project, we find tentative evidence for a positive relation between the number of observations $N_{\rm obs}$ and the probability of detecting Variable AGN (see Sect.~\ref{var_classification_res}). In particular, sources with fewer observations show a mixed population of Variable and Non-variable AGN, whereas sources with larger numbers of epochs ($N_{\rm obs}\gtrsim5$--6) are increasingly dominated by Variable AGN (Fig.~\ref{fig:bayesian}). This result highlights the importance of both large samples and adequate temporal sampling for identifying \nhlos\ variability.

    \item \textbf{Timescales and physical origin of the absorbing material.}  
    In Sect.~\ref{evol_nh_with_t}, we examine the epoch-to-epoch changes in column density as a function of time separation, $\Delta$N$_{\mathrm{H,LoS}}$ versus $\Delta$t. We find that small $\Delta$N$_{\mathrm{H,LoS}}$ variations occur on all timescales, whereas large changes ($\Delta$N$_{\mathrm{H,LoS}}$ $\gtrsim$ 5 $\times$ 10$^{23}~\mathrm{cm^{-2}}$) are preferentially associated with long time intervals ($\Delta$t $\gtrsim$ 100 days; see Fig.~\ref{fig:large_nh_large_t}). This behavior, also observed by \citetalias{torres23}, \citetalias{pizzetti2025}, and \citet{cox2025}, supports a scenario in which both BLR-scale and torus-scale absorbers contribute to the observed variability.

    \item \textbf{Parameter distributions and correlations.} The distributions of torus and continuum parameters (Figs.~\ref{fig:distr_3models} and \ref{fig:distr_3models_bestfit}) do not show statistically significant differences across AGN variability classes (see Sect.~\ref{distributions} and Table~\ref{tab:p-values_ktests}). No statistically significant difference is present between the pre-selected non-variable and variable samples distributions in terms of number of $\Delta$t pairs (see Fig.~\ref{fig:delta_t_distr}).
    We also investigate possible correlations between torus and AGN parameters. However, no statistically significant correlations are detected in the current sample (see Sect.~\ref{corr} for more details).

    \item \textbf{The case of MCG-03-34-064.}  
    MCG-03-34-064 represents a special case in our sample and was therefore analyzed separately (Sect.~\ref{mcg}). Unlike the other AGN, the assumption of constant spectral shape across epochs is not valid for this source, preventing a simultaneous multi-epoch fit. When observations are fitted independently (see Sect.~\ref{mcg} for the details on the analysis, Table~\ref{tab:spec_4} for the best-fit results and Figs.~\ref{fig:mcg_spectra} and~\ref{fig:cp_mgc} for the spectra and contours plots), MCG-03-34-064 exhibits clear variability in \nhlos\ on timescales ranging from days to years (see Table~\ref{tab:spec_4} and Fig.~\ref{fig:mcg_plots}). This variability may be due to changes in the line-of-sight absorbing column density, to the dual-AGN nature of the system, or to a combination of both effects.
\end{enumerate}

In conclusion, our analysis reveals that a substantial fraction of AGN previously classified as non-variable do, in fact, exhibit significant line-of-sight column density variability when sufficient temporal coverage is available. This result strengthens the case for a clumpy, inhomogeneous, and dynamic torus as a common feature of nearby AGN. More generally, it highlights the critical importance of long-term, multi-epoch X-ray monitoring for properly characterizing AGN obscuration. The extension of this analysis to the remaining (pre-selected) non-variable sources in the CI-CTAGN sample will allow these findings to be placed on firmer statistical grounds and will further constrain the physical structure of the AGN torus.

\begin{acknowledgments}
We thank Bert Vander Meulen for providing the \skirt\ code used in our analysis prior to its public release. This research has made use of data obtained through the High Energy Astrophysics Science Archive Research Center online service, provided by the NASA/Goddard Space Flight Center. VEG acknowledges funding under NASA contract 80NSSC24K1403. CR acknowledges support from SNSF Consolidator grant F01$-$13252, Fondecyt Regular grant 1230345, ANID BASAL project FB210003 and the China-Chile joint research fund. MG acknowledges support from the ERC Consolidator Grant \textit{BlackHoleWeather} (101086804).
\end{acknowledgments}


\appendix
\renewcommand{\thetable}{\Alph{section}.\arabic{table}}
\setcounter{table}{0}

\section{Sample}
In Table~\ref{tab:1_observations}, we report the information on the X-ray observations of the sub-sample of 11 non-variable AGN analyzed in this paper.

\startlongtable
\begin{deluxetable*}{cccccc}
\tablewidth{0pt}
\tablecaption{X-ray observation details.\label{tab:1_observations}}
\tablehead{
\colhead{\scriptsize (1)} &
\colhead{\scriptsize (2)} &
\colhead{\scriptsize (3)} &
\colhead{\scriptsize (4)} &
\colhead{\scriptsize (5)} &
\colhead{\scriptsize (6)}\\
\colhead{Source} &
\colhead{Telescope} &
\colhead{Obs. ID} &
\colhead{Obs. date} &
\colhead{Exp. time} &
\colhead{Tot. counts} \\
&
&
&
&
\colhead{[ks]}
}
\startdata
ESO~464-G016 &
\begin{tabular}{@{}c@{}}
\xmm \\
\nustar\\
\chandra
\end{tabular} &
\begin{tabular}{@{}c@{}}
0762920401\\
60101013002\\
27972
\end{tabular} &
\begin{tabular}{@{}c@{}}
2016-04-11\\
2016-04-13\\
2023-08-04
\end{tabular} &
\begin{tabular}{@{}c@{}}
16.2\\
22.1 (0.3\%)\\
9.9
\end{tabular} &
\begin{tabular}{@{}c@{}}
713 \\
774 (4.5\%) \\
71
\end{tabular}\\
\graytableline
IC~5063 &
\begin{tabular}{@{}c@{}}
\nustar\\
\chandra\ 1\\
\chandra\ 2\\
\chandra\ 3\\
\chandra\ 4\\
\chandra\ 5\\
\chandra\ 6\\
\chandra\ 7
\end{tabular} &
\begin{tabular}{@{}c@{}}
60061302002\\
7878\\
21467\\
21999\\
22000\\
22001\\
22002\\
21466
\end{tabular} &
\begin{tabular}{@{}c@{}}
2013-07-08\\
2007-06-15\\
2018-12-11\\
2018-12-12\\
2018-12-13\\
2018-12-15\\
2018-12-16\\
2019-07-23
\end{tabular} &
\begin{tabular}{@{}c@{}}
18.4 (0.1\%)\\
34.1\\
26.9\\
34.1\\
15.6\\
29.3\\
43.9\\
87.7
\end{tabular} &
\begin{tabular}{@{}c@{}}
7550 (8.2\%)\\
4810\\
3086\\
3612\\
1610\\
3325\\
4443\\
11971
\end{tabular} \\
\graytableline
LEDA~511869 &
\begin{tabular}{@{}c@{}}
\xmm\\
\nustar\ 1\\
\nustar\ 2
\end{tabular} &
\begin{tabular}{@{}c@{}}
0802450401\\
60201033002\\
60061252002
\end{tabular} &
\begin{tabular}{@{}c@{}}
2018-02-09\\
2016-05-25\\
2020-03-25
\end{tabular} &
\begin{tabular}{@{}c@{}}
17.3\\
21.5 (0.1\%)\\
23.5 (0.9\%)
\end{tabular} &
\begin{tabular}{@{}c@{}}
1176\\
774 (0.9\%)\\
1470 (8.7\%)
\end{tabular}\\
\graytableline
MCG-03-34-064 &
\begin{tabular}{@{}c@{}}
\xmm\ 1\\
\xmm\ 2\\
\xmm\ 3\\
\nustar\\
\chandra\ 1\\
\chandra\ 2\\
\chandra\ 3\\
\chandra\ 4\\
\chandra\ 5
\end{tabular} &
\begin{tabular}{@{}c@{}}
0206580101\\
0506340101\\
0763220201\\
60101020002\\
7373\\
23690\\
25253\\
27802\\
27803
\end{tabular} &
\begin{tabular}{@{}c@{}}
2005-01-24\\
2008-01-24\\
2016-01-17\\
2016-01-17\\
2006-07-31\\
2021-03-29\\
2023-04-19\\
2023-04-20\\
2023-04-21
\end{tabular} &
\begin{tabular}{@{}c@{}}
38.3\\
76.5\\
123.3\\
78.4 (0.3\%)\\
7.1\\
3.1\\
15.6\\
17.9\\
16.9
\end{tabular} &
\begin{tabular}{@{}c@{}}
21394\\
41441\\
62583\\
11038 (2.6\%)\\
992\\
225\\
926\\
1150\\
1003
\end{tabular} \\
\graytableline
Mrk~18 &
\begin{tabular}{@{}c@{}}
\xmm\\
\nustar\ 1\\
\nustar\ 2 {\bf $\dagger$}\\
\chandra\ {\bf $\dagger$}
\end{tabular} &
\begin{tabular}{@{}c@{}}
0312191001\\
60061088002\\
60760004001\\
23813
\end{tabular} &
\begin{tabular}{@{}c@{}}
2006-03-23\\
2013-12-15\\
2022-08-15\\
2022-08-15
\end{tabular} &
\begin{tabular}{@{}c@{}}
9.8\\
20.2 (0.2\%)\\
30.7 (1.0\%)\\
19.2
\end{tabular} &
\begin{tabular}{@{}c@{}}
1547\\
832 (15.4\%)\\
801 (13.2\%)\\
375
\end{tabular} \\
\graytableline
Mrk~1498 &
\begin{tabular}{@{}c@{}}
\xmm\\
\nustar\\
\chandra
\end{tabular} &
\begin{tabular}{@{}c@{}}
0500850501\\
60160640002\\
17085
\end{tabular} &
\begin{tabular}{@{}c@{}}
2007-06-23\\
2015-05-11\\
2016-05-15
\end{tabular} &
\begin{tabular}{@{}c@{}}
7.6\\
23.7 (0.3\%)\\
19.5
\end{tabular} &
\begin{tabular}{@{}c@{}}
4771\\
8458 (0.3\%)\\
3172
\end{tabular} \\
\graytableline
NGC~1194 &
\begin{tabular}{@{}c@{}}
\xmm\ 1\\
\xmm\ 2 {\bf $\dagger$}\\
\nustar\ 1\\
\nustar\ 2 {\bf $\dagger$}\\
\chandra\ 1\\
\chandra\ 2\\
\chandra\ 3\\
\chandra\ 4
\end{tabular} &
\begin{tabular}{@{}c@{}}
0307000701\\
0852200101\\
60061035002\\
60501011002\\
22552\\
22880\\
22881\\
23688
\end{tabular} &
\begin{tabular}{@{}c@{}}
2006-02-19\\
2020-01-16\\
2015-02-28\\
2020-01-17\\
2019-10-22\\
2019-10-25\\
2019-10-27\\
2020-11-05
\end{tabular} &
\begin{tabular}{@{}c@{}}
12.5\\
52.9\\
31.5 (0.2\%)\\
58.1 (0.3\%)\\
39.5\\
20.0\\
20.3\\
7.6
\end{tabular} &
\begin{tabular}{@{}c@{}}
1303\\
5294\\
2639 (1.8\%)\\
5160 (3.2\%)\\
1056\\
482\\
541\\
216
\end{tabular} \\
\graytableline
NGC~2655 &
\begin{tabular}{@{}c@{}}
\xmm\\
\nustar\ 1\\
\nustar\ 2 *\\
\nustar\ 2 *\\
\chandra\ 1 *\\
\chandra\ 1 *
\end{tabular} &
\begin{tabular}{@{}c@{}}
0301650301\\
60160341002\\
60160341003\\
60160341004\\
28152\\
30827
\end{tabular} &
\begin{tabular}{@{}c@{}}
2005-09-04\\
2016-11-02\\
2016-11-10\\
2016-11-10\\
2025-03-03\\
2025-03-03
\end{tabular} &
\begin{tabular}{@{}c@{}}
8.8\\
5.1 (0.2\%)\\
0.2 (0.3\%)\\
15.9 (0.3\%)\\
12.1\\
11.9
\end{tabular} &
\begin{tabular}{@{}c@{}}
2704\\
203 (6.4\%)\\
15 (20.7\%)\\
705 (1.4\%)\\
205\\
216
\end{tabular} \\
\graytableline
NGC~4785 &
\begin{tabular}{@{}c@{}}
\xmm\ 1 {\bf $\dagger$}\\
\xmm\ 2\\
\xmm\ 3\\
\nustar\ 1 {\bf $\dagger$}\\
\nustar\ 2\\
\nustar\ 3\\
\chandra
\end{tabular} &
\begin{tabular}{@{}c@{}}
0743010101\\
0762920701\\
0762920801\\
60001143002\\
60161523002\\
60161523004\\
18074
\end{tabular} &
\begin{tabular}{@{}c@{}}
2014-08-20\\
2015-07-03\\
2016-01-15\\
2014-08-20\\
2021-03-10\\
2023-06-04\\
2016-04-16
\end{tabular} &
\begin{tabular}{@{}c@{}}
22.9\\
16.1\\
16.9\\
48.8 (0.1\%)\\
18.3 (0.6\%)\\
21.5 (0.5\%)\\
9.8
\end{tabular} &
\begin{tabular}{@{}c@{}}
2066\\
1500 \\
1796\\
2764 (12.4\%)\\
147 (20.4\%)\\
675 (27.3\%)\\
245
\end{tabular} \\
\graytableline
UGC~03752 &
\begin{tabular}{@{}c@{}}
\xmm\\
\nustar\ 1\\
\nustar\ 2\\
\nustar\ 3\\
\nustar\ 4\\
\nustar\ 5\\
\chandra
\end{tabular} &
\begin{tabular}{@{}c@{}}
0883210701\\
60061072002\\
60061072004\\
60702053002\\
60702053004\\
60702053006\\
30709
\end{tabular} &
\begin{tabular}{@{}c@{}}
2022-03-28\\
2013-12-03\\
2020-04-17\\
2021-10-22\\
2021-11-08\\
2022-03-01\\
2025-01-05
\end{tabular} &
\begin{tabular}{@{}c@{}}
18.3\\
23.6 (0.2\%)\\
31.3 (0.5\%)\\
23.8 (0.4\%)\\
29.7 (0.5\%)\\
26.7 (0.6\%)\\
9.9
\end{tabular} &
\begin{tabular}{@{}c@{}}
2147\\
871 (4.1\%)\\
788 (3.9\%)\\
535 (16.3\%)\\
1038 (9.8\%)\\
2239 (8.6\%)\\
93
\end{tabular} \\
\graytableline
WISE~J144850 &
\begin{tabular}{@{}c@{}}
\xmm\\
\nustar\\
\chandra
\end{tabular} &
\begin{tabular}{@{}c@{}}
0720280101\\
60463049002\\
22226
\end{tabular} &
\begin{tabular}{@{}c@{}}
2014-02-16\\
2019-03-12\\
2019-05-19
\end{tabular} &
\begin{tabular}{@{}c@{}}
22.4\\
20.2 (0.7\%)\\
10.0
\end{tabular} &
\begin{tabular}{@{}c@{}}
21038\\
3471 (18.8\%)\\
1705
\end{tabular} \\
\enddata
\tablecomments{
\footnotesize {\it Columns:} 1) source name; 2) telescope used in the analysis; 3) observation ID; 4) observation date; 5) effective exposure time after reduction in ks; 6) total source counts after reduction. We note that the exposure times for XMM refer to the EPIC-pn spectra. Meanwhile, the reported exposure times and total counts for \nustar\ are averaged between FPMA and FPMB, with the range of percentage variance between the two in parentheses. \\
{\bf $\dagger$}: the marked observations are simultaneous and will be fitted accordingly, i.e., assuming no variability in flux and line-of-sight column density. \\
{\bf *}: the \nustar\ observations 60160341003 and 60160341004, as well as the \chandra\ observations 28152 and 30827, are consecutive in time; therefore, each pair is treated as a single \nustar\ and a single \chandra\ observation, respectively, during the spectral analysis.
}
\end{deluxetable*}

\section{X-ray fitting results}\label{bestfit} 
\begin{figure*}
\centering
    \includegraphics[width=.78\textwidth, trim=5 10 0 0, clip]{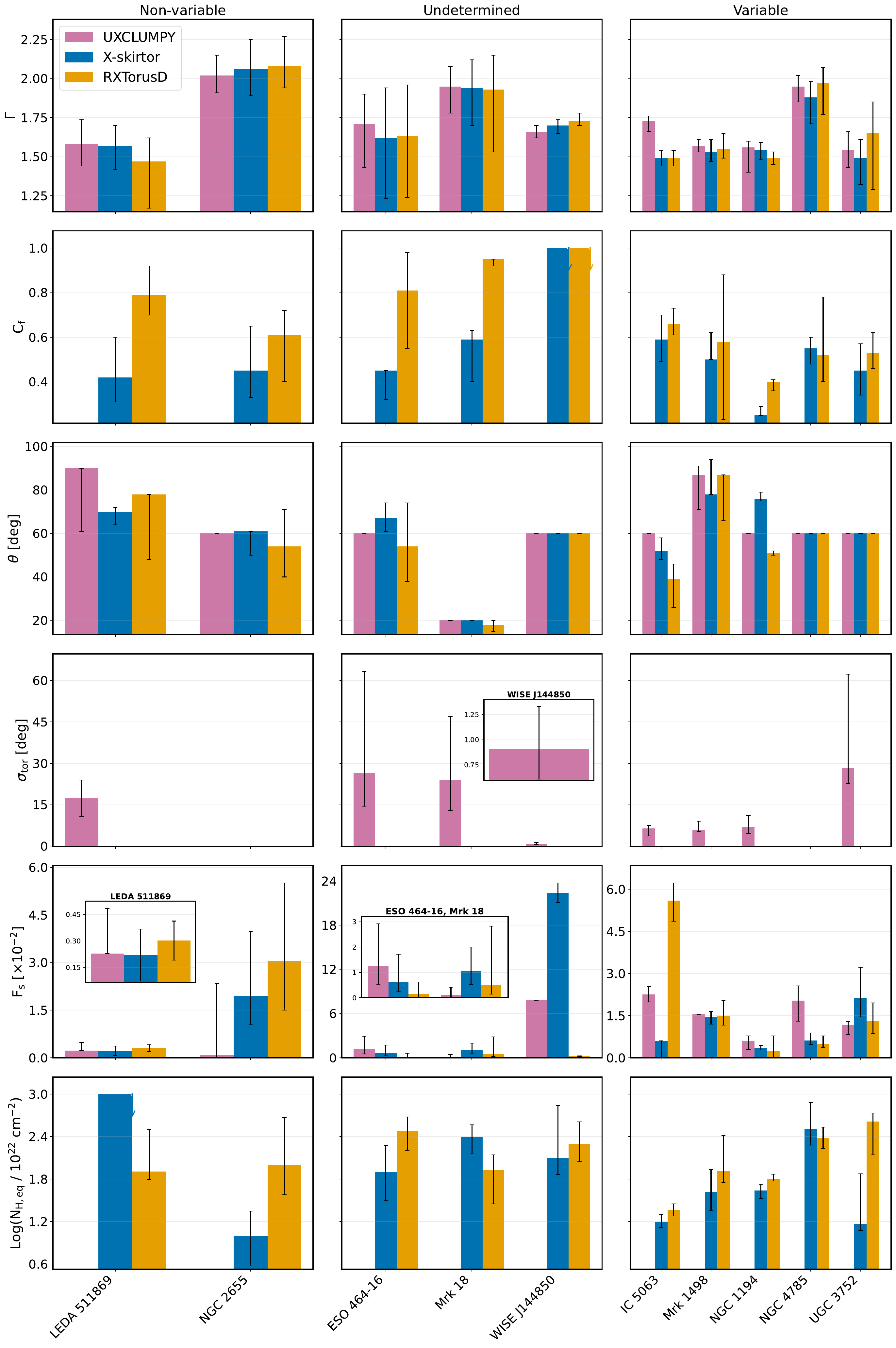}
    \caption[]
    {{Distributions of the torus parameters grouped by variability class. {\it Rows:} 1) photon index $\Gamma$; 2) covering factor C$_{\rm{f}}$; 3) inclination angle $\theta$; 4) vertical extent of the cloud population $\sigma_{\rm tor}$; 5) scattering fraction of the primary continuum F$_{\rm{s}}$; 6) equatorial column density N$_{\rm H,eq}$. Each bar shows the best-fit value of each torus models (\ux\ in pink, \skirt\ in blue, and \rxt\ in yellow) with asymmetric uncertainties at 90\% c.l.. {\it Left panels} present the Non-variable, {\it middle panels} the Undetermined, and {\it right panels} the Variable AGN samples. Insets are included to improve the visualization of AGN with low best-fit values. Within the current sample, no statistically significant differences are present among variability classes for the torus and continuum parameters.
    }}
    \label{fig:distr_3models_bestfit}
\end{figure*}

\nopagebreak
In this work, we present the best-fit results for 11 Compton-thin, local AGN with multi-epoch observations selected from the 100-month BAT catalog. We analyze 60 X-ray observations from \chandra, \xmm, and \nustar, using physical torus models (\texttt{xskirtor}, \texttt{RXTorusD}, and \texttt{UXCLUMPY}) and constraining the torus parameters and the line-of-sight column density for each observation. In Tables \ref{tab:spec_1}-\ref{tab:spec_11}, we report the best-fit values obtained from the simultaneous multi-epoch spectral analysis. Particularly, for each Table the following parameters are shown. We report the physical and geometrical properties of the torus are described by the following parameters: C$_\mathrm{f}$ is the covering factor of the torus; $\theta$ is the inclination angle, where $\theta = 0^{\circ}$ is the face-on scenario; F$_\mathrm{s}$ is the fraction of scattered continuum; and Norm is the normalization of the AGN emission. 
We note that, if the inclination results to be completely unconstrained during the fit, we fix it to $\theta = 60^{\circ}$.
In the following tables, we report the parameter \texttt{CTKcover} of the \ux\ model as C$_\mathrm{f}$. However, the reader should be aware that this quantity represents the covering factor of the inner Compton-thick ring of clouds (see Sect.~\ref{sec:uxc}), and therefore it does not physically correspond to the global torus covering factor defined in the other models. Consequently, these parameters cannot be directly compared.
For what concern \texttt{RxTorusD} and \texttt{xskirtor} models, N$_{\rm H,eq}$ is the equatorial column density of the torus. The vertical extent of the cloud population is reported as $\sigma_{\rm tor}$ for the \texttt{UXCLUMPY} model.
N$_{\rm H,inst.,num.}$ shows the LoS hydrogen column density for a given observation.
C$_{\rm inst.,num}$ represents the cross-normalization constant for each observation, with respect to the intrinsic flux of the first \chandra\ observation.
The last part of each Table presents the $\chi^2$ of the best-fit when considering: a) no variability between observations; b) no intrinsic flux variability; c) no \nhlos\ variability. In Fig.~\ref{fig:distr_3models_bestfit}, we show the distributions of the best-fit parameters for each AGN and each adopted model.

\clearpage
\newpage

\begin{deluxetable*}{lccc}
\tablecaption{ESO~464-G016: best-fit spectral analyses for each torus model.\label{tab:spec_1}}
\tablewidth{\textwidth}
\tablehead{
\colhead{Parameter} & \colhead{\texttt{xskirtor}} & \colhead{\texttt{RXTorusD}} & \colhead{\texttt{UXCLUMPY}}
}
\startdata
\multicolumn{4}{c}{\texttt{apec \footnotesize{(Thermal emission)}}} \\
kT / keV & 0.26$^{+0.06}_{-0.04}$ & 0.28$^{+0.09}_{-0.06}$ & 0.27$^{+0.06}_{-0.04}$ \\
\multicolumn{4}{c}{\texttt{\footnotesize{Comptonized primary continuum}}} \\
$\Gamma$ & 1.62$^{+0.32}_{-0.39}$ & 1.63$^{+0.33}_{-0.39}$ & 1.71$^{+0.19}_{-0.28}$ \\
\multicolumn{4}{c}{\texttt{\footnotesize{Neutral reflector}}} \\
C$_\mathrm{f}$ $^{\bf +}$ & 0.45$_{-0.13}$ & 0.81$^{+0.17}_{-0.26}$ & 0$^f$ \\
$\theta$ / deg & 67$^{+7}_{-6}$ & 54$^{+20}_{-16}$ & 60$^f$ \\
$\sigma_{\rm tor}$ / deg & / & / & 26.49$^{+36.78}_{-11.97}$ \\
F$_\mathrm{s}$ / $10^{-2}$ & 0.61$^{+1.11}_{-0.37}$ & 0.14$^{+0.48}_{-0.11}$ & 1.24$^{+1.68}_{-0.71}$ \\
norm / $10^{-3}$ & 1.23$^{+2.28}_{-0.88}$ & 5.55$^{+1.55}_{-4.28}$ & 2.11$^{+1.33}_{-1.14}$ \\
N$_{\rm H,eq}$ / 10$^{22}$ cm$^{-2}$ & 79.43$^{+110.27}_{-47.81}$ & 304.72$^{+170.28}_{-143.72}$ & / \\
\multicolumn{4}{c}{\texttt{N$_{\rm H,inst.,num.}$ \footnotesize{(LoS hydrogen column density)}}} \\
N$_{\rm H}^{Ch}$ / 10$^{22}$ cm$^{-2}$ & 78.79$^{+24.09}_{-23.84}$ & 112.21$^{+34.82}_{-30.68}$ & 107.36$^{+49.60}_{-31.40}$ \\
N$_{\rm H}^{XMM}$ / 10$^{22}$ cm$^{-2}$ & 71.21$^{+17.83}_{-20.97}$ & 102.66$^{+33.23}_{-27.28}$ & 95.02$^{+40.88}_{-21.95}$ \\
N$_{\rm H}^{NuS}$ / 10$^{22}$ cm$^{-2}$ & 65.60$^{+19.85}_{-16.80}$ & 98.52$^{+33.69}_{-13.93}$ & 79.27$^{+20.74}_{-10.37}$ \\
\multicolumn{4}{c}{\texttt{C$_{\rm inst.,num}$ \footnotesize{(Cross-normalization constant)}}} \\
C$_{Ch}$ & 1$^f$ & 1$^f$ & 1$^f$ \\
C$_{XMM}$ & 0.92$^{+0.49}_{-0.31}$ & 0.90$^{+0.48}_{-0.30}$ & 0.94$^{+0.48}_{-0.35}$ \\
C$_{NuS}$ & 1.00$^{+0.74}_{-0.41}$ & 1.02$^{+0.83}_{-0.42}$ & 0.94$^{+0.50}_{-0.28}$ \\
\graytableline
\multicolumn{4}{c}{\texttt{Statistic}} \\ 
C-stat/d.o.f. & 268/215 & 259/215 & 239/217 \\
{\it T} & 3.61$\sigma$ & 3.00$\sigma$ & 1.49$\sigma$ \\
p-value & 0.91 & 0.95 & 0.71 \\
\multicolumn{4}{c}{\texttt{Statistic \footnotesize{(No variability)}}} \\
C-stat/d.o.f. & 278/219 & 278/219 & 249/221 \\
{\it T} & 3.99$\sigma$ & 3.99$\sigma$ & 1.88$\sigma$ \\
\multicolumn{4}{c}{\texttt{Statistic \footnotesize{(No flux variability)}}} \\
C-stat/d.o.f. & 269/217 & 270/217 & 240/219 \\
{\it T} & 3.53$\sigma$ & 3.60$\sigma$ & 1.42$\sigma$ \\
\multicolumn{4}{c}{\texttt{Statistic \footnotesize{(No \nh\ variability)}}} \\
C-stat/d.o.f. & 269/217 & 268/217 & 241/219 \\
{\it T} & 3.53$\sigma$ & 3.46$\sigma$ & 1.49$\sigma$ \\
\enddata
\tablecomments{$^{\bf +}$: For \ux, C$_\mathrm{f}$ represents the inner Compton-thick ring covering factor and it cannot be directly compared with the covering factors of \skirt\ and \rxt, see Sect.~\ref{sec:uxc}.}
\end{deluxetable*}

\begin{figure}
\centering
    \subfloat[][]{
    \includegraphics[width=0.4\textwidth, trim=0 180 380 200, clip]{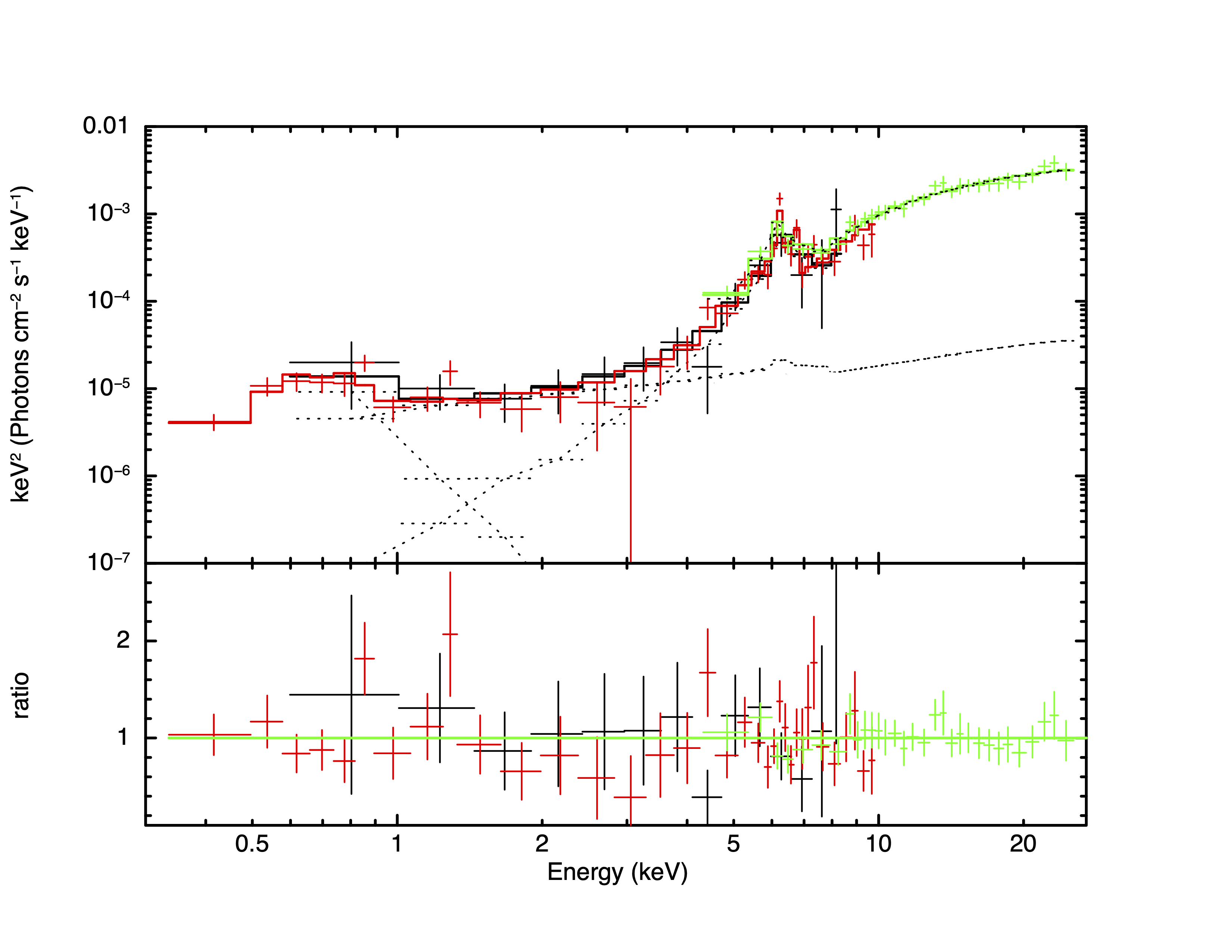}} \\
    \subfloat[][]{
    \includegraphics[width=0.4\textwidth, trim=10 10 13 0, clip]{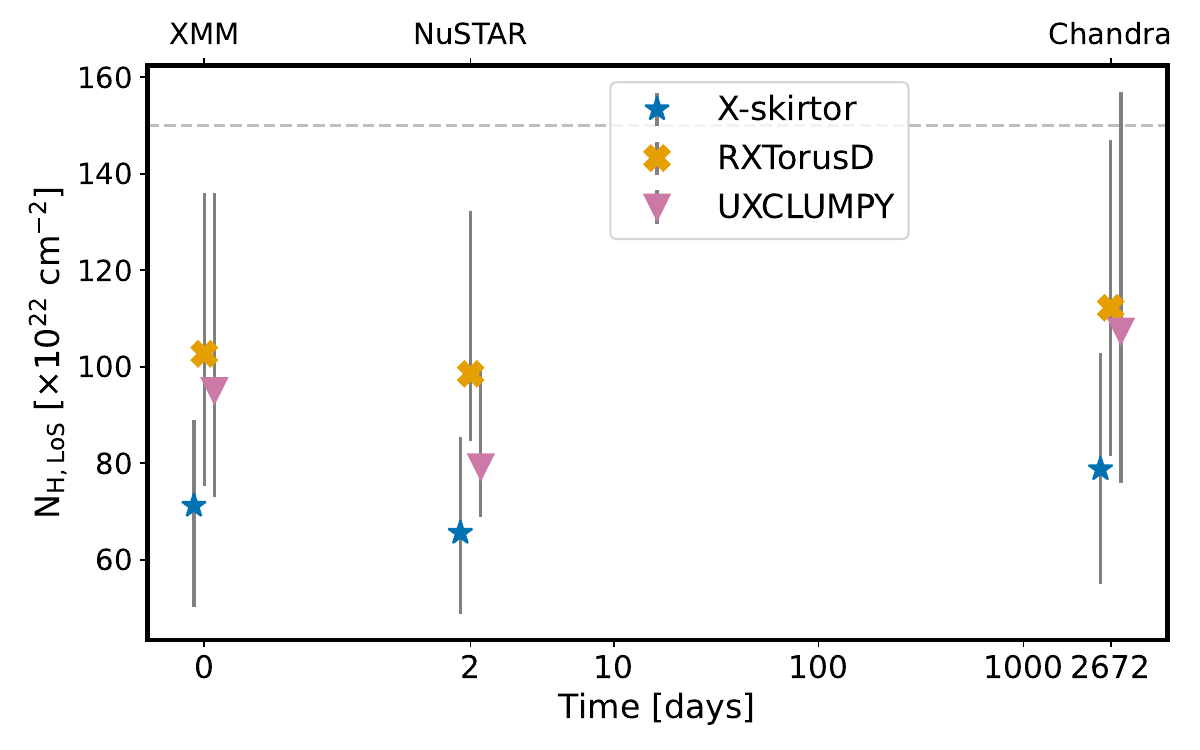}}\\
    \subfloat[][]{
    \includegraphics[width=0.4\textwidth, trim=5 5 6 0, clip]{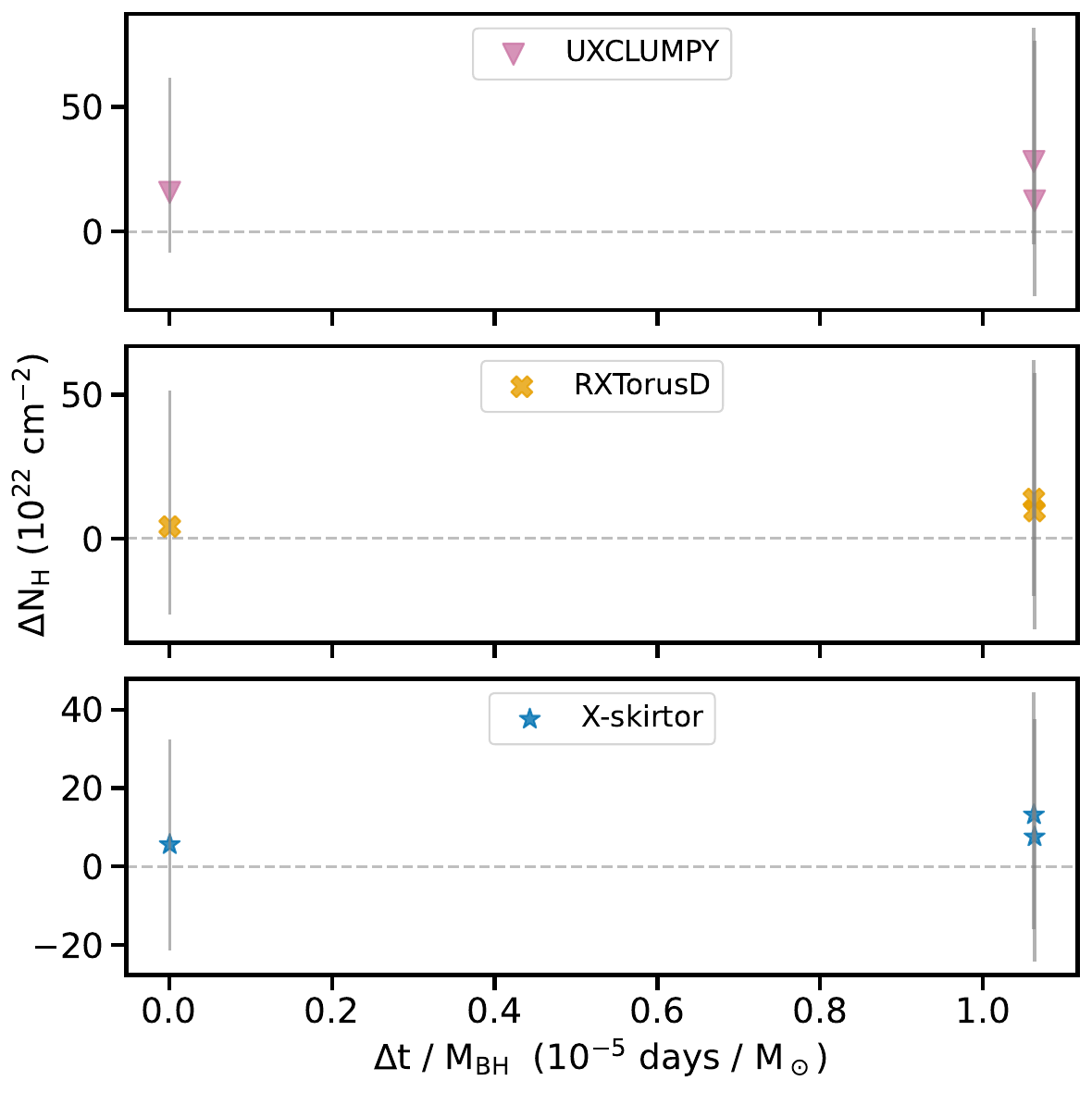}}
    \caption[]{{ESO~464-G016. {\it Panel a:} Best-fit with the \texttt{UXCLUMPY} torus model. The \chandra\ (in black), \xmm\ (in red), and \nustar\ (FPMA and FPMB spectra are grouped together in green) spectra are shown. {\it Panel b:} evolution of the \nhlos\ derived from the different torus models (\skirt\ in yellow stars, \rxt\ in green crosses, and \ux\ in purple triangles). The x-axis is in logarithmic scale to clearly present the timescale between observations. {\it Panel c:} Variations of \nhlos\ as a function of the time separation between observations, normalized by the black-hole mass, for three adopted torus models (\ux\ in pink, \rxt\ in yellow, and \skirt\ in light blue). }}
    \label{fig:eso_plots}
\end{figure}

\clearpage
\newpage

\renewcommand{\arraystretch}{1.5}
\begin{deluxetable}{lccc}
\tablecaption{IC~5063: spectral analyses for each torus model.\label{tab:spec_2}}
\tabletypesize{\scriptsize}
\tablewidth{\textwidth}
\tablehead{
\colhead{Parameter} & \colhead{\texttt{xskirtor}} & \colhead{\texttt{RXTorusD}} & \colhead{\texttt{UXCLUMPY}}
}
\startdata
\multicolumn{4}{c}{\texttt{apec \footnotesize{(Thermal emission)}}} \\
kT / keV & 0.74$^{+0.07}_{-0.08}$ & 0.75$^{+0.07}_{-0.08}$ & 0.71$^{+0.12}_{-0.15}$ \\
\multicolumn{4}{c}{\texttt{\footnotesize{Comptonized primary continuum}}} \\
$\Gamma$ & 1.49$\pm$0.05 & 1.49$\pm$0.05 & 1.73$^{+0.03}_{-0.07}$ \\
\multicolumn{4}{c}{\texttt{\footnotesize{Neutral reflector}}} \\
C$_\mathrm{f}$ $^{\bf +}$ & 0.59$^{+0.11}_{-0.10}$ & 0.66$^{+0.07}_{-0.05}$ & 0.60$_{-0.05}$ \\
$\theta$ / deg & 52$^{+6}_{-4}$ & 39$^{+7}_{-13}$ & 60$^{f}$ \\
$\sigma_{\rm tor}$ / deg & / & / & 6.50$^{+1.00}_{-2.78}$ \\
F$_\mathrm{s}$ / $\times 10^{-2}$ & 0.59$^{+0.01}_{-0.9}$ & 5.59$^{+0.63}_{-0.73}$ & 2.25$^{+0.28}_{-0.26}$ \\
norm / $\times 10^{-3}$ & 3.47$^{+0.46}_{-0.41}$ & 3.44$^{+0.43}_{-0.38}$ & 7.10$^{+1.07}_{-1.24}$ \\
N$_{\rm H,eq}$ / 10$^{22}$ cm$^{-2}$ & 15.55$^{+4.25}_{-2.43}$ & 22.99$^{+5.15}_{-3.93}$ & / \\
\multicolumn{4}{c}{\texttt{N$_{\rm H,inst.,num.}$ \footnotesize{(LoS hydrogen column density)}}} \\
N$_{\rm H}^{Ch,1}$ / 10$^{22}$ cm$^{-2}$ & 23.19$^{+1.21}_{-1.17}$ & 26.56$^{+1.45}_{-1.37}$ & 21.41$^{+1.18}_{-1.14}$ \\
N$_{\rm H}^{Ch,2}$ / 10$^{22}$ cm$^{-2}$ & 35.83$^{+2.15}_{-2.18}$ & 41.07$^{+2.81}_{-2.55}$ & 32.44$^{+1.59}_{-1.94}$ \\
N$_{\rm H}^{Ch,3}$ / 10$^{22}$ cm$^{-2}$ & 39.04$^{+2.47}_{-2.46}$ & 44.82$^{+3.10}_{-2.92}$ & 34.36$^{+1.37}_{-1.60}$ \\
N$_{\rm H}^{Ch,4}$ / 10$^{22}$ cm$^{-2}$ & 35.69$^{+3.10}_{-3.06}$ & 40.79$^{+4.01}_{-3.48}$ & 32.00$^{+2.20}_{-2.66}$ \\
N$_{\rm H}^{Ch,5}$ / 10$^{22}$ cm$^{-2}$ & 36.78$^{+2.20}_{-2.23}$ & 42.16$^{+2.85}_{-2.61}$ & 33.04$^{+1.51}_{-1.82}$ \\
N$_{\rm H}^{Ch,6}$ / 10$^{22}$ cm$^{-2}$ & 40.46$^{+2.61}_{-2.65}$ & 46.48$^{+3.23}_{-3.10}$ & 34.66$^{+1.27}_{-1.41}$ \\
N$_{\rm H}^{Ch,7}$ / 10$^{22}$ cm$^{-2}$ & 30.72$^{+1.20}_{-1.22}$ & 35.16$^{+1.51}_{-1.41}$ & 27.25$^{+0.51}_{-0.64}$ \\
N$_{\rm H}^{NuS}$ / 10$^{22}$ cm$^{-2}$ & 21.48$^{+2.14}_{-2.18}$ & 24.08$^{+2.35}_{-2.39}$ & 20.16$^{+2.25}_{-1.53}$ \\
\multicolumn{4}{c}{\texttt{C$_{\rm inst.,num}$ \footnotesize{(Cross-normalization constant)}}} \\
C$_{Ch,1}$ & 1$^f$ & 1$^f$ & 1$^f$ \\
C$_{Ch,2}$ & 1.34$^{+0.12}_{-0.11}$ & 1.34$^{+0.12}_{-0.11}$ & 1.28$^{+0.14}_{-0.12}$ \\
C$_{Ch,3}$ & 1.35$\pm$0.12 & 1.34$^{+0.12}_{-0.11}$ & 1.26$^{+0.13}_{-0.10}$ \\
C$_{Ch,4}$ & 1.19$^{+0.14}_{-0.13}$ & 1.18$^{+0.14}_{-0.13}$ & 1.12$^{+0.15}_{-0.13}$ \\
C$_{Ch,5}$ & 1.36$\pm$0.12 & 1.36$\pm$0.11 & 1.29$^{+0.14}_{-0.11}$ \\
C$_{Ch,6}$ & 1.34$^{+0.12}_{-0.11}$ & 1.33$^{+0.12}_{-0.11}$ & 1.22$^{+0.13}_{-0.09}$ \\
C$_{Ch,7}$ & 1.45$\pm$0.09 & 1.45$^{+0.10}_{-0.09}$ & 1.34$^{+0.11}_{-0.08}$ \\
C$_{NuS}$ & 1.28$^{+0.11}_{-0.10}$ & 1.25$\pm$0.10 & 1.36$^{+0.10}_{-0.19}$ \\
\graytableline
\multicolumn{4}{c}{\texttt{Statistic}} \\
C-stat/d.o.f. & 753/681 & 762/681 & 770/682 \\
{\it T} & 2.76$\sigma$ & 3.10$\sigma$ & 3.47$\sigma$ \\
p-value & 0 & 0 & 0 \\
\multicolumn{4}{c}{\texttt{Statistic \footnotesize{(No variability)}}} \\
C-stat/d.o.f. & 2283/695 & 2261/695 & 2487/696 \\
{\it T} & 60.24$\sigma$ & 59.40$\sigma$ & 67.89$\sigma$ \\
\multicolumn{4}{c}{\texttt{Statistic \footnotesize{(No flux variability)}}} \\
C-stat/d.o.f. & 845/688 & 868/688 & 981/689 \\
{\it T} & 5.99$\sigma$ & 6.86$\sigma$ & 11.12$\sigma$ \\
\multicolumn{4}{c}{\texttt{Statistic \footnotesize{(No \nh\ variability)}}} \\
C-stat/d.o.f. & 1055/688 & 1069/688 & 1249/689 \\
{\it T} & 13.99$\sigma$ & 14.53$\sigma$ & 21.33$\sigma$ \\
\enddata
\tablecomments{Refer to note in Table~\ref{tab:spec_1}.}
\end{deluxetable}

\begin{figure}
\centering
    \subfloat[][]{
    \includegraphics[width=0.4\textwidth, trim=0 180 380 200, clip]{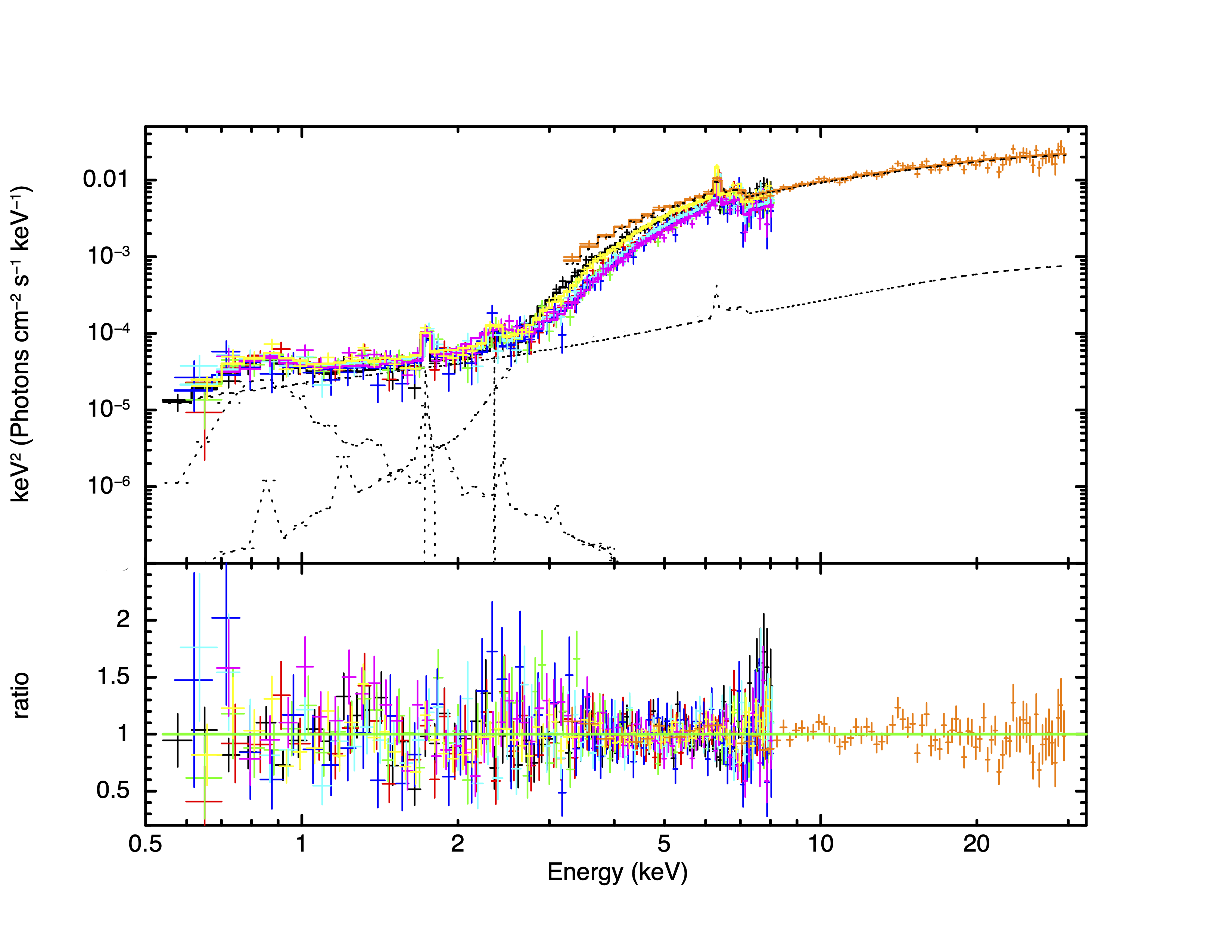}} \\
    \subfloat[][]{\includegraphics[width=0.5\textwidth, trim=10 10 0 0, clip]{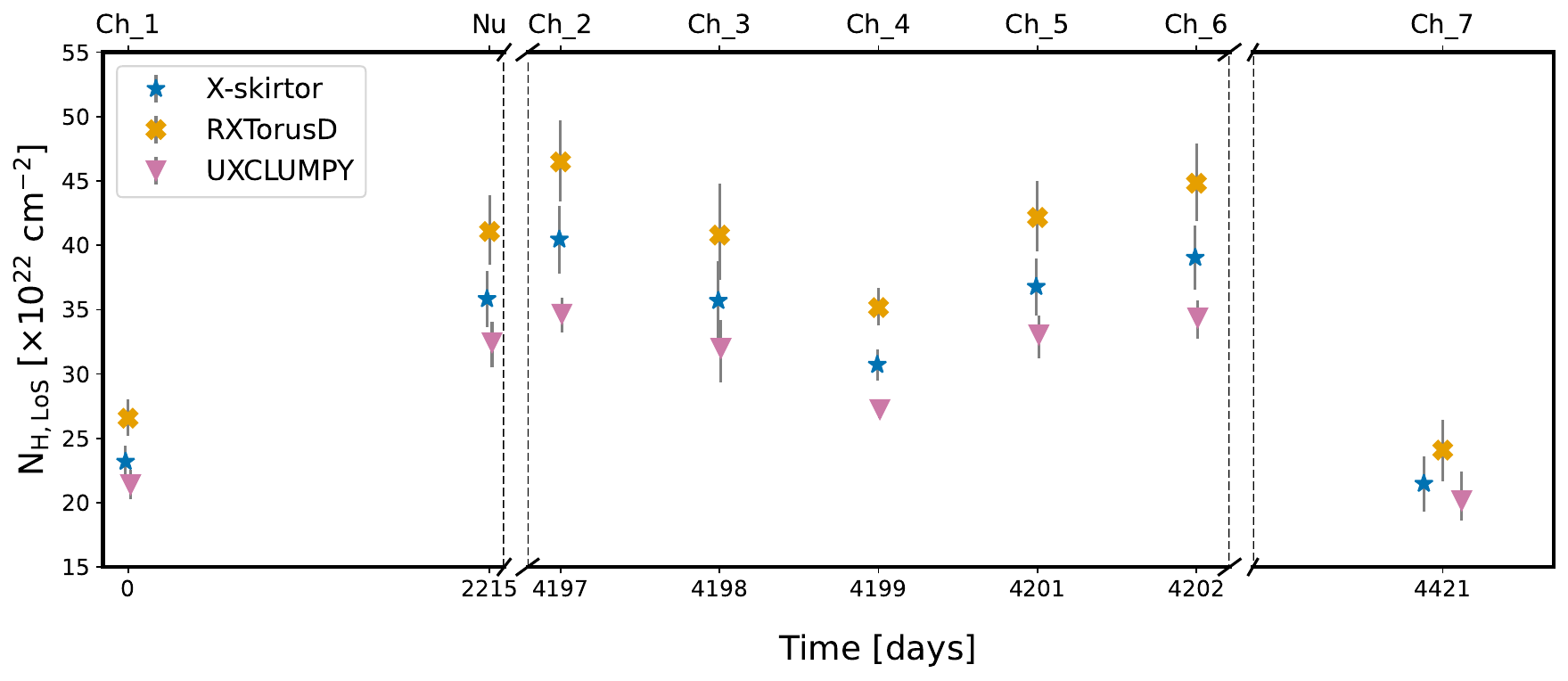}}\\
    \subfloat[][]{\includegraphics[width=0.4\textwidth, trim=5 5 6 0, clip]{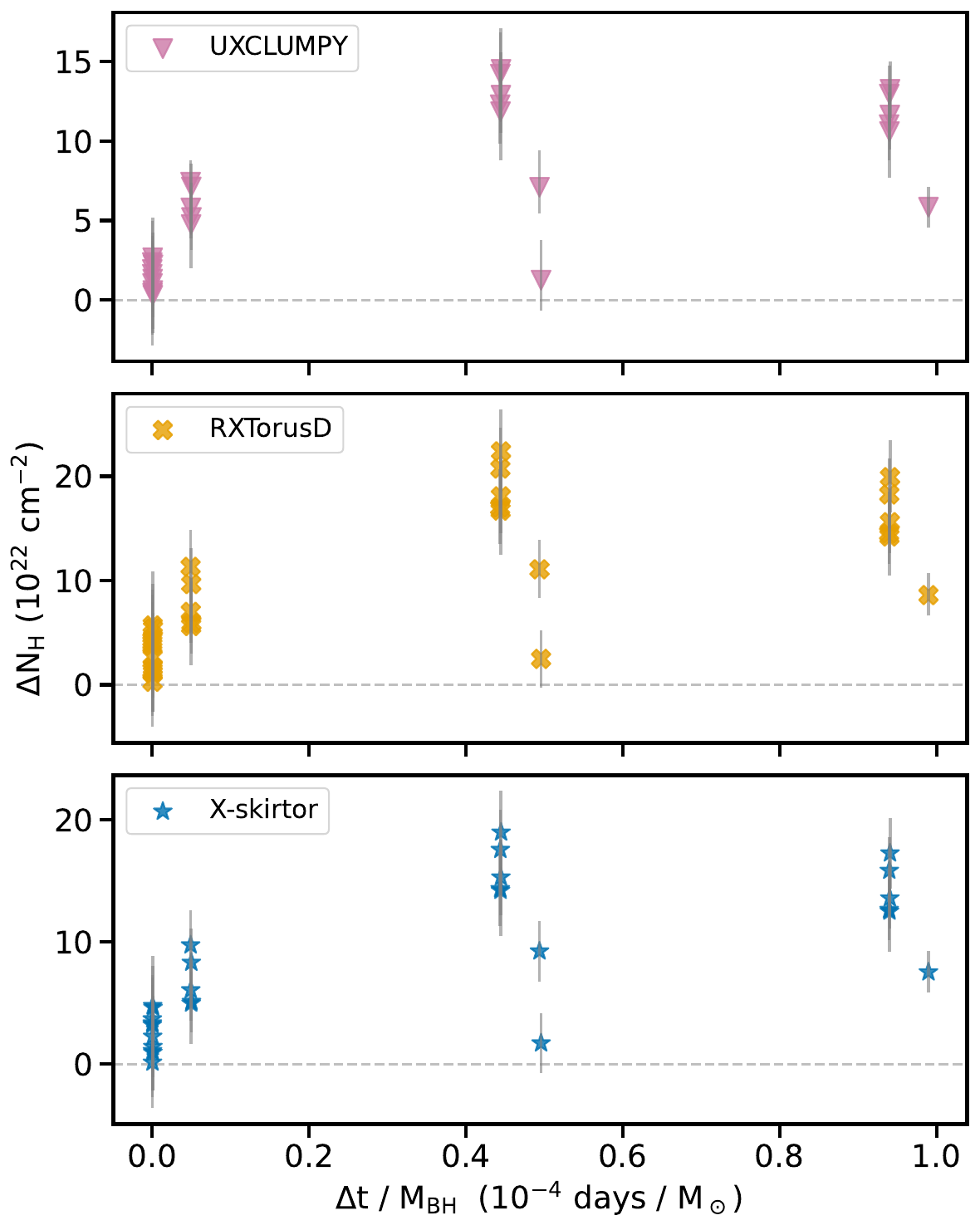}}
    \caption[]
    {{IC~5063. {\it Panel a:} best-fit with the \texttt{UXCLUMPY} torus model. The \chandra\ (in black, red, blue, cyan, pink, and yellow) and \nustar\ (FPMA and FPMB spectra are grouped together in orange) spectra are shown. {\it Panel b:} evolution of the \nhlos. The x-axis is in logarithmic scale for the left and right panel, linear for the middle panel, with a visual break to compress the long timescale between observations. {\it Panel c:} Variations of \nhlos\ as a function of the time separation between observations, normalized by the black-hole mass. Panels b and c adopt the same color code as Fig.~\ref{fig:eso_plots}. }}
    \label{fig:ic_plots}
\end{figure}

\newpage
\clearpage

\begin{table}
\centering
\caption{LEDA~511869: best-fit spectral analyses for each torus model.}
\label{tab:spec_3}
\begin{tabular}{ccccccc}
\hline
{Parameter} & \texttt{xskirtor} & \texttt{RXTorusD} & \texttt{UXCLUMPY}  \\
\hline\hline
\multicolumn{4}{c}{\texttt{apec \footnotesize{(Thermal emission)}}}\\
kT / keV & 0.63$^{+0.51}_{-0.31}$ & 0.69$^{+0.38}_{-0.64}$ & 0.55$^{+0.32}_{-0.43}$ \\
\multicolumn{4}{c}{\texttt{\footnotesize{Comptonized primary continuum}}} \\
$\Gamma$ & 1.57$^{+0.13}_{-0.15}$ & 1.47$^{+0.15}_{-0.30}$ & 1.58$^{+0.16}_{-0.14}$ \\
\multicolumn{4}{c}{\texttt{\footnotesize{Neutral reflector}}} \\
C$_\mathrm{f}$ $^{\bf +}$ & 0.42$^{+0.18}_{-0.11}$ & 0.79$^{+0.13}_{-0.09}$ & 0.30$^{+0.12}_{-0.17}$ \\
$\theta$ / deg & 70$^{+2}_{-6}$ & 78$_{-30}$ & 90$_{-29}$ \\
$\sigma_{\rm tor}$ / deg & /  & / & 17.33$^{+6.65}_{-6.53}$ \\
F$_\mathrm{s}$ / $10^{-3}$ & 2.19$^{+1.48}_{-1.46}$ & 3.02 $\pm$ 1.11 & 2.29$^{+2.56}$ \\
norm / $10^{-3}$ & 2.72$^{+0.99}_{-0.91}$ & 1.19$^{+0.55}_{-0.61}$ & 3.64$^{+1.53}_{-1.16}$ \\
N$_{\rm H,eq}$ / 10$^{22}$ cm$^{-2}$ & 1000$_{-76}$ & 80.94$^{+237.17}_{-18.27}$ & / \\
\multicolumn{4}{c}{\texttt{N$_{\rm H,inst.,num.}$ \footnotesize{(LoS hydrogen column density)}}} \\
N$_{\rm H}^{XMM}$ / 10$^{22}$ cm$^{-2}$ & 74.54$^{+7.59}_{-8.14}$ & 69.79$^{+18.24}_{-9.64}$  & 74.71$^{+7.90}_{-7.81}$ \\
N$_{\rm H}^{NuS,1}$ / 10$^{22}$ cm$^{-2}$ & 70.76$^{+10.33}_{-10.92}$  & 76.91$^{+15.09}_{-17.88}$  & 72.74$^{+12.04}_{-12.23}$ \\
N$_{\rm H}^{NuS,2}$ / 10$^{22}$ cm$^{-2}$ & 53.71$^{+7.27}_{-7.85}$ & 48.28$^{+15.45}_{-9.55}$  & 52.70$^{+9.97}_{-7.85}$ \\
\multicolumn{4}{c}{\texttt{C$_{\rm inst.,num}$ \footnotesize{(Cross-normalization constant)}}} \\
C$_{XMM}$ & 1$^f$ & 1$^f$ & 1$^f$ \\
C$_{NuS,1}$ & 0.71$^{+0.18}_{-0.14}$ & 0.59$^{+0.37}_{-0.15}$ & 0.51$^{+0.16}_{-0.13}$  \\
C$_{NuS,2}$ & 0.63$^{+0.16}_{-0.20}$ & 0.70$^{+0.15}_{-0.14}$  & 0.62$^{+0.17}_{-0.13}$  \\
\graytableline
\multicolumn{4}{c}{\texttt{Statistic}} \\ \smallskip
$\chi^2$/d.o.f. & 186/212 & 186/212 & 187/212
\\
{\it T} & 1.79$\sigma$ & 1.79$\sigma$ & 1.72$\sigma$ \\
p-value & 0.12 & 0.39 & 0.20 \\
\multicolumn{4}{c}{\texttt{Statistic \footnotesize{(No variability)}}} \\
$\chi^2$/d.o.f. & 422/216 & 423/216 & 424/217
\\
{\it T} & 14.02$\sigma$ & 14.08$\sigma$ & 14.05$\sigma$ \\
\multicolumn{4}{c}{\texttt{Statistic \footnotesize{(No flux variability)}}} \\
$\chi^2$/d.o.f. & 197/214 & 199/214 & 197/214
\\
{\it T} & 1.16$\sigma$ & 1.03$\sigma$ & 1.16$\sigma$ \\
\multicolumn{4}{c}{\texttt{Statistic \footnotesize{(No \nh\ variability)}}} \\
$\chi^2$/d.o.f. & 206/214 & 204/214 & 205/214
\\
{\it T} & 0.55$\sigma$ & 0.68$\sigma$ & 0.62$\sigma$ \\
\hline
\end{tabular}
\tablecomments{\footnotesize Refer to note in Table~\ref{tab:spec_1}.}
\end{table}

\begin{figure}[h!]
\centering
    \subfloat[][]{
    \includegraphics[width=0.4\textwidth, trim=0 30 80 80, clip]{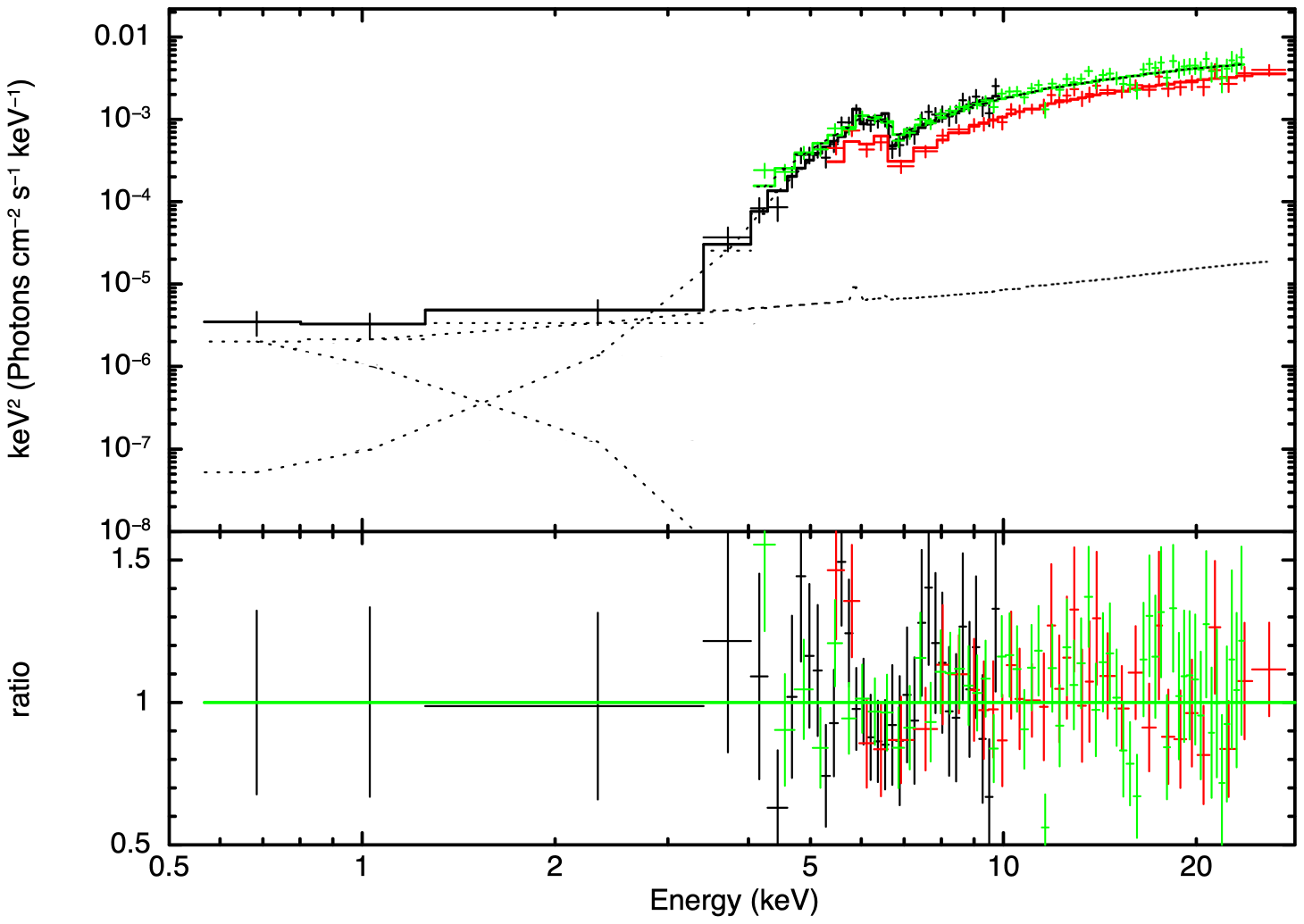}} \\
    \subfloat[][]{
    \includegraphics[width=0.4\textwidth, trim=10 10 13 0, clip]{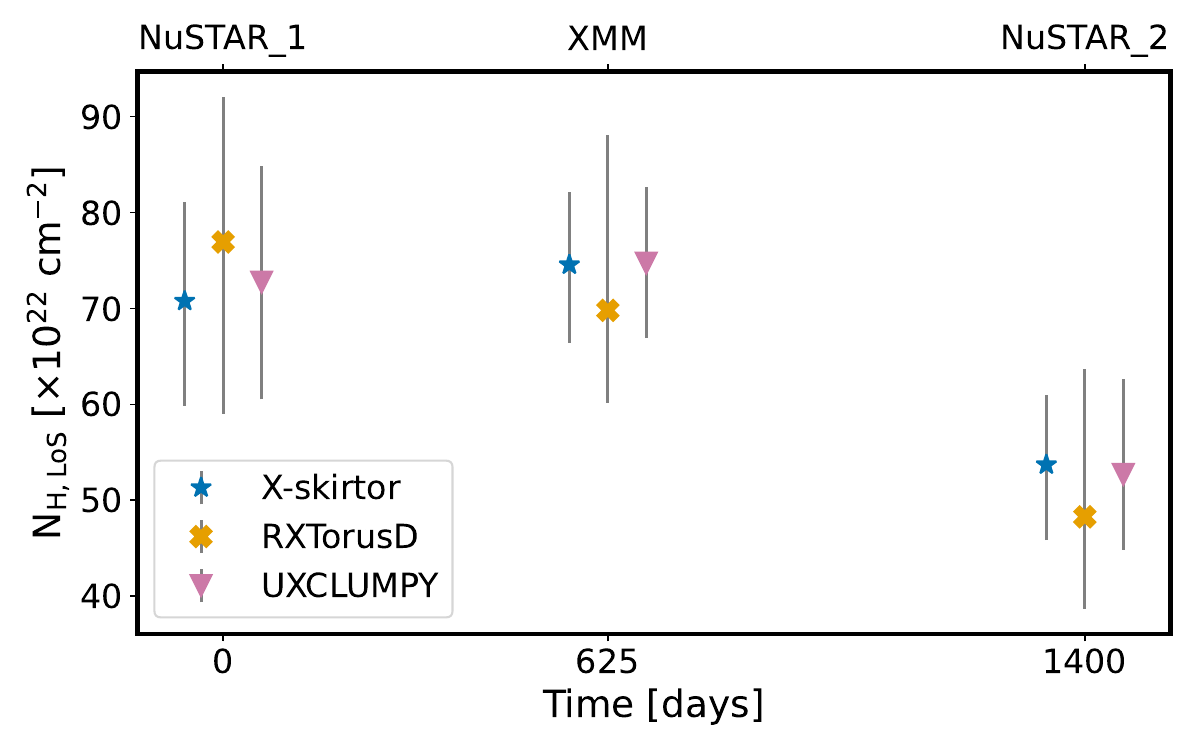}}\\
    \subfloat[][]{\includegraphics[width=0.4\textwidth, trim=5 5 6 0, clip]{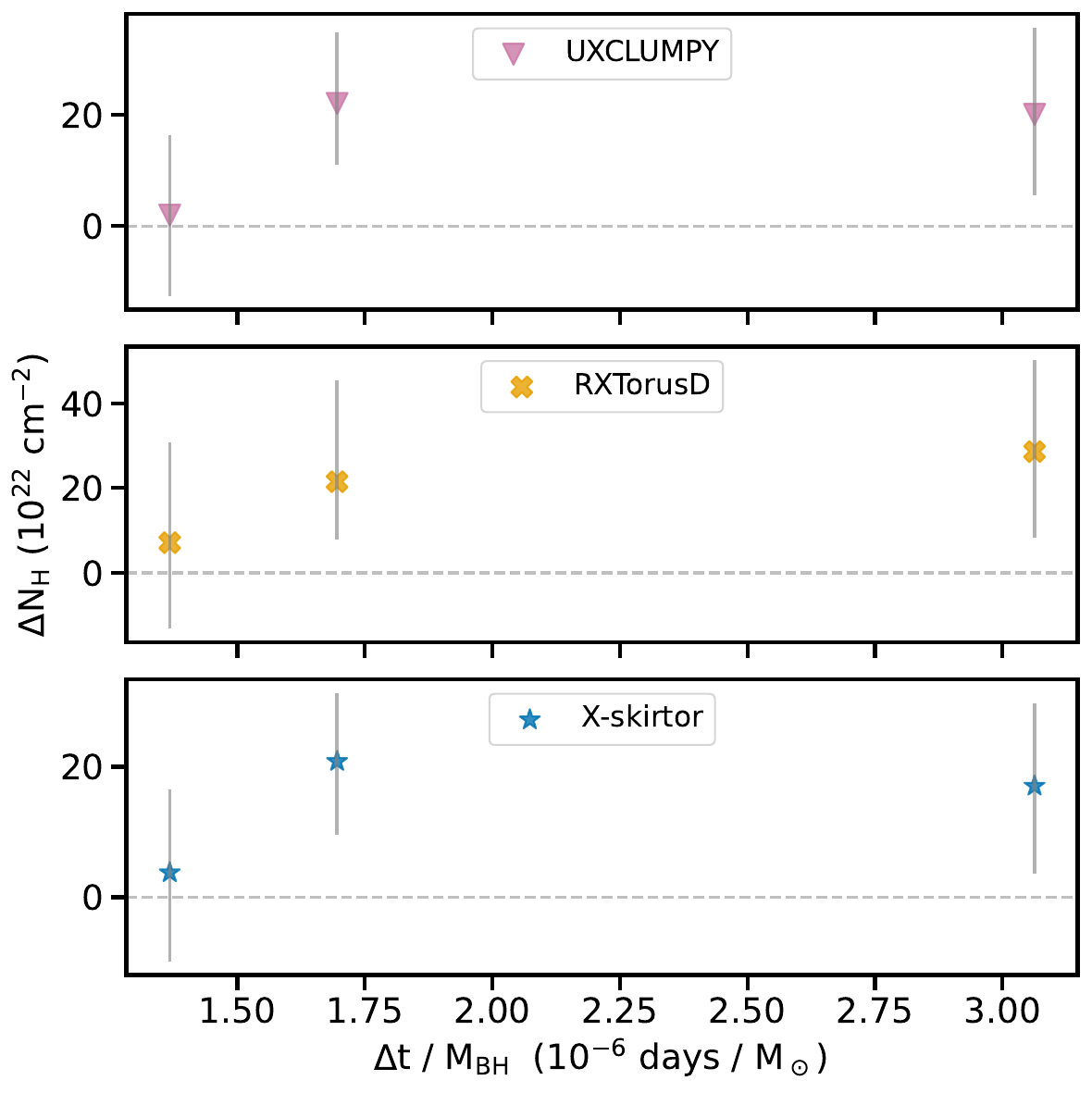}}
    \caption[]
    {{LEDA~511869. {\it Panel a:} Best-fit with the \texttt{UXCLUMPY} torus model. The \xmm\ (in black), \nustar\ 1 (in red) and \nustar\ 2 (in green) spectra are shown. FPMA and FPMB spectra are grouped together for each \nustar\ observation. {\it Panel b:} evolution of the \nhlos. The x-axis is in linear scale. {\it Panel c:} Variations of \nhlos\ as a function of the time separation between observations, normalized by the black-hole mass. Panels b and c adopt the same color code as Fig.~\ref{fig:eso_plots}.}}
    \label{fig:leda_plots}
\end{figure}

\newpage
\clearpage

\renewcommand{\arraystretch}{1.8}
\begin{deluxetable*}{lccc}
\tablecaption{Mrk~18: best-fit spectral analyses for each torus model.\label{tab:spec_5}}
\tabletypesize{\scriptsize}
\tablewidth{\textwidth}
\tablehead{
\colhead{Parameter} & \colhead{\texttt{xskirtor}} & \colhead{\texttt{RXTorusD}} & \colhead{\texttt{UXCLUMPY}}
}
\startdata
\multicolumn{4}{c}{\texttt{apec \footnotesize{(Thermal emission)}}} \\
kT / keV & 0.74$_{-0.31}^{+0.28}$ & 0.74$^{+0.18}_{-0.22}$ & 0.74$^{+0.19}_{-0.22}$ \\
\multicolumn{4}{c}{\texttt{\footnotesize{Comptonized primary continuum}}} \\
$\Gamma$ & 1.94$^{+0.18}_{-0.24}$ & 1.93$^{+0.22}_{-0.40}$ & 1.95$^{+0.13}_{-0.17}$ \\
\multicolumn{4}{c}{\texttt{\footnotesize{Neutral reflector}}} \\
C$_\mathrm{f}$ $^{\bf +}$ & 0.59$_{-0.19}^{+0.04}$ & 0.95$_{-0.03}$ & 0.30$_{-0.22}^{+0.26}$ \\
$\theta$ / deg & 20$^{f}$ & 18$^{+2}_{-3}$ & 20$^{f}$ \\
$\sigma_{\rm tor}$ / deg & / & / & 24.03$_{-11.09}^{+22.98}$ \\
F$_\mathrm{s}$ / $10^{-2}$ & 1.06$^{+0.94}_{-0.54}$ & 0.50$^{+2.33}_{-0.35}$ & 0.11$^{+0.31}_{-0.09}$ \\
norm / $10^{-3}$ & 0.47$\pm$0.18 & 0.44$^{+0.20}_{-0.24}$ & 0.62$^{+0.21}_{-0.18}$ \\
N$_{\rm H,eq}$ / 10$^{22}$ cm$^{-2}$ & 246.94$^{+123.93}_{-103.35}$ & 85.21$^{+53.24}_{-56.87}$ & / \\
\multicolumn{4}{c}{\texttt{N$_{\rm H,inst.,num.}$ \footnotesize{(LoS hydrogen column density)}}} \\
N$_{\rm H}^{Ch}$ / 10$^{22}$ cm$^{-2}$ & 9.87$^{+2.21}_{-1.91}$ & 12.72$^{+3.12}_{-5.94}$ & 8.99$^{+2.00}_{-1.68}$ \\
N$_{\rm H}^{XMM}$ / 10$^{22}$ cm$^{-2}$ & 14.18$^{+3.00}_{-2.66}$ & 19.50$^{+4.70}_{-4.23}$ & 13.29$^{+2.89}_{-2.14}$ \\
N$_{\rm H}^{NuS,1}$ / 10$^{22}$ cm$^{-2}$ & 15.86$^{+4.97}_{-4.43}$ & 21.64$^{+7.11}_{-8.67}$ & 15.65$^{+4.36}_{-4.61}$ \\
\multicolumn{4}{c}{\texttt{C$_{\rm inst.,num}$ \footnotesize{(Cross-normalization constant)}}} \\
C$_{Ch}$ & 1$^f$ & 1$^f$ & 1$^f$ \\
C$_{XMM}$ & 2.25$^{+0.55}_{-0.42}$ & 2.31$^{+0.56}_{-0.50}$ & 2.30$^{+0.57}_{-0.41}$ \\
C$_{NuS,1}$ & 2.01$^{+0.58}_{-0.39}$ & 2.04$^{+0.55}_{-0.47}$ & 2.09$^{+0.59}_{-0.43}$ \\
\graytableline
\multicolumn{4}{c}{\texttt{Statistic}} \\
$\chi^2$/d.o.f. & 222/193 & 227/192 & 231/193 \\
{\it T} & 2.08$\sigma$ & 2.52$\sigma$ & 2.74$\sigma$ \\
p-value & 0.05 & 0.04 & 0.03 \\
\multicolumn{4}{c}{\texttt{Statistic \footnotesize{(No variability)}}} \\
$\chi^2$/d.o.f. & 356/199 & 364/198 & 367/199 \\
{\it T} & 11.12$\sigma$ & 11.80$\sigma$ & 11.91$\sigma$ \\
\multicolumn{4}{c}{\texttt{Statistic \footnotesize{(No flux variability)}}} \\
$\chi^2$/d.o.f. & 294/196 & 225/195 & 302/196 \\
{\it T} & 7.00$\sigma$ & 2.15$\sigma$ & 7.57$\sigma$ \\
\multicolumn{4}{c}{\texttt{Statistic \footnotesize{(No \nh\ variability)}}} \\
$\chi^2$/d.o.f. & 234/196 & 235/195 & 242/196 \\
{\it T} & 2.71$\sigma$ & 2.86$\sigma$ & 3.29$\sigma$ \\
\enddata
\tablecomments{\footnotesize Refer to note in Table~\ref{tab:spec_1}.}
\end{deluxetable*}

\begin{figure}[h!]
\centering
    \subfloat[][]{\includegraphics[width=0.4\textwidth, trim=0 30 80 80, clip]{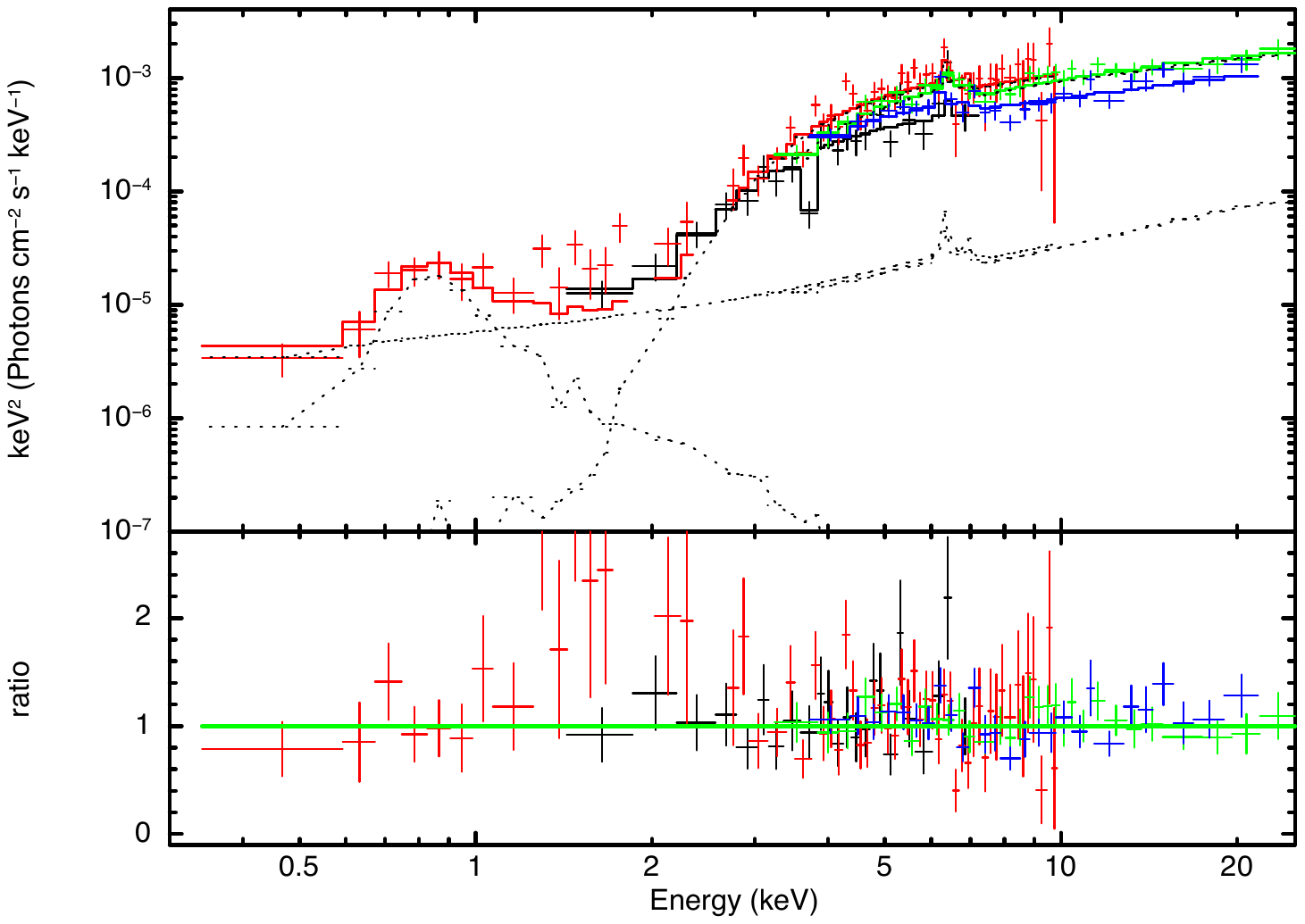}} \\
    \subfloat[][]{\includegraphics[width=0.45\textwidth, trim=10 10 10 0, clip]{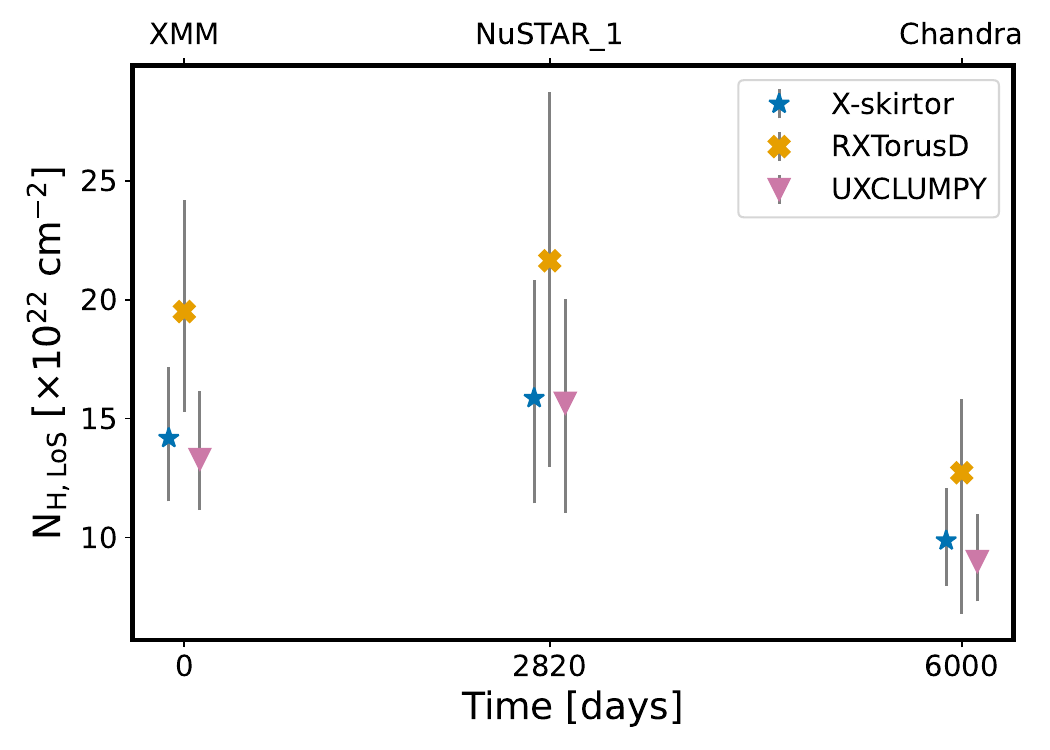}}\\
    \subfloat[][]{\includegraphics[width=0.45\textwidth, trim=5 5 6 0, clip]{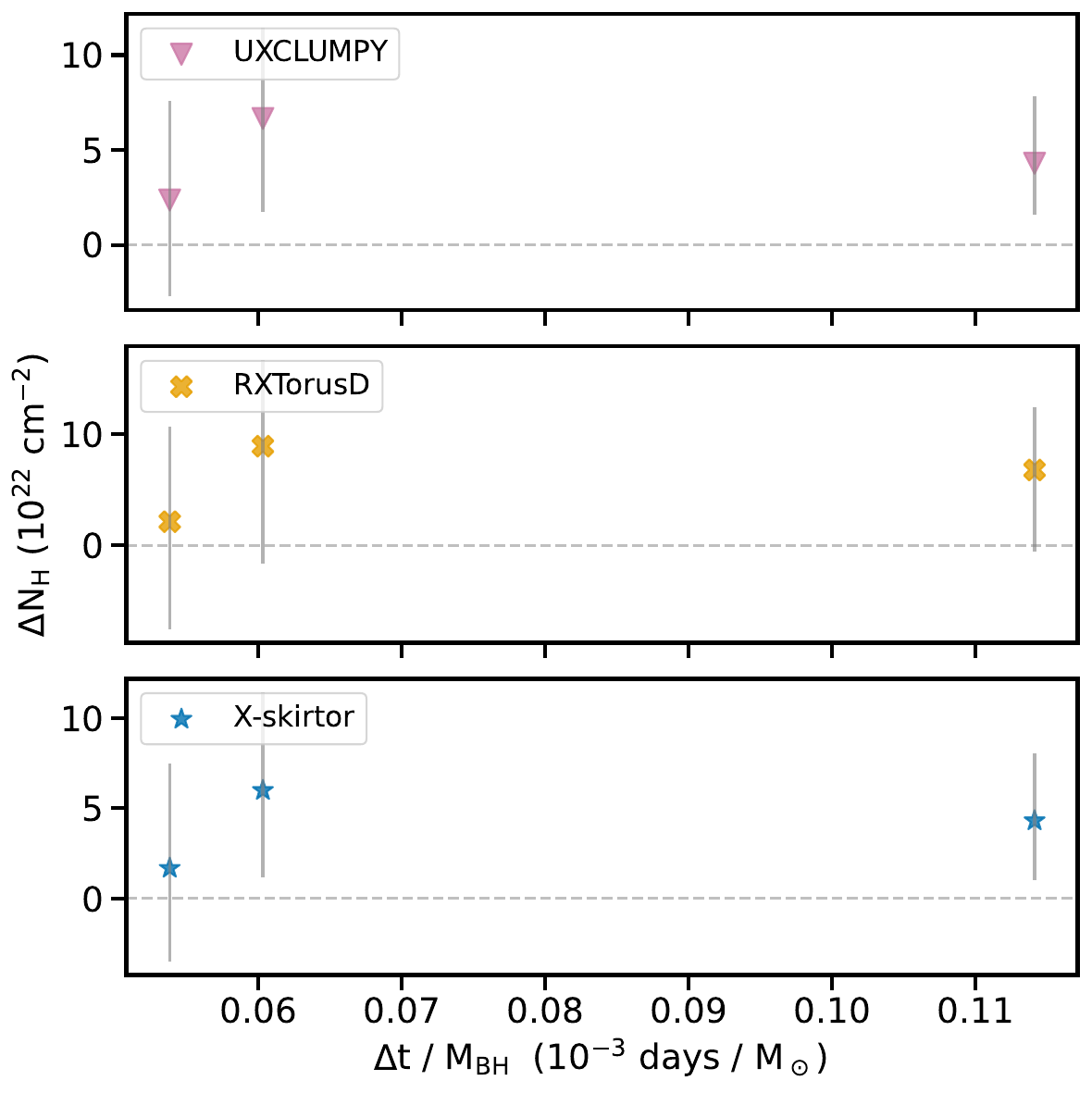}}
    \caption[]
    {{Mrk~18. {\it Panel a:} Best-fit with the \texttt{UXCLUMPY} torus model. The \chandra\ (in black), XMM (in red) and \nustar\ (FPMA and FPMB spectra grouped together in green and blue) spectra are shown.
    {\it Panel b:} evolution of the \nhlos. The left part of the x-axis is in linear scale with a visual break, the right part is not in scale to clearly visualize the \chandra\ and \nustar\ 2 observations, which are only few hours apart. {\it Panel c:} Variations of \nhlos\ as a function of the time separation between observations, normalized by the black-hole mass. Panels b and c adopt the same color code as Fig.~\ref{fig:eso_plots}.}}
    \label{fig:mrk18_plots}
\end{figure}

\renewcommand{\arraystretch}{1.9}
\begin{deluxetable*}{lccc}
\tablecaption{Mrk~1498: best-fit spectral analyses for each torus model.\label{tab:spec_6}}
\tabletypesize{\scriptsize}
\tablewidth{\textwidth}
\tablehead{
\colhead{Parameter} & \colhead{\texttt{xskirtor}} & \colhead{\texttt{RXTorusD}} & \colhead{\texttt{UXCLUMPY}}
}
\startdata
\multicolumn{4}{c}{\texttt{apec \footnotesize{(Thermal emission)}}} \\
kT / keV & 0.10$^{+0.03}_{-0.02}$ & 0.11$\pm$0.02 & 0.10$^{+0.03}_{-0.02}$ \\
\multicolumn{4}{c}{\texttt{\footnotesize{Comptonized primary continuum}}} \\
$\Gamma$ & 1.53$^{+0.08}_{-0.06}$ & 1.55$^{+0.10}_{-0.06}$ & 1.57$\pm$0.04 \\
\multicolumn{4}{c}{\texttt{\footnotesize{Neutral reflector}}} \\
C$_\mathrm{f}$ $^{\bf +}$ & 0.50$^{+0.12}$ & 0.58$^{+0.30}_{-0.35}$ & 0$^f$ \\
$\theta$ / deg & 78$^{+16}$ & 87$_{-21}$ & 87$^{+4}_{-16}$ \\
$\sigma_{\rm tor}$ / deg & / & / & 5.97$^{+3.04}_{-0.61}$ \\
F$_\mathrm{s}$ / $10^{-2}$ & 1.44$^{+0.21}_{-0.24}$ & 1.48$^{+0.55}_{-0.32}$ & 1.55$\pm$0.01 \\
norm / $10^{-3}$ & 2.98$^{+0.61}_{-0.37}$ & 2.87$^{+0.81}_{-0.79}$ & 4.79$\pm$0.88 \\
N$_{\rm H,eq}$ & 41.77$^{+43.95}_{-19.11}$ & 82.58$^{+177.25}_{-26.42}$ & / \\
\multicolumn{4}{c}{\texttt{N$_{\rm H,inst.,num.}$ \footnotesize{(LoS hydrogen column density)}}} \\
N$_{\rm H}^{Ch}$ / 10$^{22}$ cm$^{-2}$ & 17.26$^{+1.14}_{-1.03}$ & 19.04$^{+1.81}_{-1.77}$ & 16.72$^{+1.06}_{-1.15}$ \\
N$_{\rm H}^{XMM}$ / 10$^{22}$ cm$^{-2}$ & 16.17$^{+1.24}_{-1.36}$ & 17.34$^{+2.08}_{-2.05}$ & 15.75$^{+1.32}_{-1.41}$ \\
N$_{\rm H}^{NuS}$ / 10$^{22}$ cm$^{-2}$ & 10.87$^{+1.51}_{-1.43}$ & 10.75$^{+2.28}_{-2.51}$ & 10.70$^{+1.37}_{-1.32}$ \\
\multicolumn{4}{c}{\texttt{C$_{\rm inst.,num}$ \footnotesize{(Cross-normalization constant)}}} \\
C$_{Ch}$ & 1$^f$ & 1$^f$ & 1$^f$ \\
C$_{XMM}$ & 1.01$\pm$0.08 & 0.99$^{+0.08}_{-0.07}$ & 1.02$\pm$0.08 \\
C$_{NuS}$ & 0.99$^{+0.08}_{-0.07}$ & 0.96$\pm$0.07 & 1.03$^{+0.09}_{-0.26}$ \\
\graytableline
\multicolumn{4}{c}{\texttt{Statistic}} \\
C-stat/d.o.f. & 346/297 & 321/297 & 334/298 \\
{\it T} & 2.84$\sigma$ & 1.39$\sigma$ & 2.09$\sigma$ \\
p-value & 1.12e-03 & 0.01 & 2.14e-03 \\
\multicolumn{4}{c}{\texttt{Statistic \footnotesize{(No variability)}}} \\
C-stat/d.o.f. & 536/301 & 522/301 & 527/302 \\
{\it T} & 12.95$\sigma$ & 12.74$\sigma$ & 13.55$\sigma$ \\
\multicolumn{4}{c}{\texttt{Statistic \footnotesize{(No flux variability)}}} \\
C-stat/d.o.f. & 346/299 & 322/299 & 334/300 \\
{\it T} & 2.72$\sigma$ & 1.33$\sigma$ & 1.96$\sigma$ \\
\multicolumn{4}{c}{\texttt{Statistic \footnotesize{(No \nh\ variability)}}} \\
C-stat/d.o.f. & 393/299 & 378/299 & 379/300 \\
{\it T} & 5.44$\sigma$ & 4.57$\sigma$ & 4.56$\sigma$ \\
\enddata
\tablecomments{\footnotesize Refer to note in Table~\ref{tab:spec_1}.}
\end{deluxetable*}

\begin{figure}[h!]
\centering
    \subfloat[][]{
    \includegraphics[width=0.4\textwidth, trim=0 30 80 80, clip]{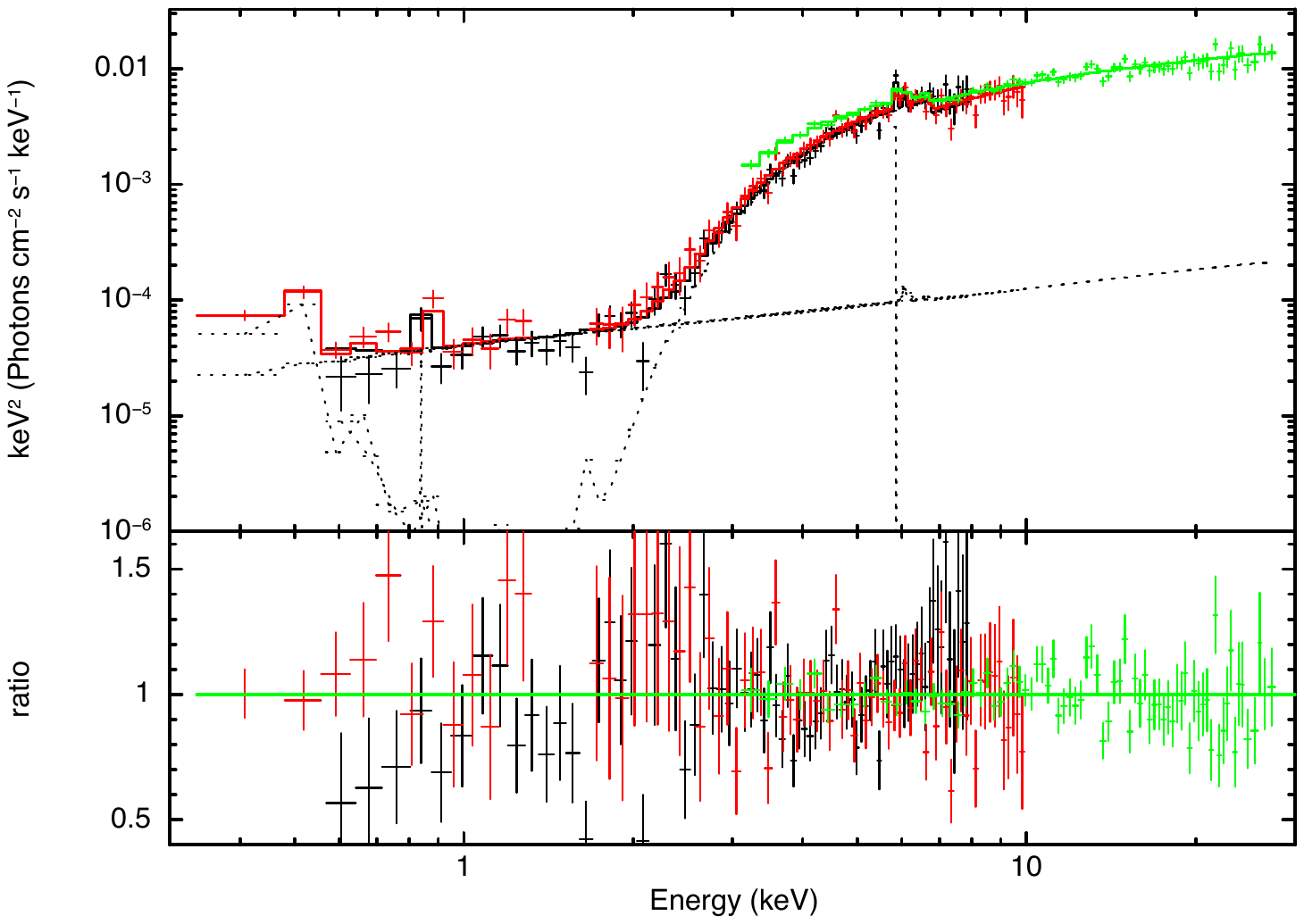}} \\
    \subfloat[][]{
    \includegraphics[width=0.4\textwidth, trim=0 0 0 0, clip]{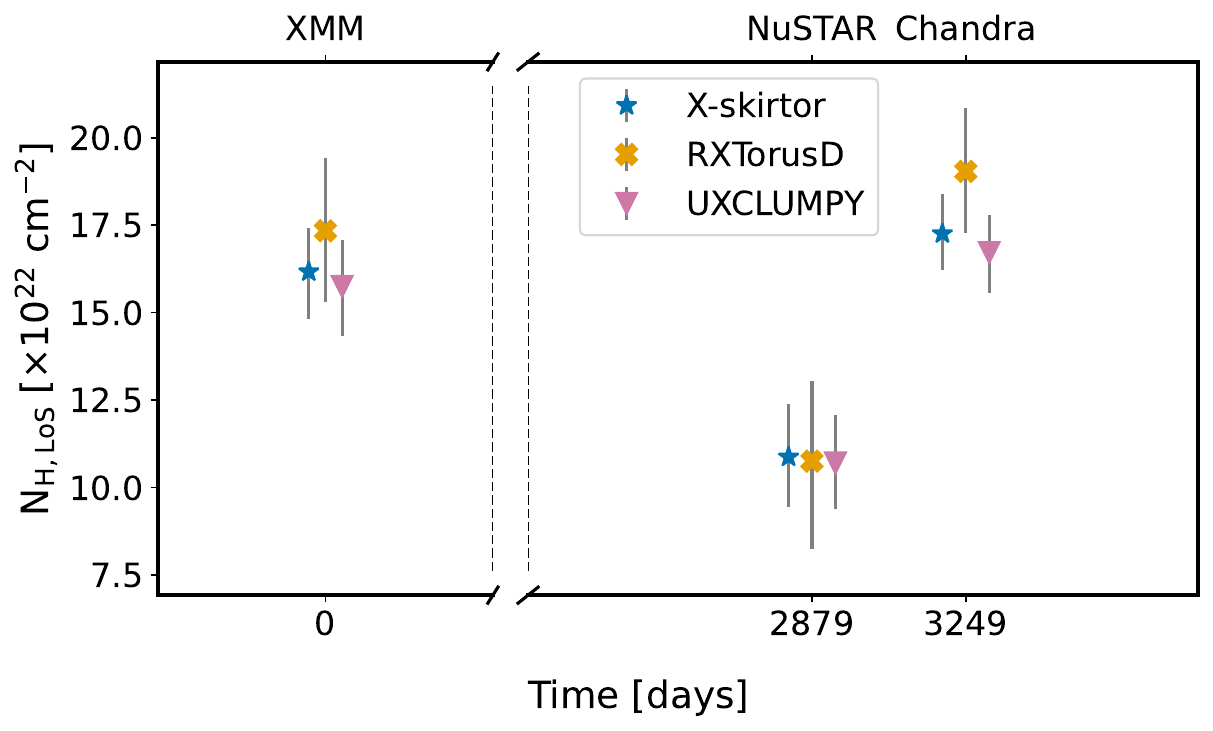}} \\
    \subfloat[][]{\includegraphics[width=0.4\textwidth, trim=5 5 6 0, clip]{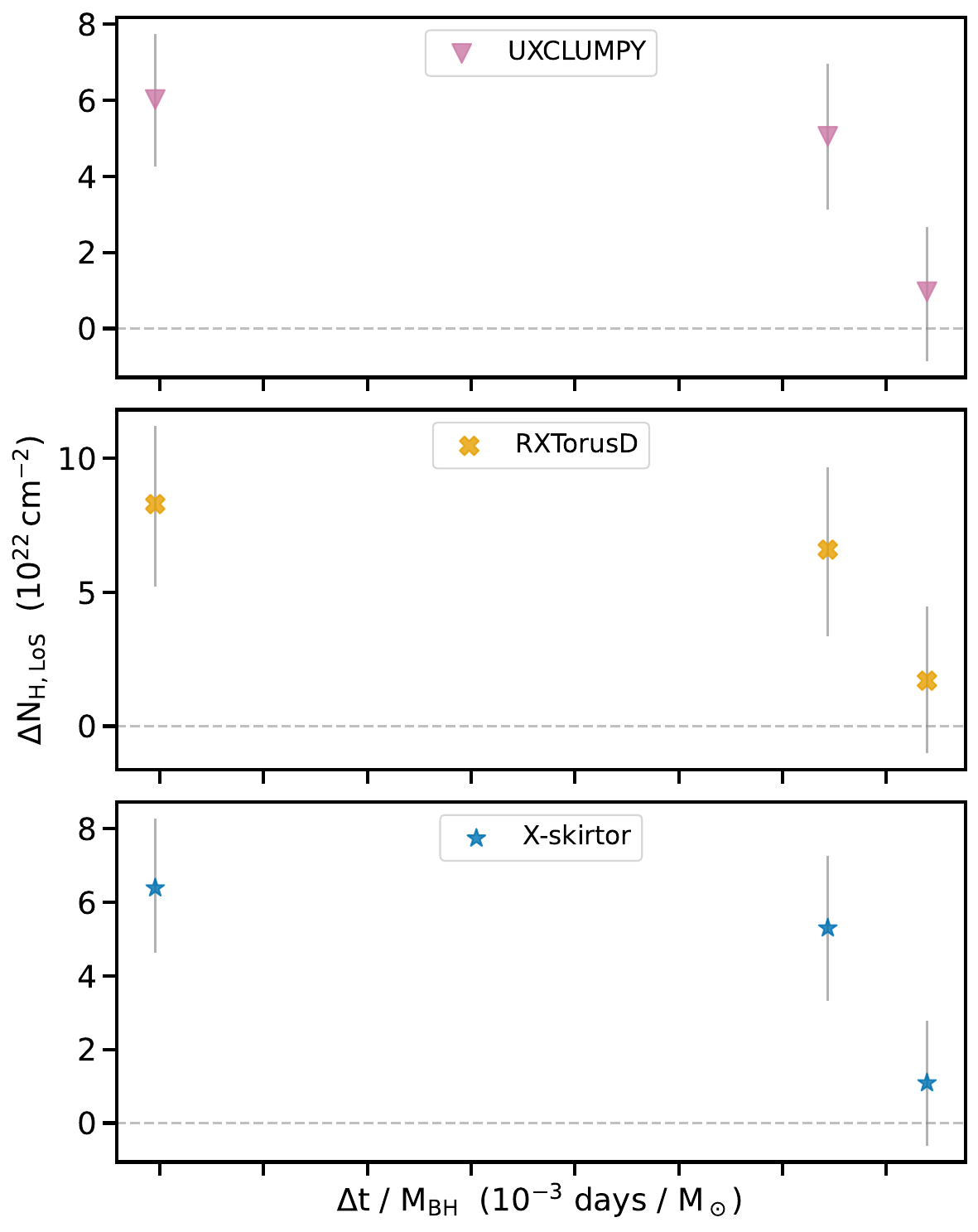}}
    \caption[]
    {{Mrk~1498. {\it Panel a:} Best-fit with the \texttt{UXCLUMPY} torus model. The \chandra\ (in black), \xmm\ (in red), and \nustar\ (FPMA and FPMB spectra are grouped together in green) spectra are shown.
    {\it Panel b:} evolution of \nhlos\ with time. The x-axis is in logarithmic scale with a visual break to compress the long timescale between observations. {\it Panel c:} variations of \nhlos\ as a function of the time separation between observations, normalized by the black-hole mass. Panels b and c adopt the same color code as Fig.~\ref{fig:eso_plots}.}}
    \label{fig:mrk1498_plots}
\end{figure}

\newpage
\clearpage

\renewcommand{\arraystretch}{1.5}
\begin{deluxetable*}{lccc}
\tablecaption{NGC~1194: best-fit spectral analyses for each torus model.\label{tab:spec_7}}
\tabletypesize{\scriptsize}
\tablewidth{\textwidth}
\tablehead{
\colhead{Parameter} & \colhead{\texttt{xskirtor}} & \colhead{\texttt{RXTorusD}} & \colhead{\texttt{UXCLUMPY}}
}
\startdata
\multicolumn{4}{c}{\texttt{apec \footnotesize{(Thermal emission)}}} \\
kT / keV & 1.18$^{+0.10}_{-0.09}$ & 1.18$^{+0.09}_{-0.08}$ & 1.22$^{+0.09}_{-0.07}$ \\
\multicolumn{4}{c}{\texttt{\footnotesize{Comptonized primary continuum}}} \\
$\Gamma$ & 1.54$^{+0.05}_{-0.06}$ & 1.49$^{+0.04}_{-0.04}$ & 1.56$^{+0.04}_{-0.16}$ \\
\multicolumn{4}{c}{\texttt{\footnotesize{Neutral reflector}}} \\
C$_\mathrm{f}$ $^{\bf +}$ & 0.25$^{+0.04}$ & 0.40$^{+0.01}_{-0.04}$ & 0.30$_{-0.04}^{+0.06}$ \\
$\theta$ / deg & 76$^{+3}_{-1}$ & 51$\pm$1 & 60$^f$ \\
$\sigma_{\rm tor}$ / deg & / & / & 7.00$_{-2.31}^{+4.07}$ \\
F$_\mathrm{s}$ / $10^{-2}$ & 0.33$^{+0.11}_{-0.05}$ & 0.24$^{+0.54}_{-0.53}$ & 0.60$^{+0.18}_{-0.30}$ \\
norm / $10^{-3}$ & 4.94$^{+1.03}_{-1.26}$ & 4.92$^{+0.99}_{-0.65}$ & 2.17$\pm$0.01 \\
N$_{\rm H,eq}$ / 10$^{22}$ cm$^{-2}$ & 43.51$^{+9.75}_{-9.69}$ & 63.58$^{+10.51}_{-4.20}$ & / \\
\multicolumn{4}{c}{\texttt{N$_{\rm H,inst.,num.}$ \footnotesize{(LoS hydrogen column density)}}} \\
N$_{\rm H}^{Ch,1}$ / 10$^{22}$ cm$^{-2}$ & 135.24$^{+22.96}_{-16.34}$ & 148.04$^{+19.25}_{-13.21}$ & 136.17$^{+9.08}_{-8.57}$ \\
N$_{\rm H}^{Ch,2}$ / 10$^{22}$ cm$^{-2}$ & 136.68$^{+50.08}_{-24.88}$ & 146.94$^{+37.68}_{-22.02}$ & 141.12$^{+23.06}_{-20.25}$ \\
N$_{\rm H}^{Ch,3}$ / 10$^{22}$ cm$^{-2}$ & 139.59$^{+41.65}_{-22.47}$ & 150.48$^{+33.78}_{-20.26}$ & 137.41$^{+17.44}_{-18.63}$ \\
N$_{\rm H}^{Ch,4}$ / 10$^{22}$ cm$^{-2}$ & 149.71$_{-33.97}$ & 161.87$^{+153.00}_{-32.50}$ & 149.99$^{+65.69}_{-28.12}$ \\
N$_{\rm H}^{XMM,1}$ / 10$^{22}$ cm$^{-2}$ & 136.55$^{+18.91}_{-15.49}$ & 168.64$^{+22.42}_{-16.98}$ & 130.33$^{+11.59}_{-13.77}$ \\
N$_{\rm H}^{XMM,2}$ / 10$^{22}$ cm$^{-2}$ & 138.90$^{+12.19}_{-11.91}$ & 163.38$^{+11.17}_{-9.84}$ & 140.08$^{+6.62}_{-6.23}$ \\
N$_{\rm H}^{NuS,1}$ / 10$^{22}$ cm$^{-2}$ & 208.60$^{+16.21}_{-17.31}$ & 238.54$^{+15.21}_{-14.33}$ & 256.22$^{+22.06}_{-21.96}$ \\
\multicolumn{4}{c}{\texttt{C$_{\rm inst.,num}$ \footnotesize{(Cross-normalization constant)}}} \\
C$_{Ch,1}$ & 1$^f$ & 1$^f$ & 1$^f$ \\
C$_{Ch,2}$ & 0.94$\pm$0.15 & 0.94$^{+0.14}_{-0.15}$ & 0.87$^{+0.13}_{-0.12}$ \\
C$_{Ch,3}$ & 1.04$^{+0.16}_{-0.15}$ & 1.04$^{+0.14}_{-0.15}$ & 1.04$^{+0.14}_{-0.12}$ \\
C$_{Ch,4}$ & 1.29$^{+0.25}_{-0.24}$ & 1.30$^{+0.23}_{-0.25}$ & 1.27$^{+0.23}_{-0.21}$ \\
C$_{XMM,1}$ & 0.76$^{+0.12}_{-0.10}$ & 0.92$^{+0.10}_{-0.11}$ & 0.62$^{+0.09}_{-0.08}$ \\ 
C$_{XMM,2}$ & 0.83$^{+0.10}_{-0.08}$ & 0.94$^{+0.08}_{-0.09}$ & 0.65$^{+0.06}_{-0.03}$ \\ 
C$_{NuS,1}$ & 1.29$^{+0.18}_{-0.16}$ & 1.49$^{+0.13}_{-0.16}$ & 0.77$^{+0.14}_{-0.11}$ \\
\graytableline
\multicolumn{4}{c}{\texttt{Statistic}} \\
$\chi^2$/d.o.f. & 653/594 & 653/594 & 660/595 \\
{\it T} & 2.42$\sigma$ & 2.42$\sigma$ & 2.66$\sigma$ \\
p-value & 2.11e-15 & 6.67e-9 & 0 \\
\multicolumn{4}{c}{\texttt{Statistic \footnotesize{(No variability)}}} \\
$\chi^2$/d.o.f. & 785/608 & 771/608 & 846/609 \\
{\it T} & 7.18$\sigma$ & 6.61$\sigma$ & 9.30$\sigma$ \\
\multicolumn{4}{c}{\texttt{Statistic \footnotesize{(No flux variability)}}} \\
$\chi^2$/d.o.f. & 766/601 & 758/601 & 831/602 \\
{\it T} & 6.73$\sigma$ & 6.40$\sigma$ & 9.33$\sigma$ \\
\multicolumn{4}{c}{\texttt{Statistic \footnotesize{(No \nh\ variability)}}} \\
$\chi^2$/d.o.f. & 747/601 & 790/601 & 919/602 \\
{\it T} & 5.95$\sigma$ & 7.71$\sigma$ & 12.92$\sigma$ \\
\enddata
\tablecomments{\footnotesize Refer to note in Table~\ref{tab:spec_1}.}
\end{deluxetable*}

\begin{figure}[h!]
\centering
    \subfloat[][]{\includegraphics[width=0.4\textwidth, trim=0 30 80 80, clip]{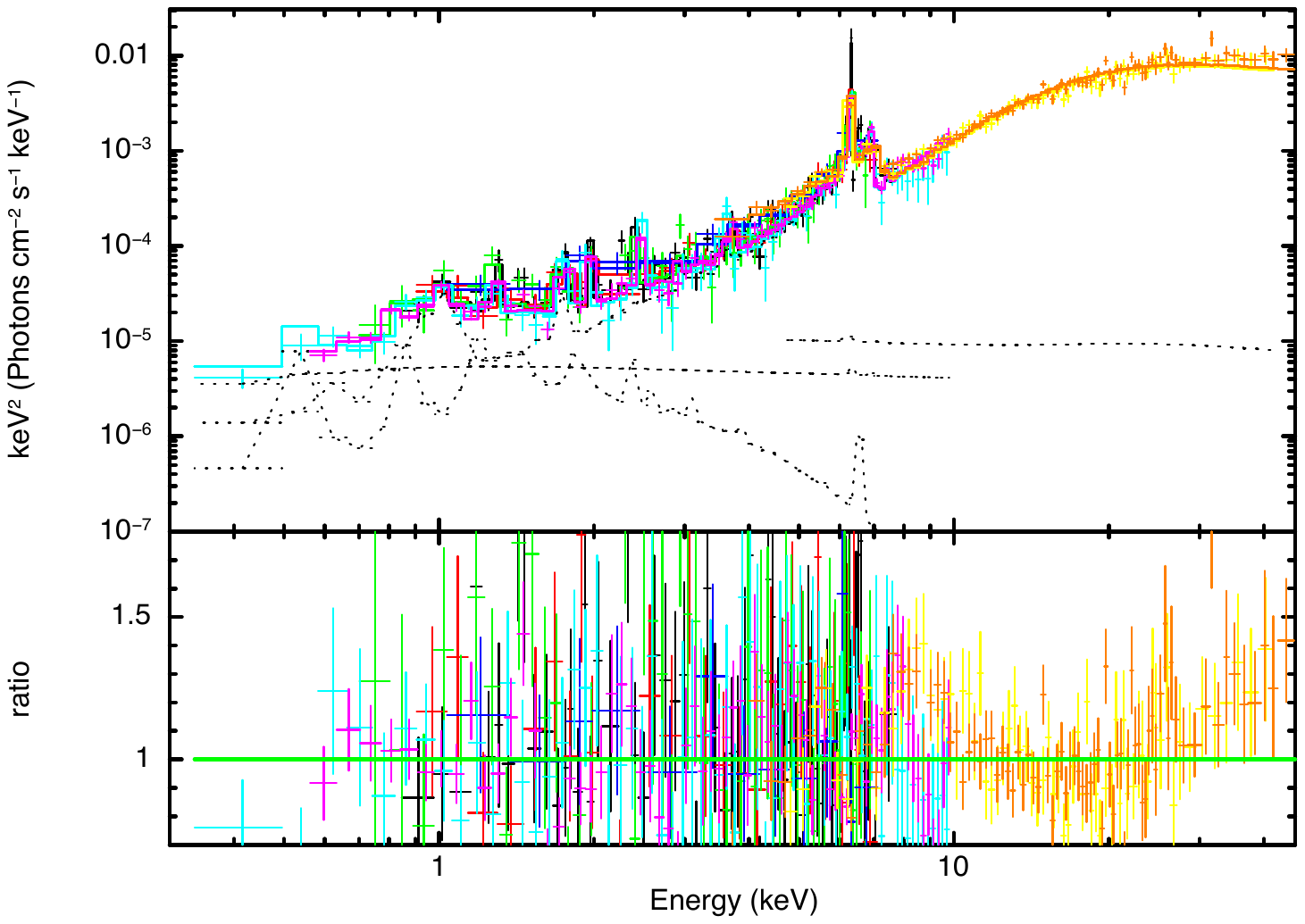}} \\
    \subfloat[][]{\includegraphics[width=0.5\textwidth, trim=10 10 10 0, clip]{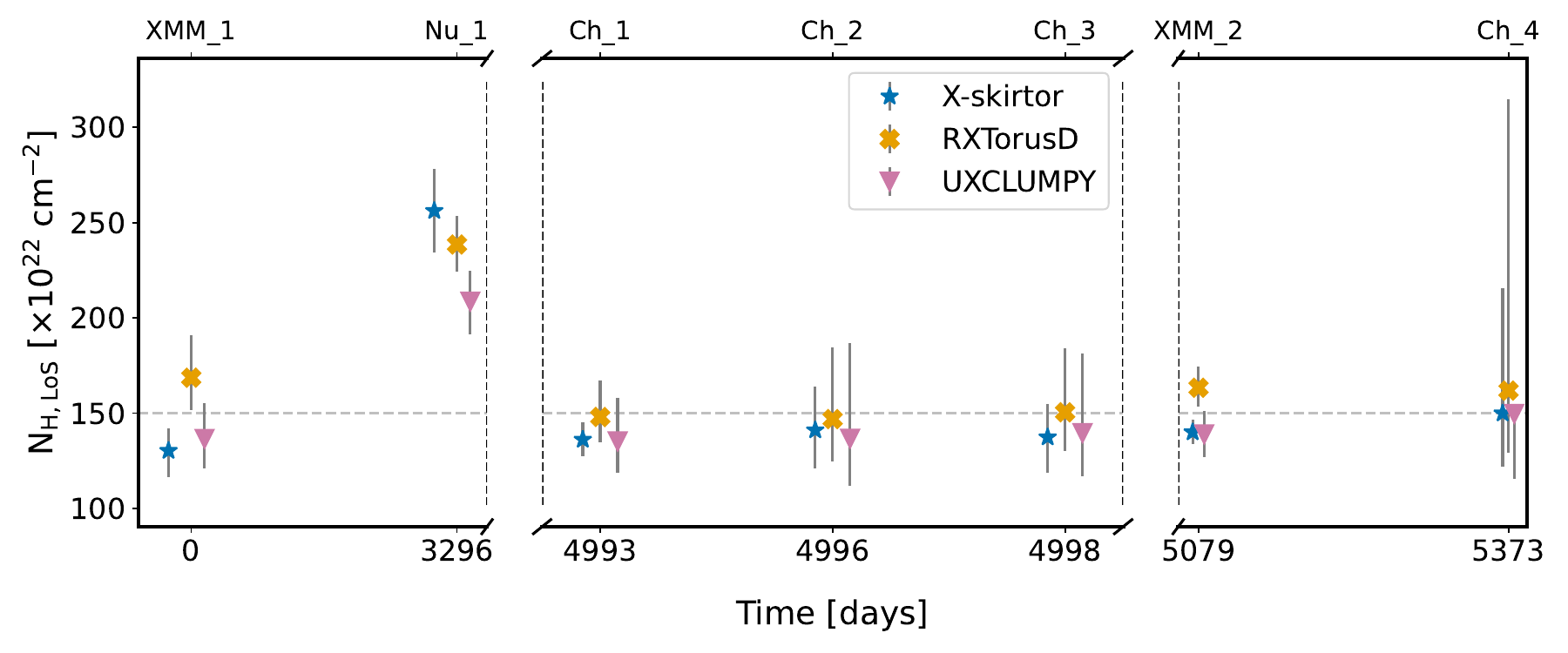}} \\
    \subfloat[][]{\includegraphics[width=0.4\textwidth, trim=5 5 6 0, clip]{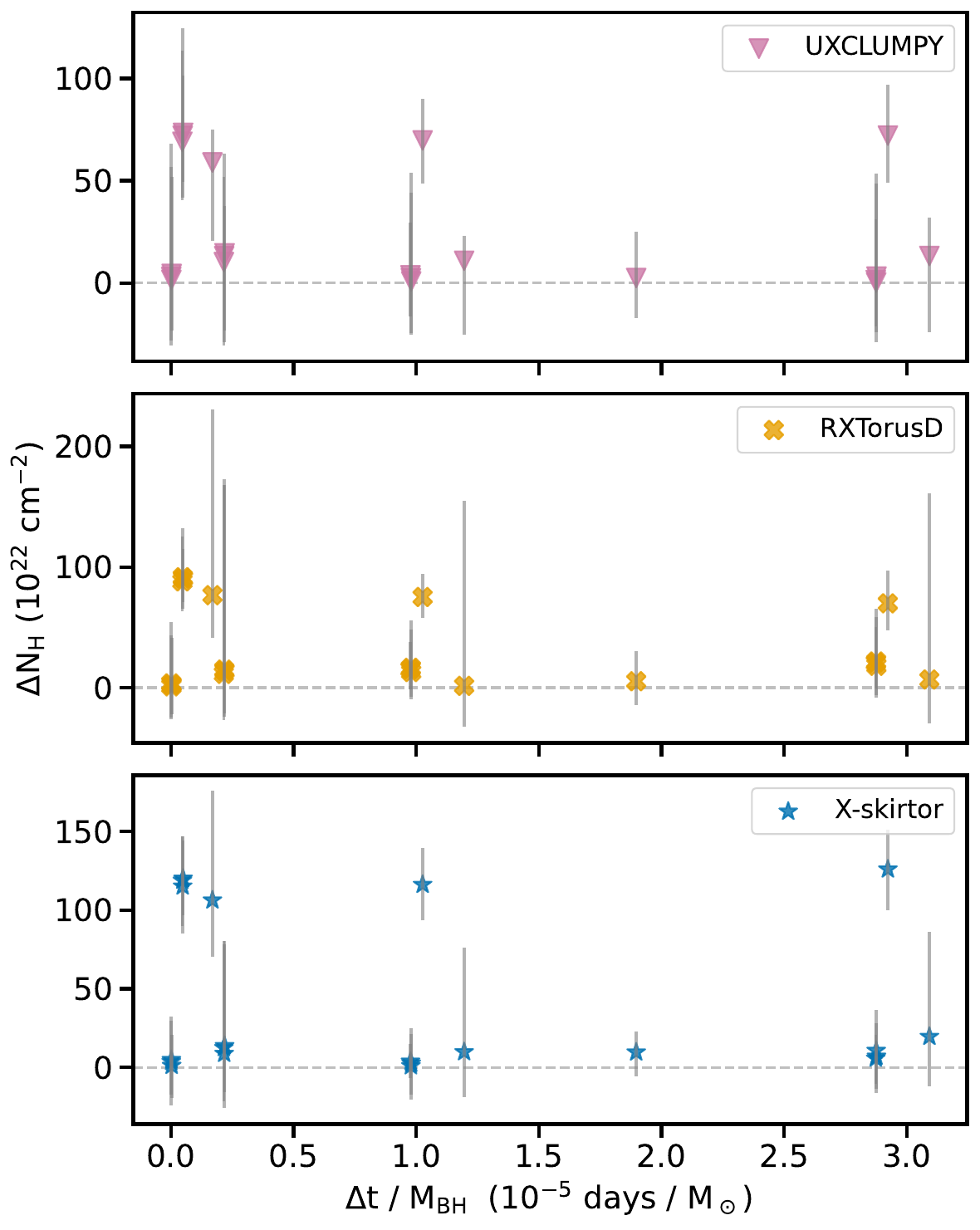}}
    \caption[]
    {{NGC~1194. {\it Panel a:} Best-fit with the \texttt{UXCLUMPY} torus model. The \chandra\ (in black, red, green, and blue), \xmm\ (in cyan and pink), and \nustar\ (FPMA and FPMB spectra are grouped together in yellow and orange) spectra are shown. See Appendix~\ref{source_note} for more information on the residuals in the fit of the \nustar\ data.
    {\it Panel b:} evolution of the \nhlos. The x-axis is in logarithmic scale for the left and right panel, linear for the middle panel, with a visual break to compress the long timescale between observations.  {\it Panel c:} Variations of \nhlos\ as a function of the time separation between observations, normalized by the black-hole mass. Panels b and c adopt the same color code as Fig.~\ref{fig:eso_plots}.}}
    \label{fig:1194_plots}
\end{figure}

\newpage
\clearpage

\renewcommand{\arraystretch}{1.8}
\begin{deluxetable*}{lccc}
\tablecaption{NGC~2655: best-fit spectral analyses for each torus model.\label{tab:spec_8}}
\tabletypesize{\scriptsize}
\tablewidth{\textwidth}
\tablehead{
\colhead{Parameter} & \colhead{\texttt{xskirtor}} & \colhead{\texttt{RXTorusD}} & \colhead{\texttt{UXCLUMPY}}
}
\startdata
\multicolumn{4}{c}{\texttt{apec \footnotesize{(Thermal emission)}}} \\
kT / keV & 0.74$\pm$0.05 & 0.74$^{+0.05}_{-0.05}$ & 0.82$\pm$0.17 \\
\multicolumn{4}{c}{\texttt{\footnotesize{Comptonized primary continuum}}} \\
$\Gamma$ & 2.06$^{+0.19}_{-0.17}$ & 2.08$^{+0.19}_{-0.14}$ & 2.02$^{+0.13}_{-0.11}$ \\
\multicolumn{4}{c}{\texttt{\footnotesize{Neutral reflector}}} \\
C$_\mathrm{f}$ $^{\bf +}$ & 0.45$^{+0.20}_{-0.12}$ & 0.61$^{-0.11}_{-0.21}$ & 0.006$^{+0.004}_{-0.007}$ \\
$\theta$ / deg & 61$_{-11}$ & 54$^{+17}_{-14}$ & 60$^f$ \\
$\sigma_{\rm tor}$ / deg & / & / & 0$^f$ \\
F$_\mathrm{s}$ / $10^{-2}$ & 1.95$^{+2.04}_{-0.91}$ & 3.05$^{+2.46}_{-1.55}$ & 0.08$^{+2.26}_{-0.78}$ \\
norm / $10^{-3}$ & 1.06$^{+0.93}_{-0.54}$ & 0.71$^{+0.94}_{-0.33}$ & 4.16$^{+3.34}_{-0.03}$ \\
N$_{\rm H,eq}$ / 10$^{22}$ cm$^{-2}$ & 10.00$^{+12.39}_{-6.24}$ & 100.02$^{+368.63}_{-62.20}$ & / \\
\multicolumn{4}{c}{\texttt{N$_{\rm H,inst.,num.}$ \footnotesize{(LoS hydrogen column density)}}} \\
N$_{\rm H}^{Ch}$ / 10$^{22}$ cm$^{-2}$ & 31.42$^{+7.27}_{-11.70}$ & 25.88$^{+22.55}_{-6.18}$ & 26.76$^{+8.47}_{-5.26}$ \\
N$_{\rm H}^{XMM}$ / 10$^{22}$ cm$^{-2}$ & 40.86$^{+11.21}_{-14.78}$ & 33.51$^{+27.48}_{-6.52}$ & 32.32$^{+7.56}_{-6.71}$ \\
N$_{\rm H}^{NuS,1}$ / 10$^{22}$ cm$^{-2}$ & 44.20$^{+21.66}_{-16.76}$ & 32.94$^{-32.94}_{-17.67}$ & 33.43$^{+15.30}_{-15.49}$ \\
N$_{\rm H}^{NuS,2}$ / 10$^{22}$ cm$^{-2}$ & 30.68$^{+9.88}_{-14.31}$ & 21.40$^{+22.68}_{-8.08}$ & 20.02$^{+9.64}_{-7.88}$ \\
\multicolumn{4}{c}{\texttt{C$_{\rm inst.,num}$ \footnotesize{(Cross-normalization constant)}}} \\
C$_{Ch}$ & 1$^f$ & 1$^f$ & 1$^f$ \\
C$_{XMM}$ & 2.28$^{+0.67}_{-0.51}$ & 2.16$^{+0.64}_{-0.42}$ & 1.96$^{+0.41}_{-0.34}$ \\
C$_{NuS,1}$ & 2.83$^{+1.73}_{-1.06}$ & 2.43$^{+5.09}_{-0.77}$ & 2.69$^{+1.28}_{-0.97}$ \\
C$_{NuS,2}$ & 2.04$^{+0.85}_{-0.69}$ & 1.92$^{+0.75}_{-0.48}$ & 2.56$^{+0.82}_{-0.88}$ \\
\graytableline
\multicolumn{4}{c}{\texttt{Statistic}} \\
C-stat/d.o.f. & 260/242 & 264/242 & 245/244 \\
{\it T} & 1.16$\sigma$ & 1.41$\sigma$ & 0.02$\sigma$ \\
p-value & 0.75 & 0.85 & 0.74 \\
\multicolumn{4}{c}{\texttt{Statistic \footnotesize{(No variability)}}} \\
C-stat/d.o.f. & 375/248 & 374/248 & 362/250 \\
{\it T} & 8.06$\sigma$ & 8.00$\sigma$ & 7.08$\sigma$ \\
\multicolumn{4}{c}{\texttt{Statistic \footnotesize{(No flux variability)}}} \\
C-stat/d.o.f. & 349/245 & 303/245 & 317/247 \\
{\it T} & 6.64$\sigma$ & 3.71$\sigma$ & 4.45$\sigma$ \\
\multicolumn{4}{c}{\texttt{Statistic \footnotesize{(No \nh\ variability)}}} \\
C-stat/d.o.f. & 269/245 & 270/245 & 259/247 \\
{\it T} & 0.76$\sigma$ & 1.60$\sigma$ & 1.53$\sigma$ \\
\enddata
\tablecomments{\footnotesize Refer to note in Table~\ref{tab:spec_1}.}
\end{deluxetable*}

\begin{figure}[h!]
\centering
    \subfloat[][]{\includegraphics[width=0.4\textwidth, trim=0 30 80 80, clip]{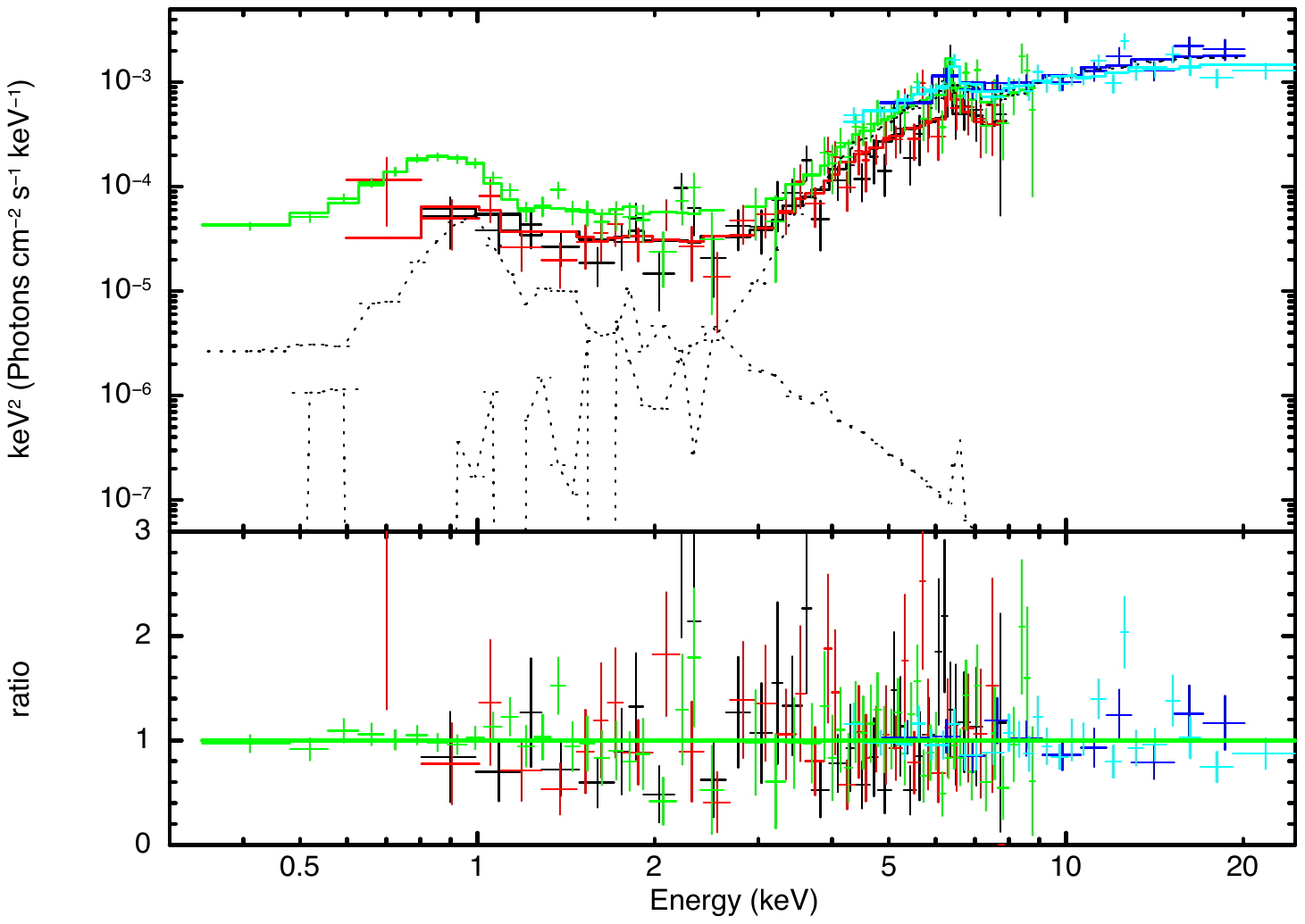}} \\
    \subfloat[][]{\includegraphics[width=0.45\textwidth, trim=10 10 10 0, clip]{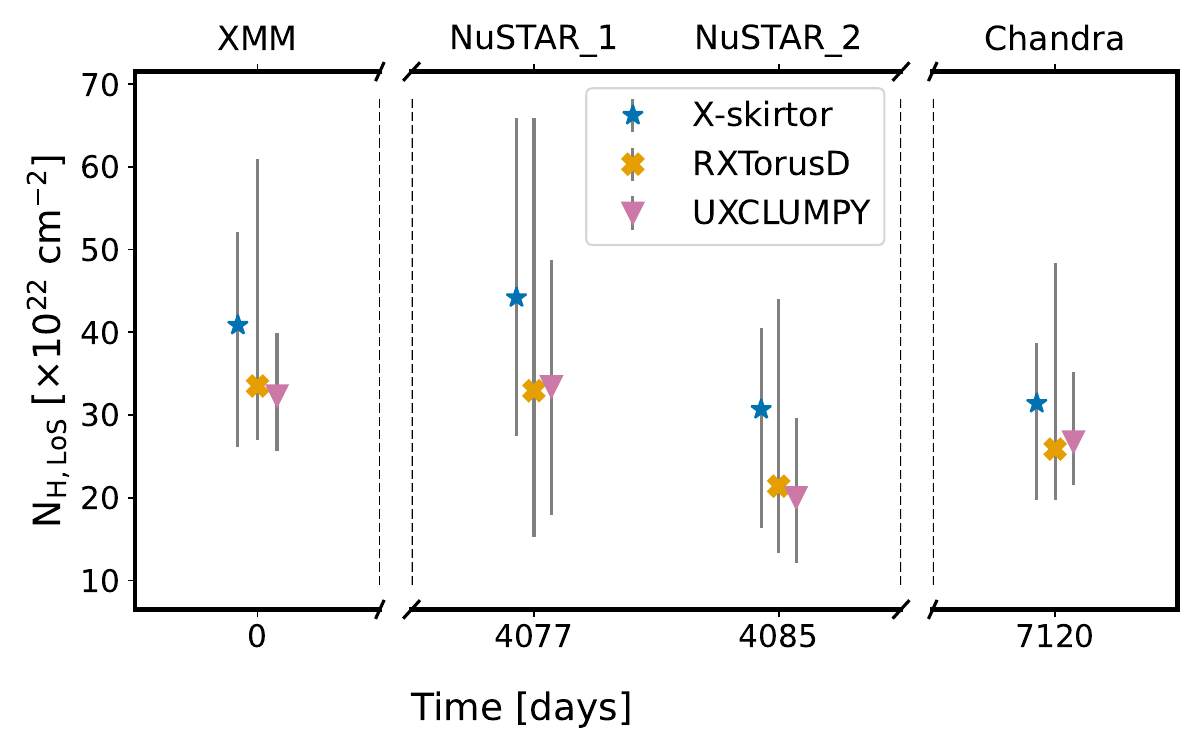}} \\
    \subfloat[][]{\includegraphics[width=0.45\textwidth, trim=5 5 7 0, clip]{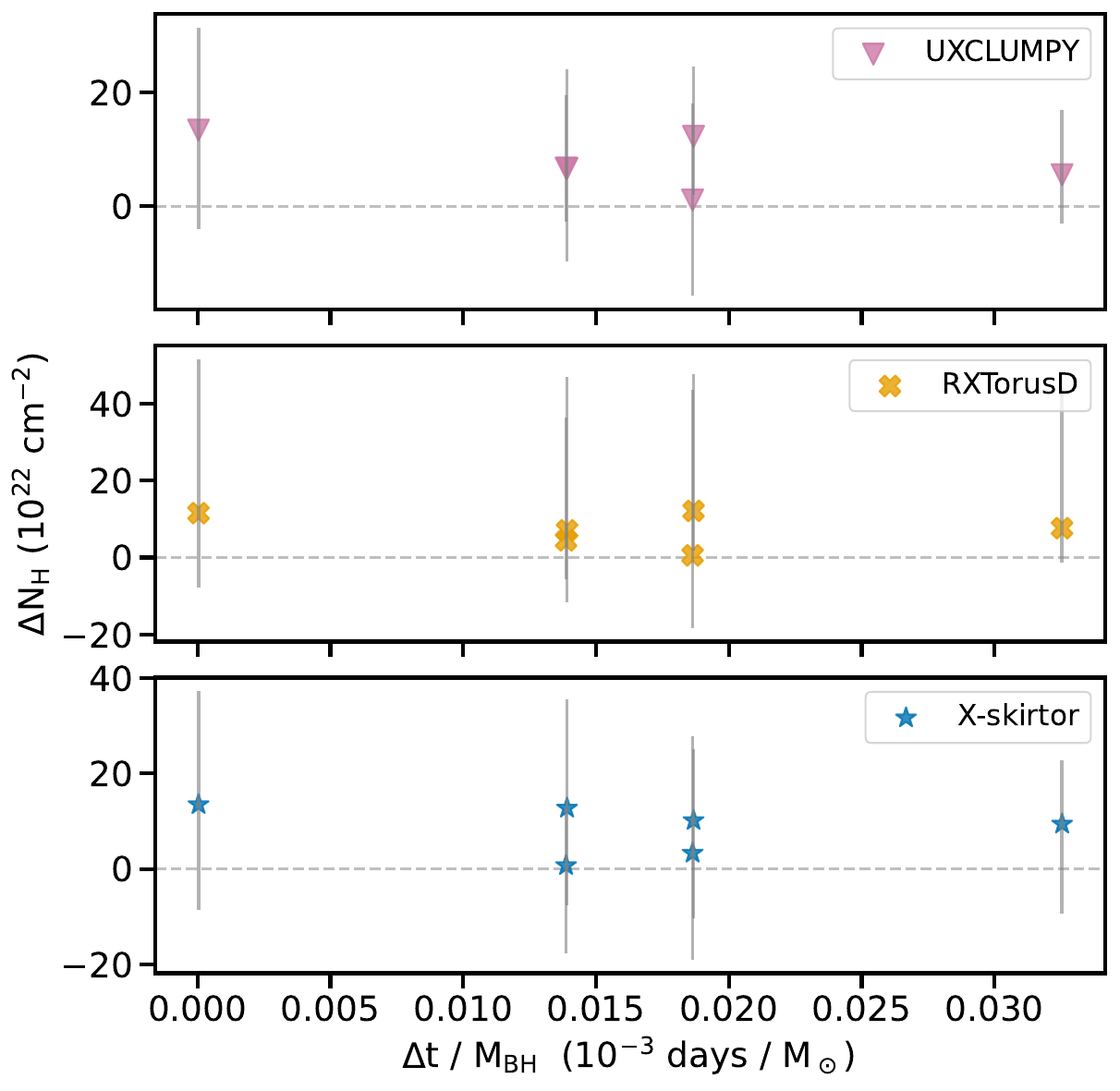}}
    \caption[]
    {{NGC~2655. {\it Panel a:} Best-fit with the \texttt{UXCLUMPY} torus model. The \chandra\ (in black and red), XMM (in green) and \nustar\ (FPMA and FPMB grouped together in cyan and blue) spectra are shown.
    {\it Panel b:} evolution of the \nhlos. The x-axis is in linear scale with a visual break to compress the long timescale between observations. {\it Panel c:} Variations of \nhlos\ as a function of the time separation between observations, normalized by the black-hole mass. Panels b and c adopt the same color code as Fig.~\ref{fig:eso_plots}.}}
    \label{fig:ngc2655_plots}
\end{figure}

\newpage
\clearpage

\renewcommand{\arraystretch}{1.6}
\begin{deluxetable*}{lccc}
\tablecaption{NGC~4785: spectral analyses for each torus model.\label{tab:spec_9}}
\tabletypesize{\scriptsize}
\tablewidth{\textwidth}
\tablehead{
\colhead{Parameter} & \colhead{\texttt{xskirtor}} & \colhead{\texttt{RXTorusD}} & \colhead{\texttt{UXCLUMPY}}
}
\startdata
\multicolumn{4}{c}{\texttt{apec \footnotesize{(Thermal emission)}}} \\
kT / keV & 0.31$^{+0.04}_{-0.03}$ & 0.30$^{+0.03}_{-0.02}$ & 0.31$^{+0.04}_{-0.03}$ \\
\multicolumn{4}{c}{\texttt{\footnotesize{Comptonized primary continuum}}} \\
$\Gamma$ & 1.88$^{+0.10}_{-0.17}$ & 1.97$^{+0.10}_{-0.20}$ & 1.95$^{+0.07}_{-0.10}$ \\
\multicolumn{4}{c}{\texttt{\footnotesize{Neutral reflector}}} \\
C$_\mathrm{f}$ $^{\bf +}$ & 0.55$^{+0.05}_{-0.07}$ & 0.52$^{+0.26}_{-0.12}$ & 0.40$^{+0.12}_{-0.05}$ \\
$\theta$ / deg & 60$^f$ & 60$^f$ & 60$^f$ \\
$\sigma_{\rm tor}$ / deg & / & / & 0$^f$ \\
F$_\mathrm{s}$ / $10^{-2}$ & 0.61$^{+0.27}_{-0.14}$ & 0.49$^{+0.29}_{-0.12}$ & 2.03$_{-0.73}^{+0.53}$ \\
norm / $10^{-3}$ & 2.11$^{+0.84}_{-0.77}$ & 2.76$\pm$0.11 & 3.12$^{+0.11}_{-0.89}$ \\
N$_{\rm H,eq}$ / 10$^{22}$ cm$^{-2}$ & 323.59$^{+434.92}_{-133.05}$ & 242.33$^{+100.58}_{-69.99}$ & / \\
\multicolumn{4}{c}{\texttt{N$_{\rm H,inst.,num.}$ \footnotesize{(LoS hydrogen column density)}}} \\
N$_{\rm H}^{Ch}$ / 10$^{22}$ cm$^{-2}$ & 36.75$^{+6.25}_{-5.52}$ & 43.28$^{+7.58}_{-6.63}$ & 37.03$^{+2.67}_{-6.20}$ \\
N$_{\rm H}^{XMM,1}$ / 10$^{22}$ cm$^{-2}$ & 60.07$^{+6.57}_{-5.53}$ & 72.05$^{+10.07}_{-8.17}$ & 60.76$^{+7.51}_{-6.60}$ \\
N$_{\rm H}^{XMM,2}$ / 10$^{22}$ cm$^{-2}$ & 49.18$^{+6.35}_{-5.35}$ & 55.97$^{+7.76}_{-6.42}$ & 49.81$^{+4.46}_{-5.41}$ \\
N$_{\rm H}^{XMM,3}$ / 10$^{22}$ cm$^{-2}$ & 40.63$^{+5.20}_{-4.43}$ & 48.14$^{+6.35}_{-5.52}$ & 39.53$^{+5.12}_{-3.54}$ \\
N$_{\rm H}^{NuS,2}$ / 10$^{22}$ cm$^{-2}$ & 80.42$^{+33.26}_{-24.79}$ & 97.17$^{+104.86}_{-36.39}$ & 83.43$^{+27.43}_{-25.71}$ \\
N$_{\rm H}^{NuS,3}$ / 10$^{22}$ cm$^{-2}$ & 100.72$^{+21.82}_{-15.36}$ & 134.40$^{+135.03}_{-32.70}$ & 104.02$^{+19.24}_{-13.58}$ \\
\multicolumn{4}{c}{\texttt{C$_{\rm inst.,num}$ \footnotesize{(Cross-normalization constant)}}} \\
C$_{Ch}$ & 1$^f$ & 1$^f$ & 1$^f$ \\
C$_{XMM,1}$ & 1.16$^{+0.35}_{-0.25}$ & 1.09$^{+0.32}_{-0.23}$ & 1.19$^{+0.38}_{-0.25}$ \\
C$_{XMM,2}$ & 0.96$^{+0.30}_{-0.22}$ & 0.88$^{+0.27}_{-0.19}$ & 0.97$^{+0.33}_{-0.22}$ \\
C$_{XMM,3}$ & 0.83$^{+0.26}_{-0.19}$ & 0.64$^{+0.25}_{-0.18}$ & 0.79$^{+0.27}_{-0.17}$ \\
C$_{NuS,2}$ & 0.39$^{+0.22}_{-0.14}$ & 0.34$^{+0.36}_{-0.11}$ & 0.39$^{+0.24}_{-0.14}$ \\
C$_{NuS,3}$ & 1.24$^{+0.50}_{-0.33}$ & 0.76$^{+0.50}_{-0.29}$ & 1.33$^{+0.51}_{-0.37}$ \\
\graytableline
\multicolumn{4}{c}{\texttt{Statistic}} \\
$\chi^2$/d.o.f. & 414/372 & 430/372 & 431/373 \\
{\it T} & 2.18$\sigma$ & 3.00$\sigma$ & 3.00$\sigma$ \\
p-value & 3.35e-07 & 2.91e-08 & 0 \\
\multicolumn{4}{c}{\texttt{Statistic \footnotesize{(No variability)}}} \\
$\chi^2$/d.o.f. & 1839/385 & 1558/385 & 1608/386 \\
{\it T} & 74.10$\sigma$ & 59.78$\sigma$ & 62.20$\sigma$ \\
\multicolumn{4}{c}{\texttt{Statistic \footnotesize{(No flux variability)}}} \\
$\chi^2$/d.o.f. & 448/378 & 468/378 & 467/379 \\
{\it T} & 3.60$\sigma$ & 4.63$\sigma$ & 4.52$\sigma$ \\
\multicolumn{4}{c}{\texttt{Statistic \footnotesize{(No \nh\ variability)}}} \\
$\chi^2$/d.o.f. & 515/378 & 523/378 & 505/379 \\
{\it T} & 7.05$\sigma$ & 7.46$\sigma$ & 6.47$\sigma$ \\
\enddata
\tablecomments{\footnotesize Refer to note in Table~\ref{tab:spec_1}.}
\end{deluxetable*}

\begin{figure}[h!]
\centering
    \subfloat[][]{\includegraphics[width=0.4\textwidth, trim=0 30 80 80, clip]{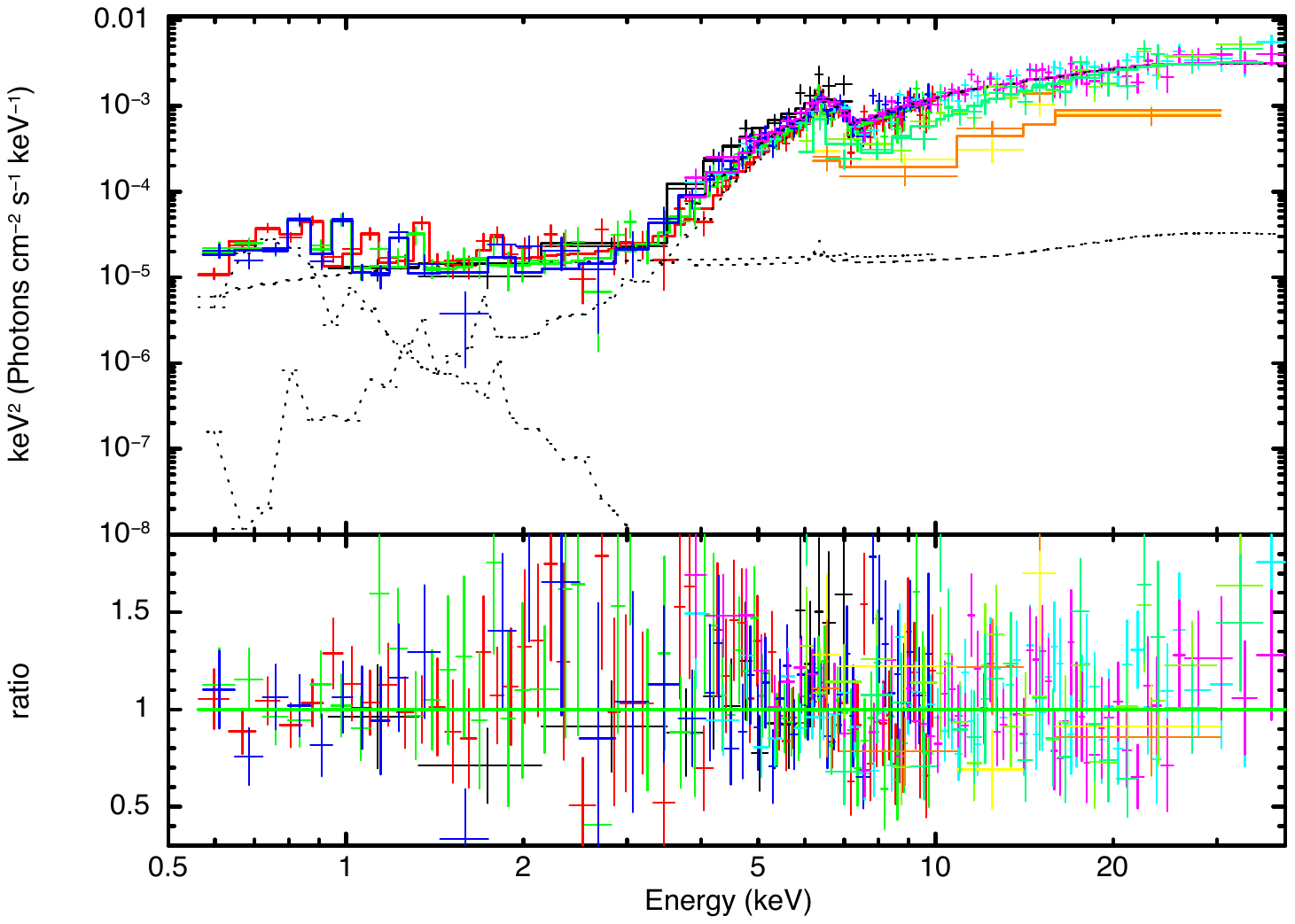}} \\
    \subfloat[][]{\includegraphics[width=0.5\textwidth, trim=10 10 10 0, clip]{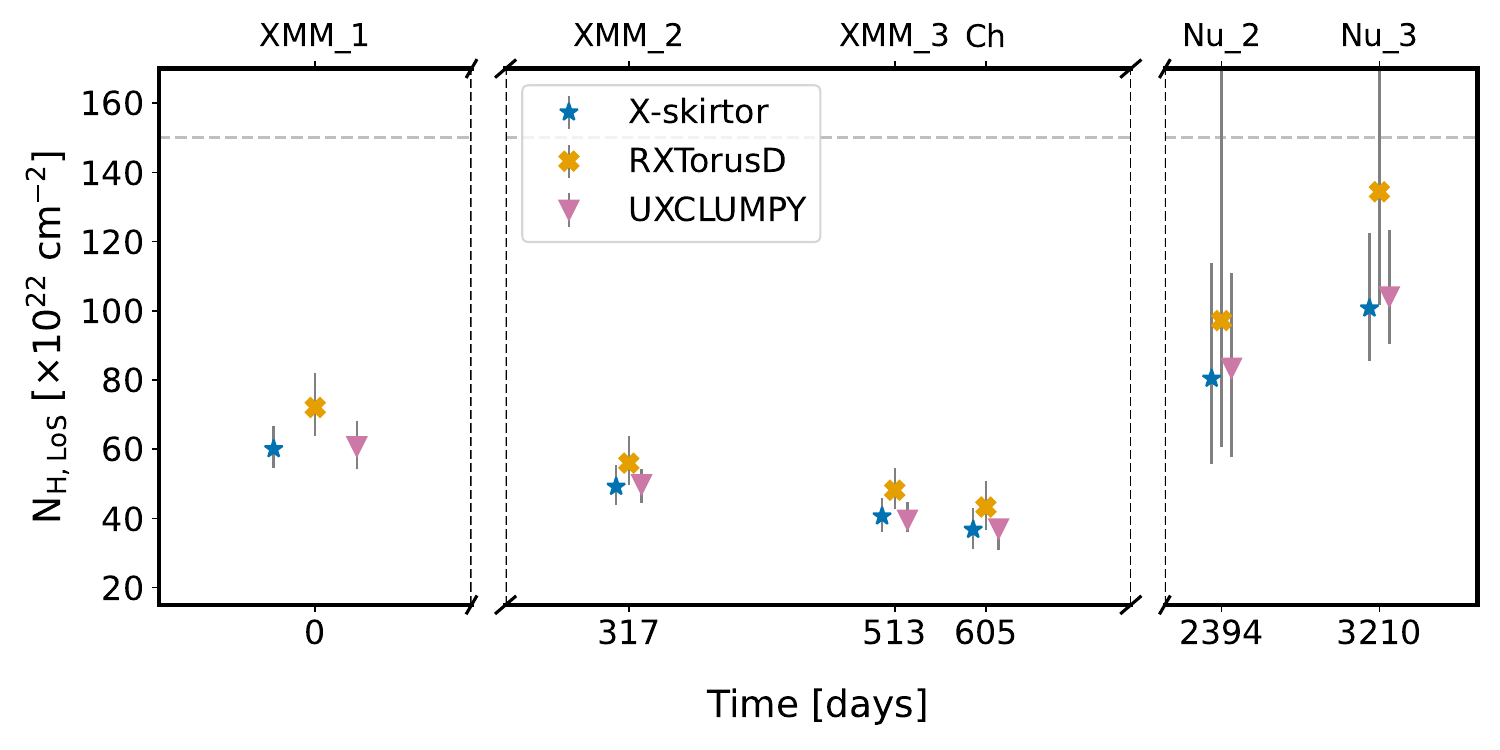}} \\
    \subfloat[][]{\includegraphics[width=0.4\textwidth, trim=5 5 7 0, clip]{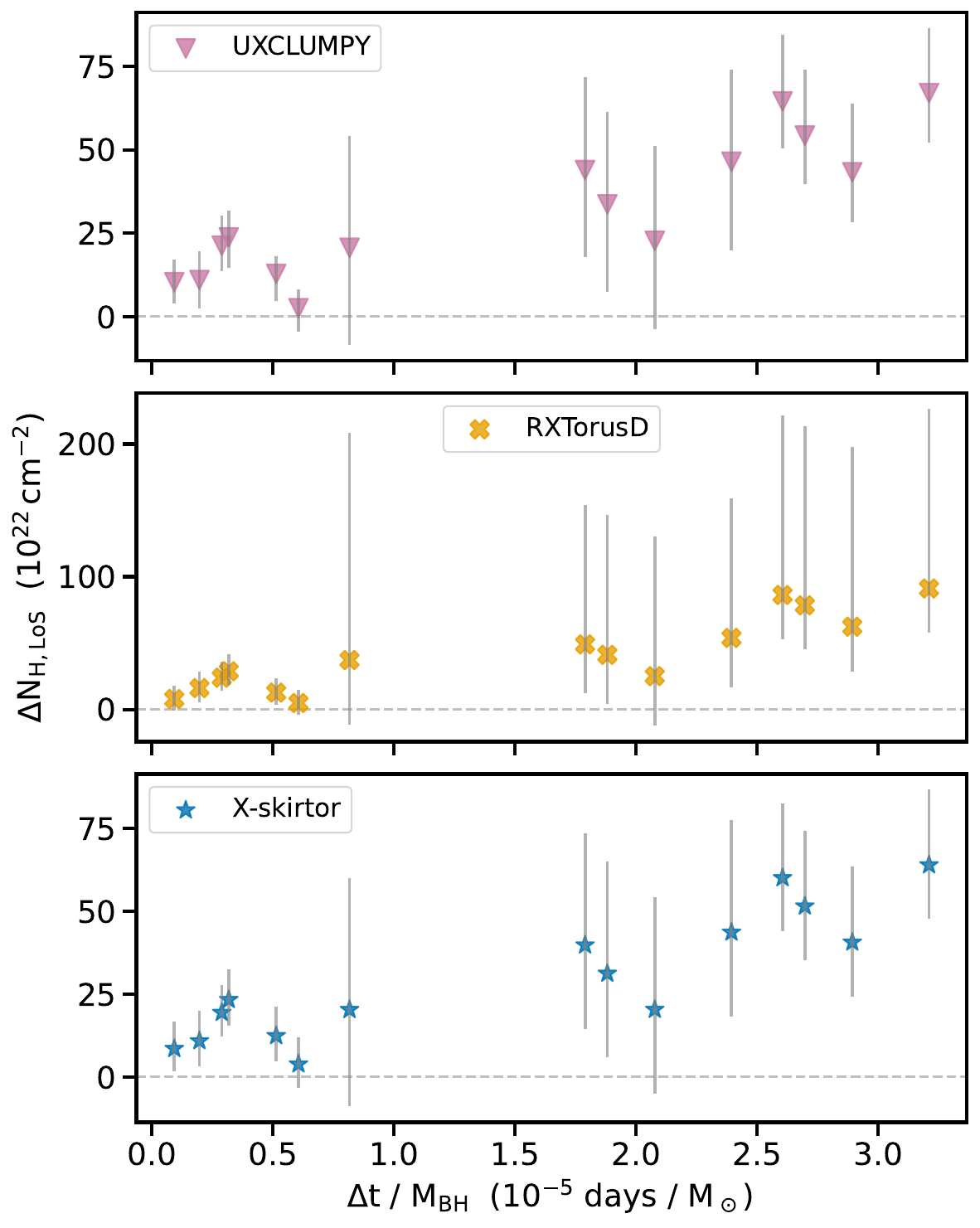}}
    \caption[]
    {{NGC~4785. {\it Panel a:} Best-fit with the \texttt{UXCLUMPY} torus model. The \chandra\ (in black), XMM (in red, green, and blue) and \nustar\ (FPMA and FPMB grouped together in cyan, blue, pink, yellow, and orange) spectra are shown.
    {\it Panel b:} evolution of the \nhlos. The x-axis is in logarithmic scale with a visual break to compress the long timescale between observations. The gray dashed line shows the CTK threshold. {\it Panel c:} Variations of \nhlos\ as a function of the time separation between observations, normalized by the black-hole mass. Panels b and c adopt the same color code as Fig.~\ref{fig:eso_plots}.}}
    \label{fig:NGC_4785_plots}
\end{figure}

\newpage
\clearpage

\renewcommand{\arraystretch}{1.5}
\begin{deluxetable*}{lccc}
\tablecaption{UGC~3752: best-fit spectral analyses for each torus model.\label{tab:spec_10}}
\tabletypesize{\scriptsize}
\tablewidth{\textwidth}
\tablehead{
\colhead{Parameter} & \colhead{\texttt{xskirtor}} & \colhead{\texttt{RXTorusD}} & \colhead{\texttt{UXCLUMPY}}
}
\startdata
\multicolumn{4}{c}{\texttt{apec \footnotesize{(Thermal emission)}}} \\
kT$_1$ / keV & 0.13$^{+0.03}_{-0.08}$ & 0.13$^{+0.04}_{-0.09}$ & 0.13$^{+0.03}_{-0.07}$ \\
kT$_2$ / keV & 0.69$\pm$0.06 & 0.62$^{+0.06}_{-0.06}$ & 0.70$\pm$0.06 \\
\multicolumn{4}{c}{\texttt{\footnotesize{Comptonized primary continuum}}} \\
$\Gamma$ & 1.49$^{+0.12}_{-0.17}$ & 1.65$^{+0.20}_{-0.36}$ & 1.54$^{+0.12}_{-0.11}$ \\
\multicolumn{4}{c}{\texttt{\footnotesize{Neutral reflector}}} \\
C$_\mathrm{f}$ $^{\bf +}$ & 0.45$^{+0.12}_{-0.11}$ & 0.53$^{+0.09}_{-0.07}$ & 0.46$^{+0.17}_{-0.12}$ \\
$\theta$ / deg & 60$^f$ & 60$^f$ & 60$^f$ \\
$\sigma_{\rm tor}$ / deg & / & / & 28.16$^{+34.07}_{-5.53}$ \\
F$_\mathrm{s}$ / $10^{-2}$ & 2.14$^{+1.08}_{-0.69}$ & 1.30$^{+0.65}_{-0.43}$ & 1.17$^{+0.12}_{-0.34}$ \\
norm / $10^{-3}$ & 0.52$^{+0.37}_{-0.24}$ & 0.88$^{+0.61}_{-0.38}$ & 0.46$^{+0.33}_{-0.20}$ \\
N$_{\rm H,eq}$ / 10$^{22}$ cm$^{-2}$ & 14.79$^{+60.26}_{-2.88}$ & 412.13$^{+131.87}_{-273.79}$ & / \\
\multicolumn{4}{c}{\texttt{N$_{\rm H,inst.,num.}$ \footnotesize{(LoS hydrogen column density)}}} \\
N$_{\rm H}^{Ch}$ / 10$^{22}$ cm$^{-2}$ & 71.24$^{+40.08}_{-22.51}$ & 88.39$^{+57.57}_{-28.43}$ & 52.27$^{+17.01}_{-15.43}$ \\
N$_{\rm H}^{XMM}$ / 10$^{22}$ cm$^{-2}$ & 58.43$^{+10.26}_{-8.77}$ & 73.57$^{+82.64}_{-14.43}$ & 47.29$^{+8.65}_{-8.23}$ \\
N$_{\rm H}^{NuS,1}$ / 10$^{22}$ cm$^{-2}$ & 89.22$^{+19.31}_{-16.26}$ & 133.52$^{+61.60}_{-26.35}$ & 95.67$^{+15.88}_{-17.37}$ \\
N$_{\rm H}^{NuS,2}$ / 10$^{22}$ cm$^{-2}$ & 180.79$^{+41.00}_{-32.89}$ & 200.20$^{+58.12}_{-42.16}$ & 171.42$^{+41.17}_{-34.22}$ \\
N$_{\rm H}^{NuS,3}$ / 10$^{22}$ cm$^{-2}$ & 221.09$^{+135.36}_{-105.38}$ & 188.33$^{+22.15}_{-17.71}$ & 131.77$^{+89.28}_{-43.28}$ \\
N$_{\rm H}^{NuS,4}$ / 10$^{22}$ cm$^{-2}$ & 106.50$^{+18.74}_{-26.05}$ & 133.61$^{+61.36}_{-58.03}$ & 108.85$^{+36.85}_{-14.43}$ \\
N$_{\rm H}^{NuS,5}$ / 10$^{22}$ cm$^{-2}$ & 34.75$^{+6.75}_{-7.71}$ & 26.62$^{+9.89}_{-20.06}$ & 35.30$^{+6.65}_{-5.45}$ \\
\multicolumn{4}{c}{\texttt{C$_{\rm inst.,num}$ \footnotesize{(Cross-normalization constant)}}} \\
C$_{Ch}$ & 1$^f$ & 1$^f$ & 1$^f$ \\
C$_{XMM}$ & 1.66$^{+0.86}_{-0.54}$ & 1.62$^{+0.71}_{-0.46}$ & 1.94$^{+1.14}_{-0.61}$ \\
C$_{NuS,1}$ & 2.00$^{+1.74}_{-1.06}$ & 2.32$^{+1.80}_{-0.99}$ & 3.42$^{+2.52}_{-1.44}$ \\
C$_{NuS,2}$ & 2.34$^{+1.11}_{-0.71}$ & 0.68$^{+1.86}_{-1.72}$ & 3.98$^{+2.45}_{-1.55}$ \\
C$_{NuS,3}$ & 2.36$^{+1.16}_{-0.75}$ & 0.99$^{+1.62}_{-0.98}$ & 2.85$^{+1.77}_{-1.47}$ \\
C$_{NuS,4}$ & 2.12$^{+1.43}_{-0.94}$ & 1.12$^{+1.86}_{-1.70}$ & 3.58$^{+2.26}_{-1.50}$ \\
C$_{NuS,5}$ & 2.17$^{+3.19}_{-5.04}$ & 0.80$^{+1.30}_{-1.32}$ & 3.42$^{+2.20}_{-1.26}$ \\
\graytableline
\multicolumn{4}{c}{\texttt{Statistic}} \\
C-stat/d.o.f. & 422/423 & 453/423 & 431/423 \\
{\it T} & 0.05$\sigma$ & 1.46$\sigma$ & 0.39$\sigma$ \\
p-value & 0 & 0 & 0 \\
\multicolumn{4}{c}{\texttt{Statistic \footnotesize{(No variability)}}} \\
C-stat/d.o.f. & 2647/435 & 2598/435 & 2630/435 \\
{\it T} & 106.06$\sigma$ & 98.92$\sigma$ & 105.24$\sigma$ \\
\multicolumn{4}{c}{\texttt{Statistic \footnotesize{(No flux variability)}}} \\
C-stat/d.o.f. & 474/429 & 470/429 & 462/429 \\
{\it T} & 2.17$\sigma$ & 1.98$\sigma$ & 1.59$\sigma$ \\
\multicolumn{4}{c}{\texttt{Statistic \footnotesize{(No \nh\ variability)}}} \\
C-stat/d.o.f. & 711/429 & 656/429 & 721/429 \\
{\it T} & 13.62$\sigma$ & 10.96$\sigma$ & 14.10$\sigma$ \\
\enddata
\tablecomments{\footnotesize Refer to note in Table~\ref{tab:spec_1}.}
\end{deluxetable*}

\begin{figure}[h!]
\centering
    \subfloat[][]{\includegraphics[width=0.4\textwidth, trim=0 30 80 80, clip]{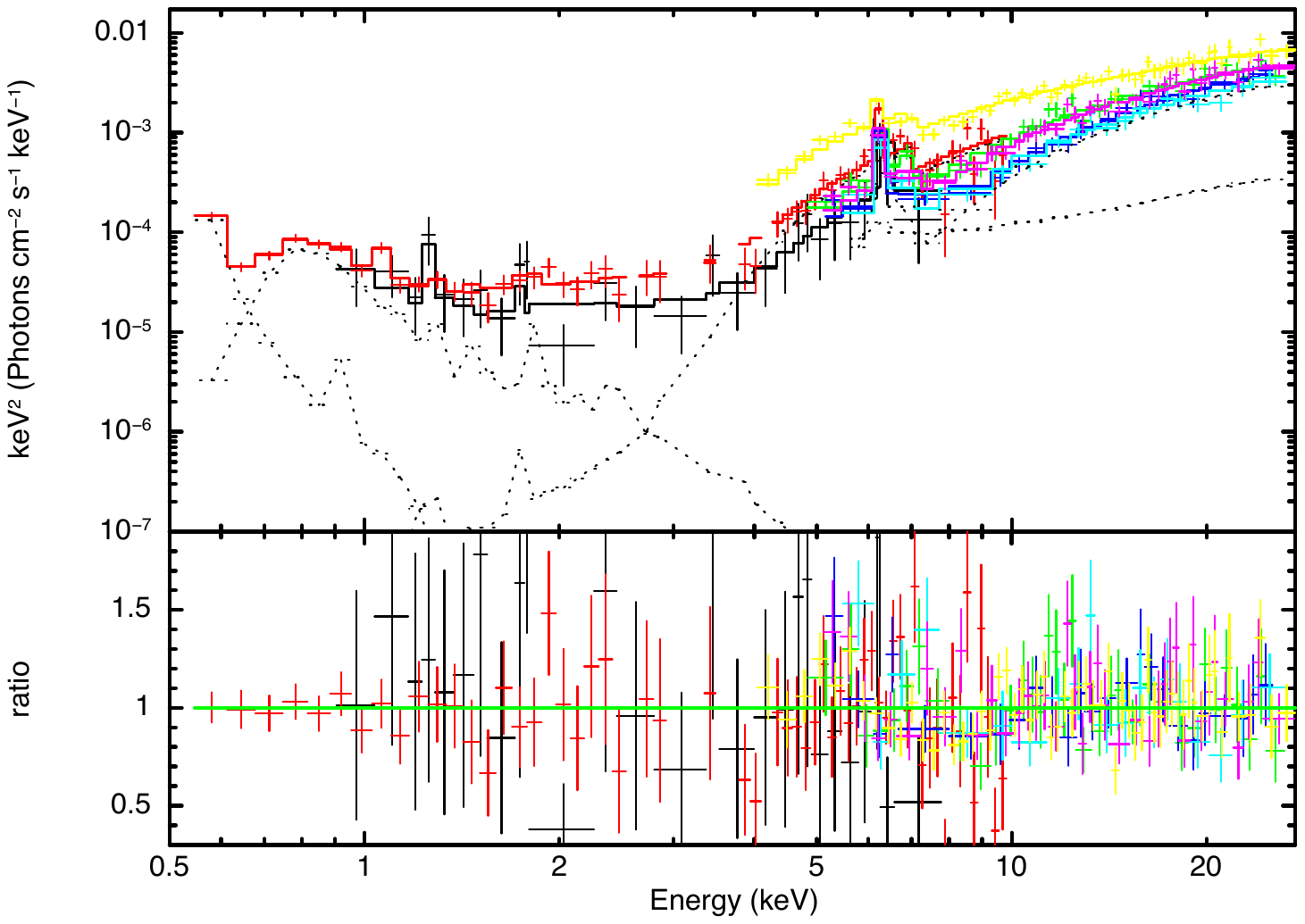}} \\
    \subfloat[][]{\includegraphics[width=0.5\textwidth, trim=0 10 10 0, clip]{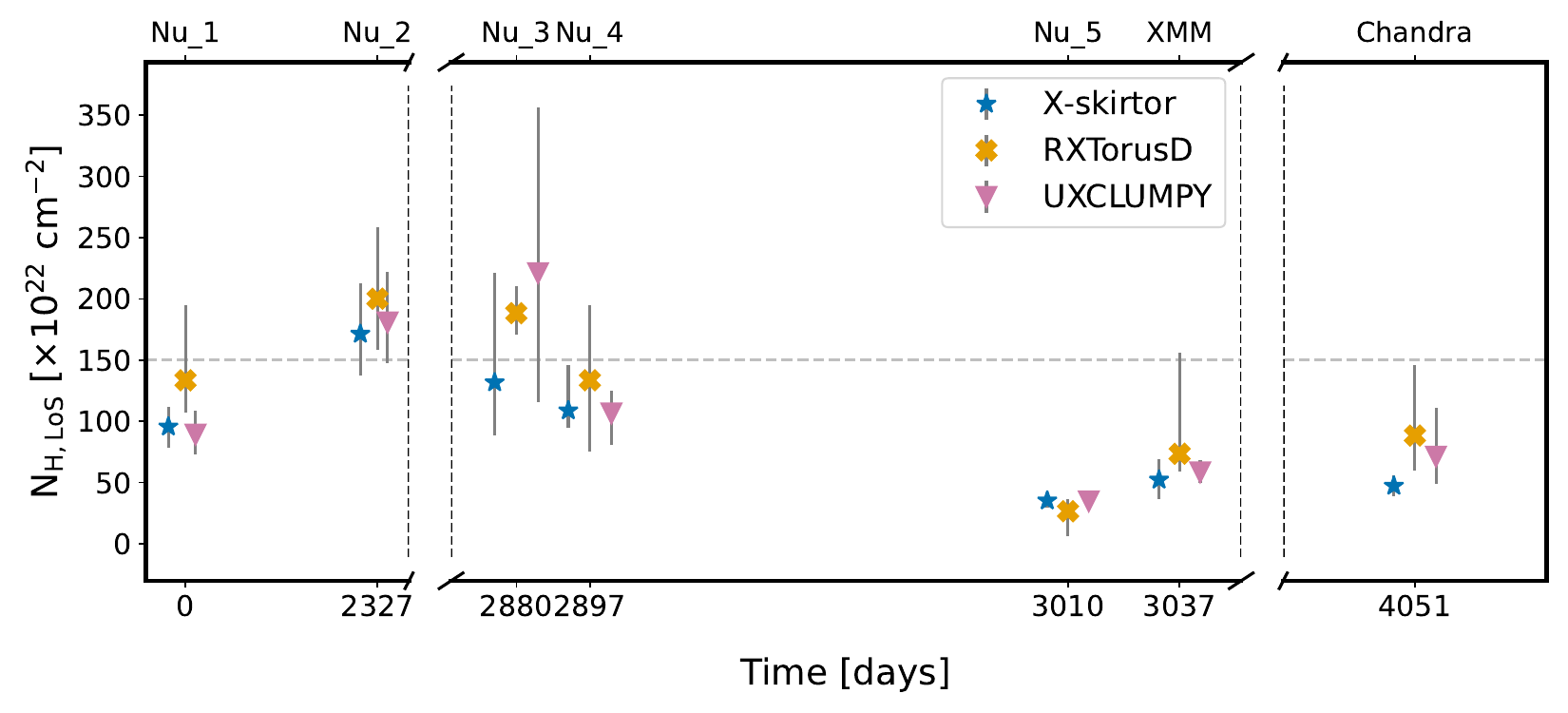}} \\
    \subfloat[][]{\includegraphics[width=0.4\textwidth, trim=5 5 7 0, clip]{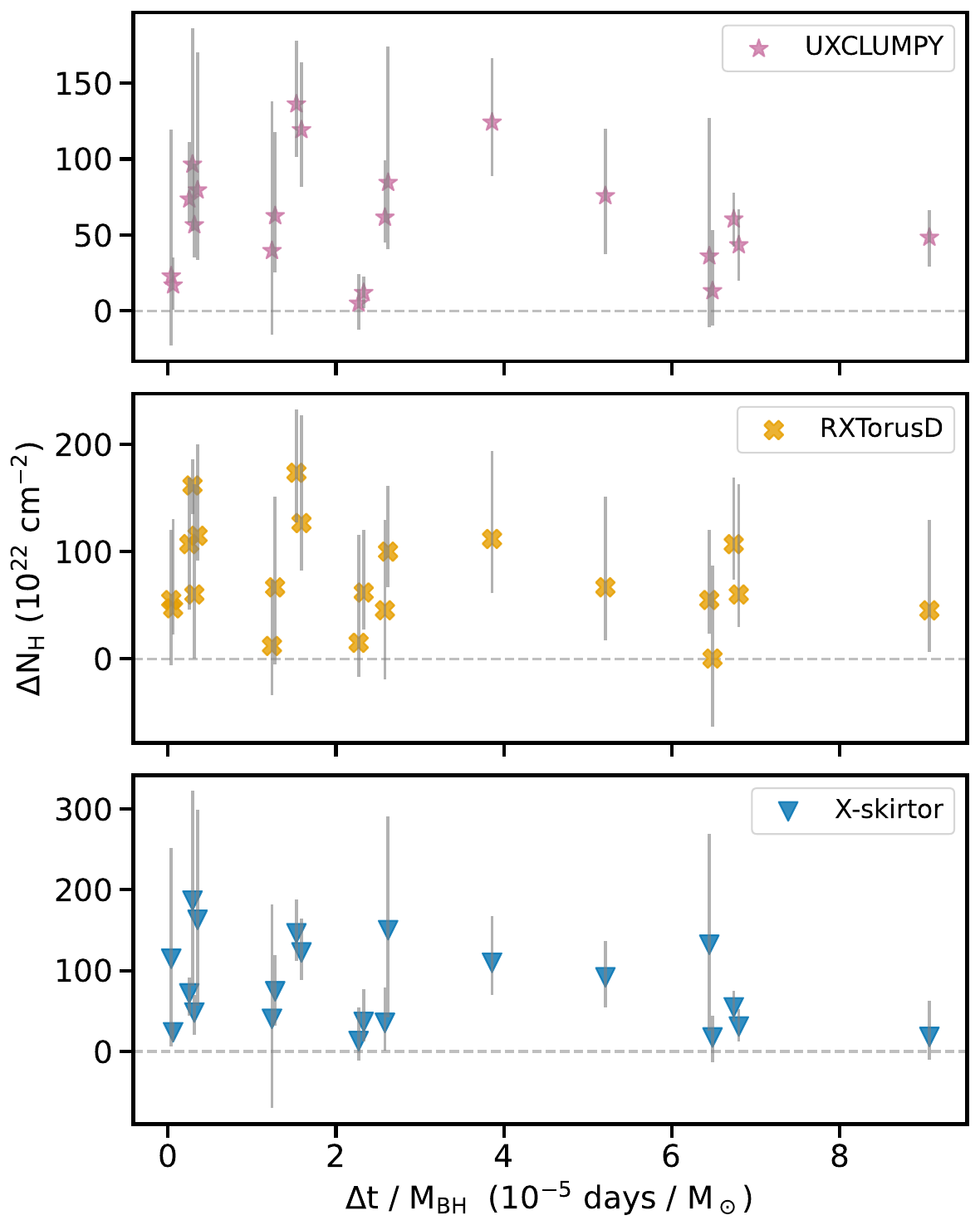}}
    \caption[]
    {{UGC~3752. {\it Panel a:} Best-fit with the \texttt{UXCLUMPY} torus model. The \chandra\ (in black), XMM (in red) and \nustar\ (FPMA and FPMB grouped together in green, blue cyan, pink, and yellow) spectra are shown.
    {\it Panel b:} evolution of the \nhlos\. The x-axis is in logarithmic scale for the left and middle panels, linear for the right panel, with visual breaks to compress the long timescale between observations. The gray dashed line shows the CTK threshold. {\it Panel c:} Variations of \nhlos\ as a function of the time separation between observations, normalized by the black-hole mass. Panels b and c adopt the same color code as Fig.~\ref{fig:eso_plots}.}}
    \label{fig:UGC_plots}
\end{figure}

\newpage
\clearpage

\begin{table}
\centering
\caption{WISE~J144850.99-400845.6: best-fit spectral analyses for each torus model.}
\label{tab:spec_11}
\begin{tabular}{ccccccc}
\hline
{Parameter} & \texttt{xskirtor} & \texttt{RXTorusD} & \texttt{UXCLUMPY}  \\
\hline\hline
\multicolumn{4}{c}{\texttt{apec \footnotesize{(Thermal emission)}}}\\
kT / keV & 0.20$\pm$0.01 & 0.19$\pm$0.01 & 0.21$\pm$0.01 \\
\multicolumn{4}{c}{\texttt{\footnotesize{Comptonized primary continuum}}} \\
$\Gamma$ & 1.70$^{+0.04}_{-0.05}$ & 1.73$^{+0.05}_{-0.03}$ & 1.66$\pm$0.04 \\
\multicolumn{4}{c}{\texttt{\footnotesize{Neutral reflector}}} \\
C$_\mathrm{f}$ $^{\bf +}$ & 0.65$_{-0.32}$ & 0.69$_{-0.18}$ & 0.34$^{+0.02}_{-0.03}$ \\
$\theta$ / deg & 60$^{f}$ & 60$^{f}$ & 60$^{f}$ \\
$\sigma_{\rm tor}$ / deg & / & / & 0.91$^{+0.42}_{-0.30}$ \\
F$_\mathrm{s}$ / $10^{-2}$ & 22.37$^{+1.37}_{-1.31}$ & 0.21$^{+0.03}_{-0.03}$ & 7.80$^{+0.02}_{-0.01}$ \\
norm / $10^{-3}$ & 0.84$^{+0.09}_{-0.08}$ & 0.90$^{+0.06}_{-0.05}$ & 2.47$^{+0.38}_{-0.40}$ \\
N$_{\rm H,eq}$ / 10$^{22}$ cm$^{-2}$ & 126.60$^{+561.41}_{-53.02}$ & 197.92$^{+207.99}_{-86.57}$ & / \\
\multicolumn{4}{c}{\texttt{N$_{\rm H,inst.,num.}$ \footnotesize{(LoS hydrogen column density)}}} \\
N$_{\rm H}^{Ch}$ / 10$^{22}$ cm$^{-2}$ & 0.48$\pm$0.05 & 0.64$^{+0.08}_{-0.07}$ & 4.57$^{+0.46}_{-0.37}$ \\
N$_{\rm H}^{XMM}$ / 10$^{22}$ cm$^{-2}$ & 0.61$\pm$0.03 & 0.82$^{+0.08}_{-0.07}$ & 2.14$\pm$0.10 \\
N$_{\rm H}^{NuS}$ / 10$^{22}$ cm$^{-2}$ & 0.001$^{+0.26}$ & 0.001$^{+0.82}$ & 0.04$\pm$0.01 \\
\multicolumn{4}{c}{\texttt{C$_{\rm inst.,num}$ \footnotesize{(Cross-normalization constant)}}} \\
C$_{Ch}$ & 1$^f$ & 1$^f$ & 1$^f$ \\
C$_{XMM}$ & 1.43$\pm$0.05 & 1.43$^{+0.05}_{-0.04}$ & 2.03$^{+0.27}_{-0.24}$ \\
C$_{NuS}$ & 1.55$\pm$0.07 & 1.53$\pm$0.07 & 1.55$^{f}$ \\
\graytableline
\multicolumn{4}{c}{\texttt{Statistic}} \\ 
\smallskip
C-stat/d.o.f. & 329/259 & 336/259 & 318/259
\\
{\it T} & 4.33$\sigma$ & 4.78$\sigma$ & 3.66$\sigma$ \\
p-value & 1.11e-16 & 9.38e-7 & 0 \\
\multicolumn{4}{c}{\texttt{Statistic \footnotesize{(No variability)}}} \\
C-stat/d.o.f. & 1313/263 & 957/263 & 1350/263
\\
{\it T} & 64.77$\sigma$ & 42.94$\sigma$ & 67.01$\sigma$ \\
\multicolumn{4}{c}{\texttt{Statistic \footnotesize{(No flux variability)}}} \\
C-stat/d.o.f. & 667/261 & 624/261 & 401/261
\\
{\it T} & 25.14$\sigma$ & 22.47$\sigma$ & 8.69$\sigma$ \\
\multicolumn{4}{c}{\texttt{Statistic \footnotesize{(No \nh\ variability)}}} \\
C-stat/d.o.f. & 353/261 & 414/261 & 530/261
\\
{\it T} & 5.68$\sigma$ & 9.47$\sigma$ & 16.68$\sigma$ \\
\hline
\end{tabular}
\tablecomments{\footnotesize Refer to note in Table~\ref{tab:spec_1}.}
\end{table}

\begin{figure}[h!]
\centering
    \subfloat[][]{\includegraphics[width=0.4\textwidth, trim=0 30 80 80, clip]{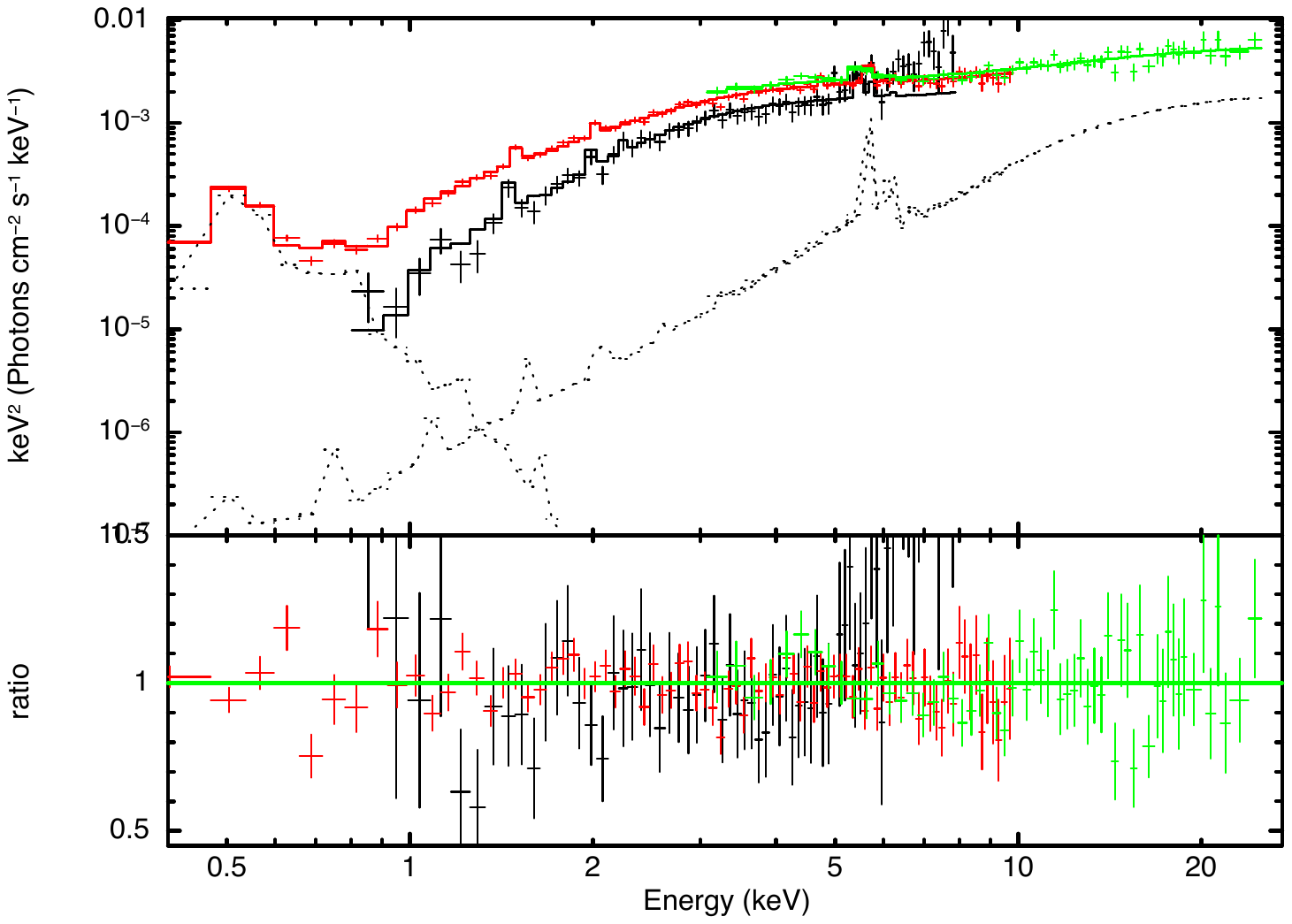}} \\
    \subfloat[][]{\includegraphics[width=0.4\textwidth, trim=10 10 10 0, clip]{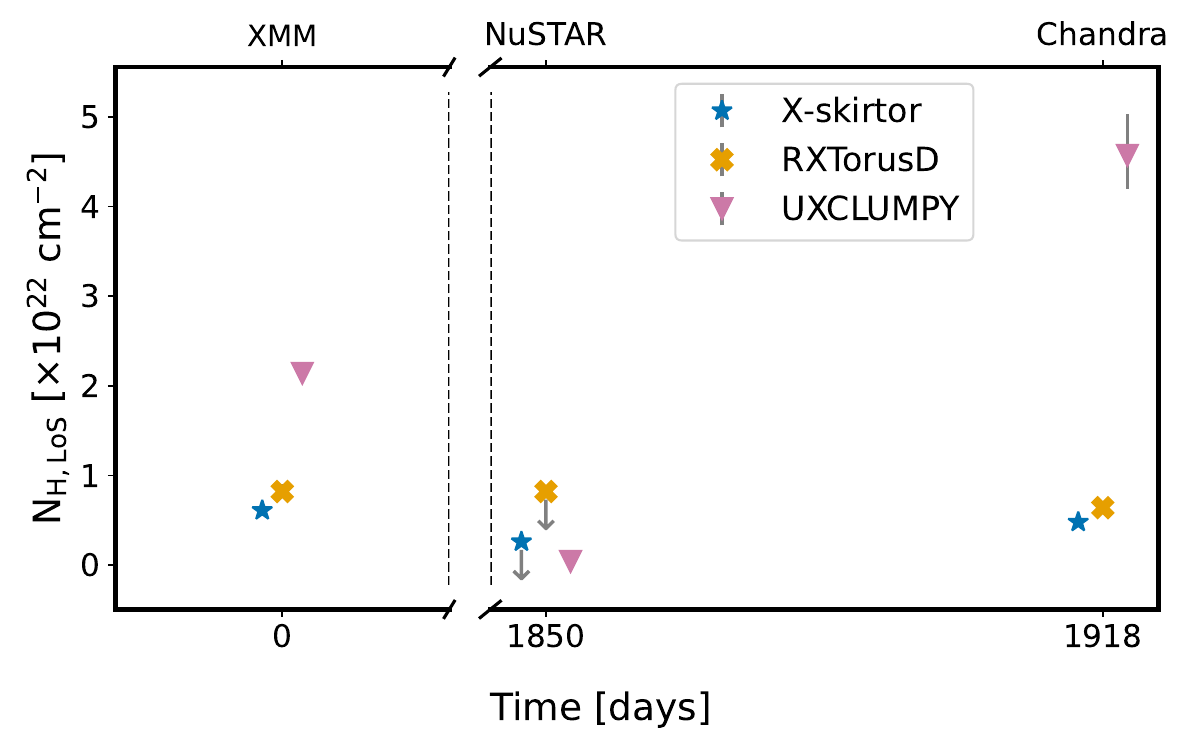}} \\
    \subfloat[][]{\includegraphics[width=0.4\textwidth, trim=5 5 7 0, clip]{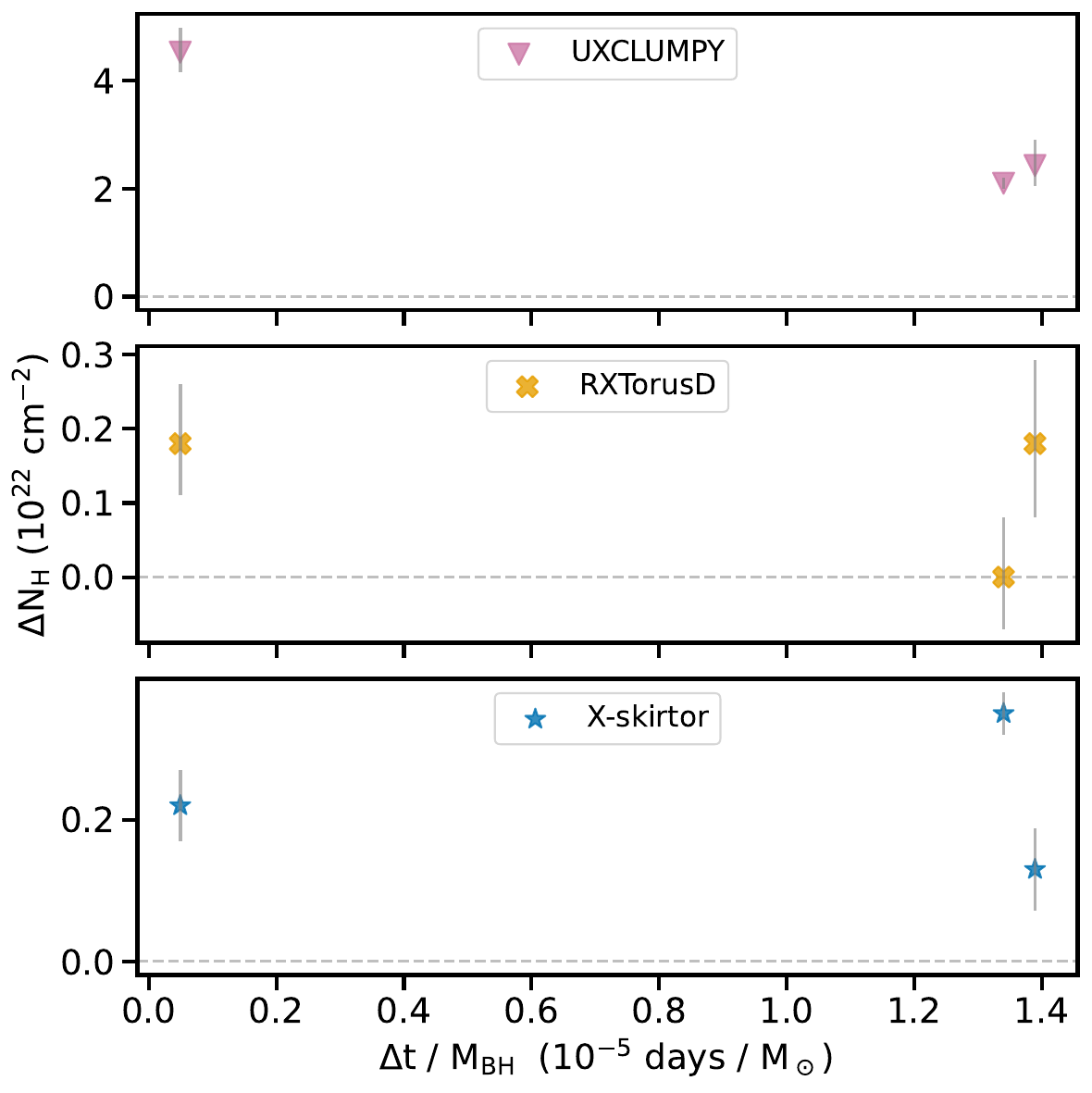}}
    \caption[]
    {{WISE~J144850.99-400845.6. {\it Panel a:} Best-fit with the \texttt{UXCLUMPY} torus model. The \chandra\ (in black), \xmm\ (in red), and \nustar\ (FPMA and FPMB spectra are grouped together in green) spectra are shown.
    {\it Panel b:} evolution of the \nhlos. The x-axis is in linear scale with a visual break to compress the long timescale between observations. The gray arrows show the upper limits obtained for the \nustar\ observation during the fit with \skirt\ and \rxt\ models. {\it Panel c:} Variations of \nhlos\ as a function of the time separation between observations, normalized by the black-hole mass. Panels b and c adopt the same color code as Fig.~\ref{fig:eso_plots}.}}
    \label{fig:wise_plots}
\end{figure}

\clearpage

\section{Notes and general details on single sources}\label{source_note}

{\it ESO~464-G016} is a Seyfert 2 galaxy \citep{chen2022} and a candidate Compton-thick \citep{marchesi2019} AGN. However, our multi-epoch analysis shows that, for only one observation of a specific model (\ux), the line-of-sight column density (\nhlos) reaches an upper limit exceeding the CTK threshold ($>$ 1.5 $\times$ 10$^{24}$ cm$^{-2}$). Based on this, we adopt a conservative approach and classify ESO~464-G016 as Compton-thin (see Sect.~\ref{ctk_cth_class}).
For the variability classification, the \nhlos\ values derived across different epochs are consistent within their uncertainties, suggesting no variability. The statistical tests yield mixed indications: the p-value test favors no variability, whereas the tension test is inconclusive, resulting in a final classification of Undetermined AGN (see Sect.~\ref{var_classification_res} and Table~\ref{tab:clas_summary}). The intrinsic flux variability tests under the no-variability hypothesis return tensions below 5, comparable to those obtained for the best-fit and the other hypotheses (see Table~\ref{tab:spec_1}). This indicates that allowing for variability does not significantly improve the fit.

\vspace{0.2cm}

{\it IC~5063} is a radio-loud \citep{Morganti2013} Seyfert 2 \citep{veron2006} AGN. Our multi-epoch analysis classify this AGN as Compton-thin (see Sect.~\ref{ctk_cth_class}).
During the fit, the residuals in the \chandra\ spectra are modeled with a Gaussian component at E = $1.33^{+0.02}_{-0.03}$ keV. 
The tension derived from the best-fit obtained with \rxt\ and \ux\ models is above 3, which is due to residuals present in \nustar/FPMB instrument. We tested the inclusion of Gaussian components that could reduce the tension below 3; however, these additional Gaussians lack any clear physical interpretation. Therefore, we chose not to include them in the best-fit and retained the value of T $>$ 3. Despite this, when testing the no \nhlos\ variability hypothesis, the resulting tensions for the two models are 14.53 and 21.33 (respectively), both significantly higher than the $T$ values obtained in the best-fit. Consequently, we classify the source as \nhlos\ Variable (see Sect.~\ref{var_classification_res} and Table~\ref{tab:clas_summary}). Moreover, the tensions associated with the no-variability and no-flux variability tests are all well above 5, indicating that flux variability is also required to adequately reproduce the spectral properties of this source (see Table~\ref{tab:spec_2}).

\vspace{0.2cm}

{\it LEDA~511869} is a Seyfert 2 galaxy \citep{2007ATel}. Our multi-epoch analysis classify this AGN as Compton-thin (see Sect.~\ref{ctk_cth_class}). Regarding the variability classification, the best-fit \nhlos\ values across different epochs are within uncertainties, indicating no significant absorption variability. This result is confirmed by both statistical tests, which classify the source as Non-Variable (see Sect.~\ref{var_classification_res} and Table~\ref{tab:clas_summary}). However, the no-variability tests show that the source cannot be reproduced under the hypothesis of constant spectral parameters (i.e., the tensions are higher than 5, see Table~\ref{tab:spec_3}). Both intrinsic flux variability and\nhlos\ variability, when allowed independently, are capable of compensating for the other and provide acceptable fits. Therefore, while variability is required to explain the data, the current analysis does not allow us to determine its physical origin.

\vspace{0.2cm}

{\it MCG-03-34-064} is a Seyfert 1.8 \citep{veron2006} AGN, candidate to host a dual system \citep{trindale2024}. 
As discussed in Sec.~\ref{mcg}, the X-ray spectrum of MCG-03-34-064 presents several significant ($>$3$\sigma$) emission and absorption lines. In the soft band, we identify (and model with Gaussian lines) prominent emission lines at $0.54 \pm 0.01$ keV, $1.07 \pm 0.01$ keV, and at $1.80 \pm 0.02$ keV. The Fe K band is known to be complex \citep{miniutti2007}, and we detect both emission and absorption features. The Fe K$\alpha$ emission line is detected at $6.39 \pm 0.01$ keV and is intrinsically broadened ($\sigma = 100^{+12}_{-11}$ eV), with a large equivalent width $\sim200$ eV. These properties indicate a possible origin in BLR-scale material or the inner regions of the obscuring torus. In addition, we detect an unresolved absorption feature at $7.23^{+0.10}_{-0.04}$ keV, consistent with blueshifted \ion{Fe}{26} Ly$\alpha$ absorption and suggestive of highly ionized outflowing gas. 
A detailed characterization of this absorber is beyond the scope of this work.
According to the p-value test applied to the best-fit \nhlos\ values (see Sect.~\ref{mcg}), \nhlos\ variability is detected in MCG-03-34-064. However, this result should be treated cautiously, as the fit is not robust and fails to reproduce the iron band.

\vspace{0.2cm}

{\it Mrk~18}. This galaxy is identified as a Seyfert 2 AGN. Our multi-epoch analysis classify this AGN as Compton-thin (see Sect.~\ref{ctk_cth_class}).
During the fitting procedure, we found that the inclination angle was unconstrained for both \skirt\ and \ux\, while it was well constrained for \rxt. Therefore, instead of adopting $60^{\circ}$ (as per methodology), we fixed the inclination to $20^{\circ}$ for \skirt\ and \ux\, consistent with the value obtained for \rxt. Moreover, we allowed the normalization of the \texttt{apec} component to vary independently between the \chandra\ (2.14$^{+4.11}_{-1.96}$ $\times$ 10$^{-5}$), \nustar\ (fixed to 0), and XMM observation (3.47 $\pm$ 0.25 $\times$ 10$^{-5}$).
This choice is motivated by the different extraction radii used for the two instruments: the larger XMM extraction region may include part of the star-forming region (SFR) emission, whereas this contribution is likely absent in the \chandra\ data due to its smaller extraction radius.
Regarding the \nhlos\ variability, the first-order analysis shows that the best-fit values for different epochs are consistent within uncertainties, indicating no significant variability. This result is supported by both statistical tests, which classify the AGN as Non-variable (see Sect.~\ref{var_classification_res} and Table~\ref{tab:clas_summary}). In contrast, the no-variability tests yield tensions above 5, indicating the presence of intrinsic flux variability (see Table~\ref{tab:spec_5}).

\vspace{0.2cm}

{\it Mrk~1498} is a Compton-thin \citep[e.g.,][]{Eguchi2009,ursini2018} Seyfert 1.9 \citep{veron2006} AGN. Our multi-epoch analysis confirms the Compton-thin nature of this AGN (see Sect.~\ref{ctk_cth_class}).
We note that the adopted models are not able to properly model the iron K band and residuals are present. We thus add an unresolved Gaussian (width fixed to 0 keV, significance $\sim$ 4$\sigma$) at energy $6.21 \pm 0.05$ keV, which is likely associated with low-ionization states of the Fe K$\alpha$ emission line. 

For this source, the tension values for the no \nhlos\ variability hypothesis are close to 5 for both \rxt\ and \ux\, and exceed 5 for \skirt. Therefore, we classify the source as Variable, even though the tension test alone would render the classification formally Undetermined. We note, however, that the $T$ values obtained under the no intrinsic flux variability hypothesis are quite low, comparable to those derived from the best-fit. Hence, we are unable to distinguish whether the observed variability arises from changes in the neutral column density or in the intrinsic flux (see Sect.~\ref{var_classification_res} and Tables~\ref{tab:clas_summary} and \ref{tab:spec_6}).

\vspace{0.2cm}

{\it NGC~1194} has been classified as a Compton-thick \citep{kuo2011,severgnini2012, marchesi2018} radio-loud Seyfert 1.9 \citep{veron2006} galaxy with megamaser emission \citep{kuo2011,fedorova2016}. Our results, considering all three models, confirm the Compton-thick nature of the source, with two observations (Chandra\_2 taken on 2019-10-27, and Chandra\_4 on 2020-11-05) showing column densities above the CTK threshold (i.e., \nhlos\ $>$ 1.5$\times$ 10$^{24}$ cm$^{-2}$) even when accounting for the lower uncertainties (see Sect.~\ref{ctk_cth_class}).
To model the significant emission features in the soft X-ray band and Fe K complex, we use three Gaussian components. Concerning the soft spectrum, emission lines at E = 1.33$^{+0.01}_{-0.02}$ keV and at E = 1.80$\pm 0.02$ are observed.
Meanwhile, a moderately broad emission line is detected at E = 6.64$^{+0.01}_{-0.02}$ with width $\sigma = 80 \pm 50$ eV and equivalent width EW = $500 \pm 100$ eV. 
Its energy is consistent with He-like \ion{Fe}{25} K$\alpha$ emission, likely originating from hot, ionized gas in the inner regions of the AGN, such as the broad-line region or circumnuclear photoionized material. 
We note that some residuals are present in the \nustar\ fit with the \ux\ model (see Fig.~\ref{fig:1194_plots}). These residuals are not seen with \skirt\ and \rxt, and the best-fit parameters are consistent across models, indicating that the \ux\ fit remains reliable.

Regarding variability, the first-order analysis reveals significant \nhlos\ variations across different epochs. This result is confirmed by both statistical tests, which consistently classify NGC~1194 as Variable (see Sect.~\ref{var_classification_res} and Table~\ref{tab:clas_summary}). In addition, the no-flux variability tests yield tensions above 5, indicating also the presence of intrinsic flux variability (see Table~\ref{tab:spec_7}).

\vspace{0.2cm}

{\it NGC~2655} is a Seyfert 2 \citep{veron2006} galaxy and low-luminosity AGN \citep{nagar2002}. Our multi-epoch analysis classify this AGN as Compton-thin (see Sect.~\ref{ctk_cth_class}). All tests indicate no variability in \nhlos, while the AGN results variable in intrinsic flux (see Sect.~\ref{var_classification_res} and Tables~\ref{tab:clas_summary} and \ref{tab:spec_8}).
 
\vspace{0.2cm}

{\it NGC~4785} is a Seyfert 2 \citep{veron2006} previously classified as Compton-thick \citep{gandhi2015} AGN. However, our multi-epoch analysis shows that, for only one observation of a specific model (\rxt), the line-of-sight column density (\nhlos) reaches an upper limit exceeding the CTK threshold ($>$ 1.5 $\times$ 10$^{24}$ cm$^{-2}$). Based on this, we adopt a conservative approach and classify NGC~4785 as a Compton-thin AGN (see Sect.~\ref{ctk_cth_class}).
Given that no direct BH mass measurement is present in the literature for NGC~4785, we adopt a value of $\log$(M$_{\rm BH}$/M$_\odot$) $\simeq$ 8.0). This estimate is inferred from the stellar velocity dispersion of the host galaxy \citep[$\sigma_\star \approx$ 187 km s$^{-1}$;][]{olivia1999} via the M–$\sigma$ relation \citep[e.g.,][]{gultekin2009} and lies within the $\sim$0.4 dex intrinsic scatter of the relation. Residuals are present in the XMM and \chandra\ spectra. We find two significant features, which are modeled with Gaussian components, at E = $1.33^{+0.05}_{-0.06}$ keV and at E = $0.98 \pm 0.03$ keV. 
Similarly to Mrk~18, during the fitting procedure we disentangled the normalization of the \texttt{apec} component between the XMM and \chandra-\nustar\ spectra for all three models. We subsequently fixed the latter normalization to zero, as it was unconstrained, while the normalization for the XMM-\textit{Newton} spectra was found to be $(1.61^{+0.31}_{-0.30}) \times 10^{-5}$. 
All tests consistently indicate the presence of variability in \nhlos. However, intrinsic flux variability cannot be firmly established, as the tension does not exceed the adopted threshold of 5 (see Sect.~\ref{var_classification_res} and Tables~\ref{tab:clas_summary} and \ref{tab:spec_9}).

\vspace{0.2cm}

{\it UGC~3752} is a Seyfert 2 \citep{veron2006} AGN. 
During the analysis, two \texttt{apec} components are used to model the soft X-ray spectrum. 
Following our spectral fitting, we find that the inferred line-of-sight column density shows significant variability across epochs and models. In particular, the \rxt\ model yields column densities whose upper limits exceed the Compton-thick threshold for the XMM–{\it Newton} observation and for several \nustar\ pointings, including the observation obtained on 2021-10-22, for which the best-fit value itself lies in the Compton-thick regime. In contrast, the \skirt\ and \ux\ models generally favor Compton-thin solutions, with Compton-thick values reached only at the upper end of the uncertainty range in a subset of \nustar\ observations, while other epochs remain fully consistent with a Compton-thin absorber. Overall, these results support the classification of UGC~3752 as a candidate changing-obscuration AGN, possibly undergoing transitions between Compton-thin and Compton-thick state (see Sect.~\ref{ctk_cth_class}). From the first order and statistical tests, \nhlos\ variability is clearly detected in UGC~3752 (see Sect.~\ref{var_classification_res} and Tables~\ref{tab:clas_summary}). The tension associated with the no-variability test is above $T$ = 5, indicating that variability is indeed required by the data. However, the no-flux variability test yields tensions well below 5, suggesting that absorption variability is predominantly driving the observed spectral variability (see Table~\ref{tab:spec_10}).

\vspace{0.2cm}

{\it WISE~J144850.99-400845.6} is a Seyfert 1.2 galaxy \citep{malizia2016}. Our multi-epoch analysis classify this AGN as Compton-thin (see Sect.~\ref{ctk_cth_class}).
During the fitting process for this source, the cross-normalization constant of the \nustar\ data in the \ux\ model, when left free to vary, tends to a nonphysical value of $\sim$17. When fixed to 1.55 (value consistent with the best-fit obtained for the other two models), the resulting fit is very similar in terms of C-stat/d.o.f. to those models. Therefore, we decided to fix the cross-normalization constant to 1.55 in the final best-fit.
WISE~J144850 is listed in Table~\ref{tab:clas_summary} as first-order undetermined, i.e., based on whether \nhlos\ values from different observations fall within each other’s uncertainties, due to the upper limits derived during \rxt\ and \skirt\ fits, which prevent a definitive classification.
The statistical tests also yield mixed indications: the p-value test favors variability, while the tension test is inconclusive, resulting in a final classification of an Undetermined AGN (see Sect.~\ref{var_classification_res} and Tables~\ref{tab:clas_summary}). The intrinsic flux variability tests under the no-variability hypothesis yield tensions $>$ 5, suggesting the presence of flux variability (see Table~\ref{tab:spec_11}).

\section{Statistical comparison of the parameter distributions}\label{distribution_test}
\begin{table*}
\centering
\caption{Statistical comparison of parameter distributions for the three adopted models across variability classes.}
\label{tab:p-values_ktests}
\renewcommand{\arraystretch}{1}
\begin{minipage}{0.48\textwidth}
\centering
\begin{tabular}{lccccc}
\hline
\multicolumn{1}{c}{\footnotesize{(1)}} & \multicolumn{1}{c}{\footnotesize{(2)}} & \multicolumn{1}{c}{\footnotesize{(3)}} & \multicolumn{1}{c}{\footnotesize{(4)}} & \multicolumn{1}{c}{\footnotesize{(5)}} & \multicolumn{1}{c}{\footnotesize{(6)}} \\
Parameter & NV & U & V & KW & KS \\
 & Median & Median & Median & p-value & p-value \\
\hline

\multicolumn{6}{c}{\ux}\\
\nhlos & 51.14 & 12.18 & 49.81 & 0.42 & NV vs U: 0.60\\
 & & & & & NV vs V: 0.95\\
 & & & & & U vs V: 0.28\\
$\sigma$(C$_{\rm inst.}$) & 0.41 & 0.26 & 0.19 & 0.87 & NV vs U: 0.90\\
 & & & & & NV vs V: 0.57\\
 & & & & & U vs V: 0.96\\
$\Gamma$ & 1.80 & 1.71 & 1.57 & 0.45 & NV vs U: 0.91\\
 & & & & & NV vs V: 0.57\\
 & & & & & U vs V: 0.47\\
C$_\mathrm{f}$ & 0.15 & 0.30 & 0.30 & 0.79 & NV vs U: 1.00\\
 & & & & & NV vs V: 0.95\\
 & & & & & U vs V: 0.86\\
$\theta$ & 75.00 & 60.00 & 60.00 & 0.22 & NV vs U: 0.69\\
 & & & & & NV vs V: 0.29\\
 & & & & & U vs V: 0.64\\
$\sigma_{\rm tor}$ & 8.66 & 24.03 & 5.97 & 0.45 & NV vs U: 0.60\\
 & & & & & NV vs V: 0.71\\
 & & & & & U vs V: 0.29\\
F$_\mathrm{s}$ & 0.15 & 1.24 & 2.03 & 0.17 & NV vs U: 0.61\\
 & & & & & NV vs V: 0.10\\
 & & & & & U vs V: 0.67\\

\hline
\multicolumn{6}{c}{\rxt}\\
\nhlos & 49.60 & 16.96 & 55.97 & 0.41 & NV vs U: 0.60\\
 & & & & & NV vs V: 0.96\\
 & & & & & U vs V: 0.29\\
$\sigma$(C$_{\rm inst.}$) & 0.31 & 0.20 & 0.22 & 0.86 & NV vs U: 0.60\\
 & & & & & NV vs V: 0.58\\
 & & & & & U vs V: 0.96\\
$\Gamma$ & 1.77 & 1.73 & 1.55 & 0.70 & NV vs U: 0.90\\
 & & & & & NV vs V: 0.81\\
 & & & & & U vs V: 0.46\\
C$_\mathrm{f}$ & 0.70 & 0.81 & 0.53 & 0.04 & NV vs U: 0.59\\
 & & & & & NV vs V: 0.28\\
 & & & & & U vs V: 0.04\\
$\theta$ & 66.00 & 54.00 & 60.00 & 0.61 & NV vs U: 0.80\\
 & & & & & NV vs V: 0.95\\
 & & & & & U vs V: 0.90\\
 F$_\mathrm{s}$ & 1.68 & 0.21 & 1.30 & 0.23 & NV vs U: 0.60\\
 & & & & & NV vs V: 1.00\\
 & & & & & U vs V: 0.29\\
N$_{\rm H,eq}$ & 90.48 & 197.92 & 82.58 & 0.58 & NV vs U: 0.60\\
 & & & & & NV vs V: 0.96\\
 & & & & & U vs V: 0.46\\

\hline
\end{tabular}
\end{minipage}
\hfill
\begin{minipage}{0.48\textwidth}
\centering
\begin{tabular}{lccccc}
\hline
\multicolumn{1}{c}{\footnotesize{(1)}} & \multicolumn{1}{c}{\footnotesize{(2)}} & \multicolumn{1}{c}{\footnotesize{(3)}} & \multicolumn{1}{c}{\footnotesize{(4)}} & \multicolumn{1}{c}{\footnotesize{(5)}} & \multicolumn{1}{c}{\footnotesize{(6)}} \\
Parameter & NV & U & V & KW & KS \\
 & Median & Median & Median & p-value & p-value \\
\hline

\multicolumn{6}{c}{\skirt}\\
\nhlos & 53.45 & 12.77 & 49.18 & 0.45 & NV vs U: 0.61\\
 & & & & & NV vs V: 0.95\\
 & & & & & U vs V: 0.29\\
$\sigma$(C$_{\rm inst.}$) & 0.36 & 0.20 & 0.19 & 0.68 & NV vs U: 0.60\\
 & & & & & NV vs V: 0.57\\
 & & & & & U vs V: 0.96\\
$\Gamma$ & 1.81 & 1.70 & 1.53 & 0.14 & NV vs U: 0.89\\
 & & & & & NV vs V: 0.29\\
 & & & & & U vs V: 0.14\\
C$_\mathrm{f}$ & 0.43 & 0.59 & 0.50 & 0.30 & NV vs U: 0.40\\
 & & & & & NV vs V: 0.43\\
 & & & & & U vs V: 0.68\\
$\theta$ & 65.50 & 60.00 & 60.00 & 0.52 & NV vs U: 0.60\\
 & & & & & NV vs V: 0.57\\
 & & & & & U vs V: 0.79\\
F$_\mathrm{s}$ & 1.08 & 1.06 & 0.61 & 0.64 & NV vs U: 0.90\\
 & & & & & NV vs V: 0.80\\
 & & & & & U vs V: 0.79\\
N$_{\rm H,eq}$ & 505.00 & 126.60 & 41.77 & 0.56 & NV vs U: 0.90\\
 & & & & & NV vs V: 0.80\\
 & & & & & U vs V: 0.14\\
\hline
\end{tabular}
\tablecomments{
We applied the Kruskal-Wallis (KW) and pairwise Kolmogorov--Smirnov (KS) tests to examine statistical differences in the fitting values across the three variability classes and models.
No significant differences are found.
The small sample sizes, however, limit the test sensitivity.
\textit{Columns:} 1) considered parameter; 2--4) medians, where NV = non-variable, U = undetermined, and V = variable; 5) KW-test p-value; 6) KS-test p-value.}
\end{minipage}
\end{table*}

\newpage
\clearpage

\section{MCG-03-34-064 fitting results}\label{mcg_res}

\renewcommand{\arraystretch}{2.2}
\begin{deluxetable*}{lccccc}
\tablecaption{MCG-03-34-064: second step best-fit spectral analysis for \skirt.\label{tab:spec_4}}
\tablehead{
\colhead{Parameter} & 
\colhead{\texttt{XMM\_1}} & 
\colhead{\texttt{XMM\_2}} & 
\colhead{\texttt{XMM\_3+\nustar}} & 
\colhead{\texttt{Ch\_1}} & 
\colhead{\texttt{Ch\_(2+3+4+5)}}
}
\startdata
\multicolumn{6}{c}{\texttt{apec \footnotesize{(Thermal emission)}}}\\
kT / keV & 0.88$\pm$0.02 & 0.89$\pm$0.02 & 0.87$\pm$0.01 & 0.93$\pm$0.13 & 0.75$^{+0.14}_{-0.29}$ \\
\multicolumn{6}{c}{\texttt{\footnotesize{Comptonized primary continuum}}}\\
$\Gamma$ & 2.56$_{-0.05}$ & 2.42$^{+0.05}_{-0.04}$ & 1.95$^{+0.03}_{-0.01}$ & 2.44$_{-0.22}$ & 2.28$^{+0.22}_{-0.21}$ \\
\multicolumn{6}{c}{\texttt{\footnotesize{Neutral reflector}}}\\
C$_\mathrm{f}$ & 0.65$_{-0.14}$ & 0.64$_{-0.02}$ & 0.65$_{-0.07}$ & 0.59$^{+0.19}_{-0.11}$ & 0.65$_{-0.04}$ \\
$\theta$ / deg & 73$^{+17}_{-13}$ & 60$^{f}$ & 56$\pm$2 & 60$^{f}$ & 60$^{f}$ \\
F$_\mathrm{s}$ / $10^{-3}$ & 7.96$^{+2.15}_{-1.91}$ & 7.47$^{+1.02}_{-1.19}$ & 24.35$^{+3.25}_{-3.84}$ & 5.82$^{+7.42}_{-3.52}$ & 11.56$^{+5.91}_{-4.03}$ \\
norm / $10^{-2}$ & 1.91$^{+0.86}_{-0.41}$ & 2.12$^{+0.39}_{-0.27}$ & 0.78$^{+0.15}_{-0.10}$ & 2.27$_{-1.37}^{+2.97}$ & 1.45$^{+1.06}_{-0.61}$ \\
N$_{\rm H,eq}$ / $10^{22}$ cm$^{-2}$ & 31.31$^{+40.34}_{-6.58}$ & 145.81$^{+47.43}_{-30.20}$ & 483.29$_{-156.60}$ & 101.55$^{+817.87}_{-67.64}$ & 1000$_{-502.08}$ \\
\multicolumn{6}{c}{\texttt{N$_{\rm H,inst.,num.}$ \footnotesize{(LoS hydrogen column density)}}}\\
N$_{\rm H}$ / $10^{22}$ cm$^{-2}$ & 54.25$^{+7.54}_{-8.16}$ & 63.52$^{+3.26}_{-2.41}$ & 91.43$^{+2.07}_{-5.89}$ &  &  \\
N$_{\rm H}^{Ch,1}$ / $10^{22}$ cm$^{-2}$ &  &  &  & 62.39$^{+14.71}_{-12.90}$ &  \\
N$_{\rm H}^{Ch,2}$ / $10^{22}$ cm$^{-2}$ &  &  &  &  & 74.70$^{+22.09}_{-16.43}$ \\
N$_{\rm H}^{Ch,3}$ / $10^{22}$ cm$^{-2}$ &  &  &  &  & 78.09$^{+15.24}_{-12.83}$ \\
N$_{\rm H}^{Ch,4}$ / $10^{22}$ cm$^{-2}$ &  &  &  &  & 68.88$^{+11.97}_{-10.35}$ \\
N$_{\rm H}^{Ch,5}$ / $10^{22}$ cm$^{-2}$ &  &  &  &  & 76.38$^{+13.35}_{-11.52}$ \\
F$_{\rm 2-10 keV}$ / $10^{-12}$ erg s$^{-1}$ cm$^{-2}$ & 2.57$^{+0.32}_{-0.74}$ & 2.84$^{+0.04}_{-0.10}$ & 1.83$_{-0.02}^{+0.12}$ & 1.79$^{+0.64}_{-1.34}$ & 1.94$^{+0.63}_{-1.49}$ \\
\multicolumn{6}{c}{\texttt{Statistic}}\\
$\chi^2$/d.o.f. & 85/79 & 86/81 & 264/236 & 56/50 & 201/183 \\
{\it T} & 0.68$\sigma$ & 0.56$\sigma$ & 1.82$\sigma$ & 0.85$\sigma$ & 1.33$\sigma$ \\
\enddata
\tablenotetext{}{p-value for the model: $1.49\times10^{-3}$}
\end{deluxetable*}

\begin{figure}[h]
\centering
    \includegraphics[width=0.4\textwidth]{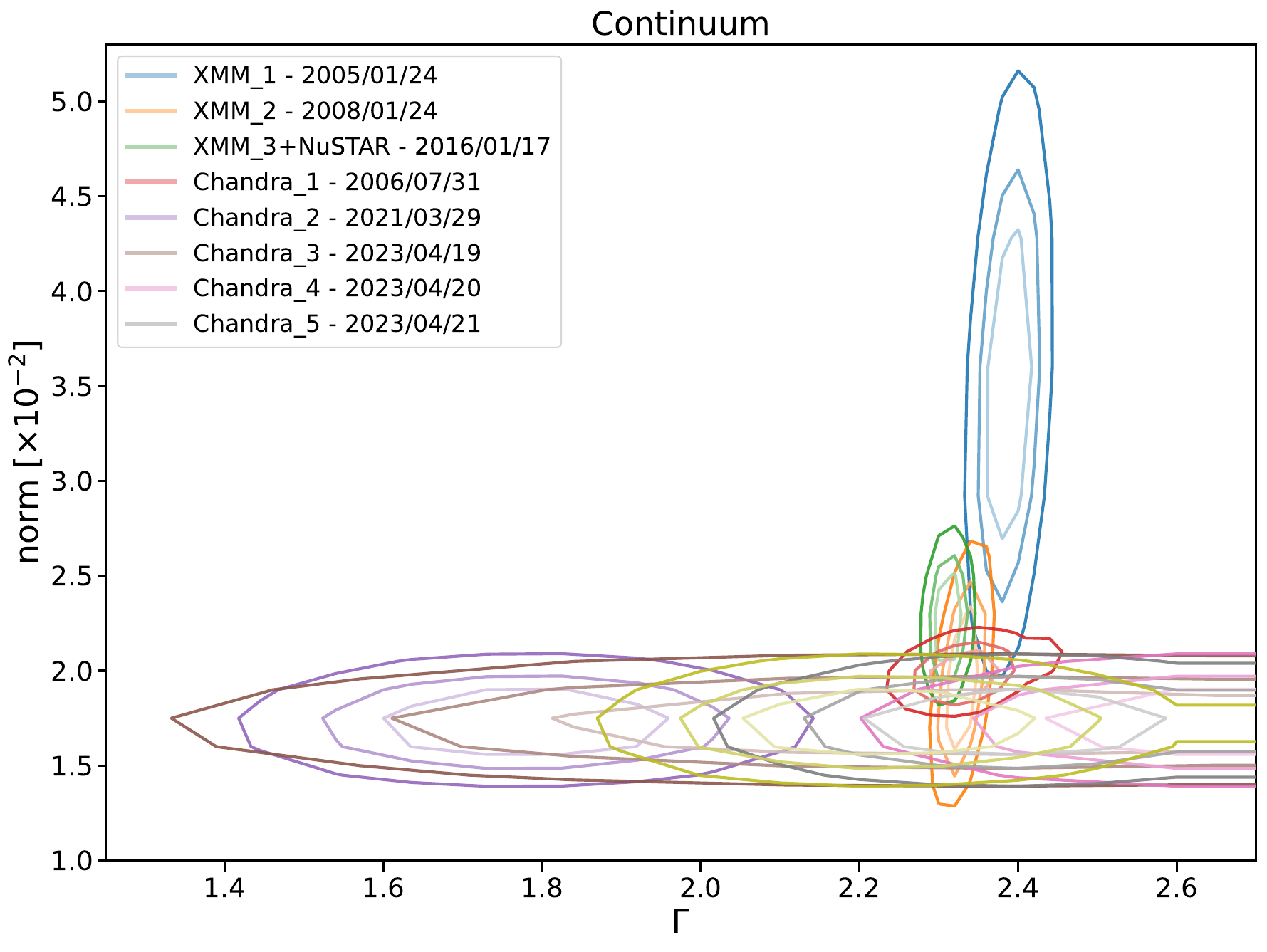}
    \includegraphics[width=0.4\textwidth]{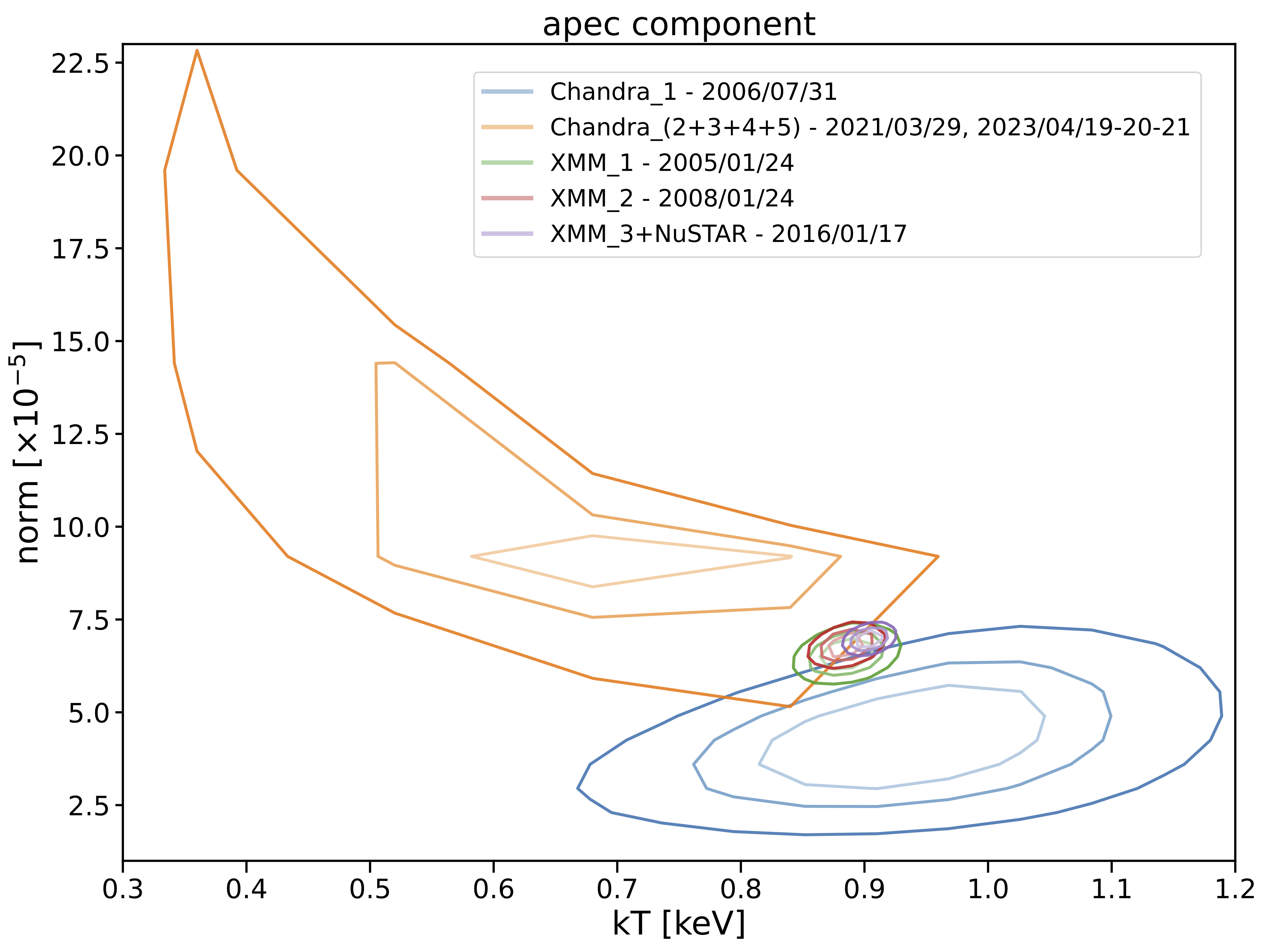}
    \includegraphics[width=0.4\textwidth]{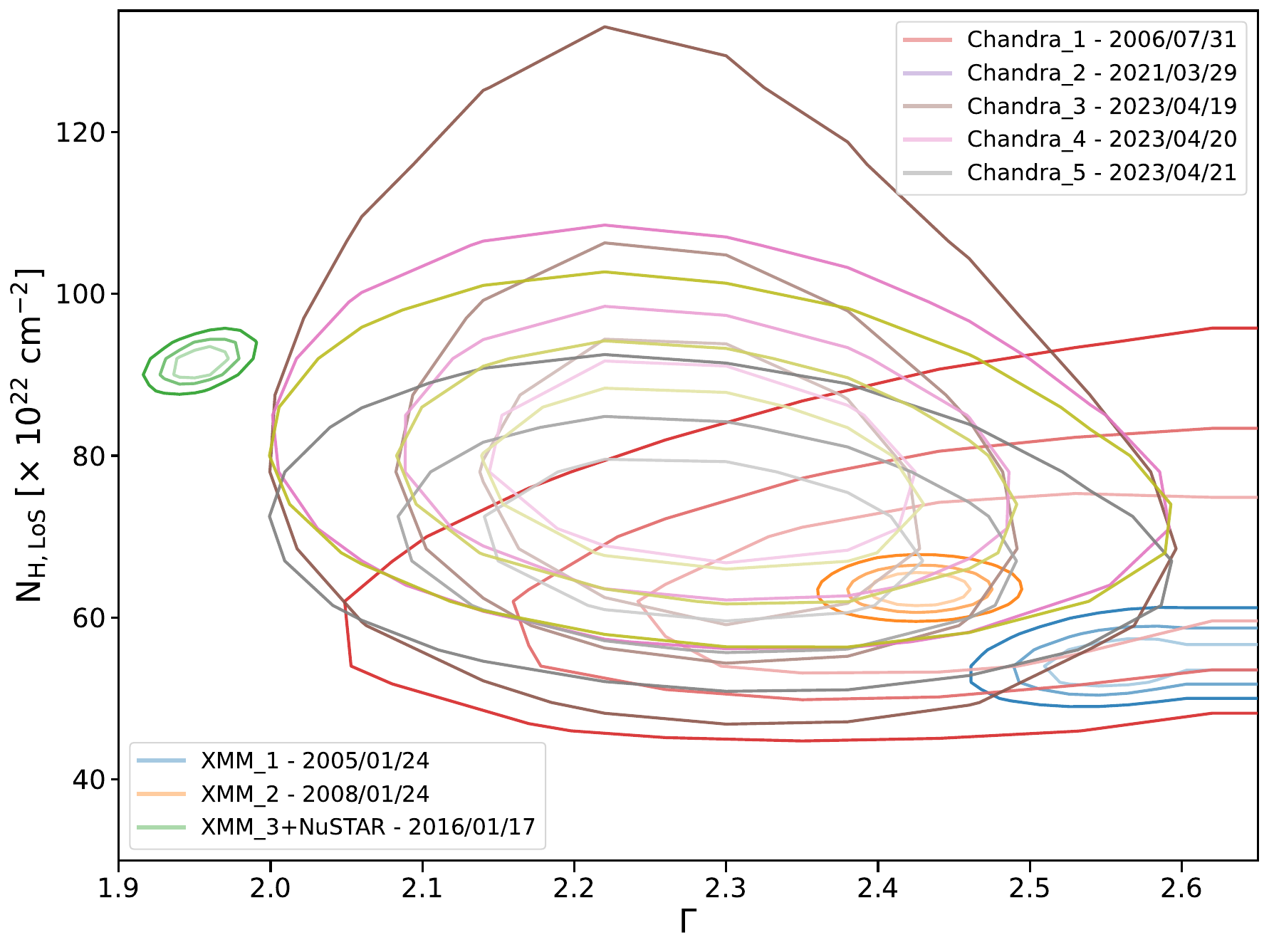}
    \caption[]
    {{MCG-03-34-064 contour plots considering the best-fit obtained with \skirt, showing that the assumption of a constant spectral shape across epochs is not valid for this AGN.
    {\it Upper panel}: photon index $\Gamma$ versus normalization of the primary continuum. {\it Middle panel:} plasma temperature versus normalization of the \texttt{apec} component. {\it Lower panel:} X-ray photon index versus \nhlos.
    The plot in the {\it upper panel} is obtained considering our first step best-fit with all observation linked together and with $\Gamma$ free to vary for each; the plots in the {\it middle} and {\it lower panels} considers the second step best-fit, where the available observations are fit independently (see Sect.~\ref{mcg} for more details).}}
    \label{fig:cp_mgc}
\end{figure}

\begin{figure}[h]
\centering
    \includegraphics[width=0.5\textwidth, trim=25 150 300 300, clip]{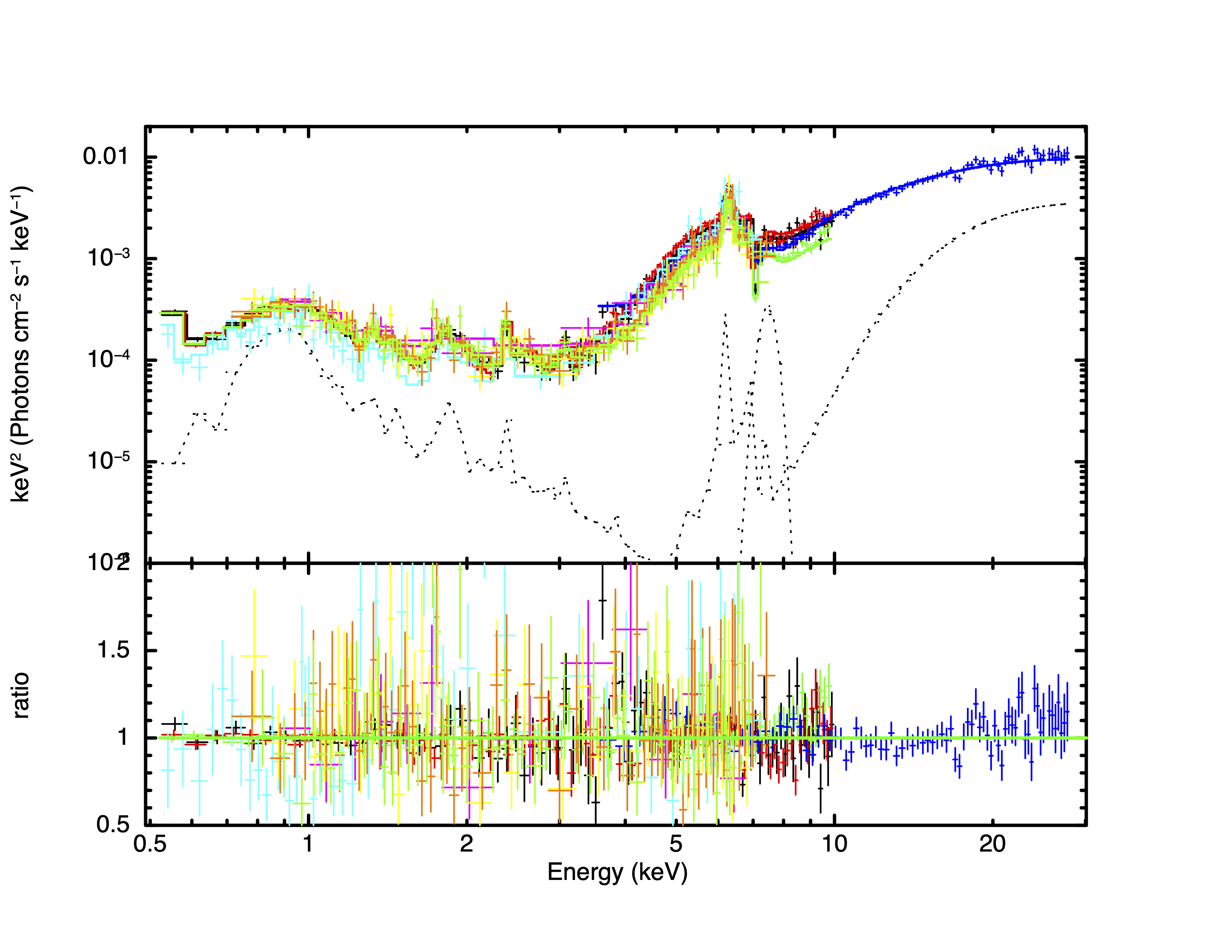}
    \caption[]
    {{MCG-03-34-064. Second step best-fit with the \skirt\ torus model, refer to the results reported in Table~\ref{tab:spec_4}.  The \chandra\ (in pink, yellow, orange, light green, and lime), \xmm\ (in black, red, and green), and \nustar\ (FPMA and FPMB spectra are grouped together in blue) spectra are shown.}}
    \label{fig:mcg_spectra}
\end{figure}

\begin{figure}[h!]
\centering
    \includegraphics[width=0.7\textwidth, trim=250 0 20 20, clip]{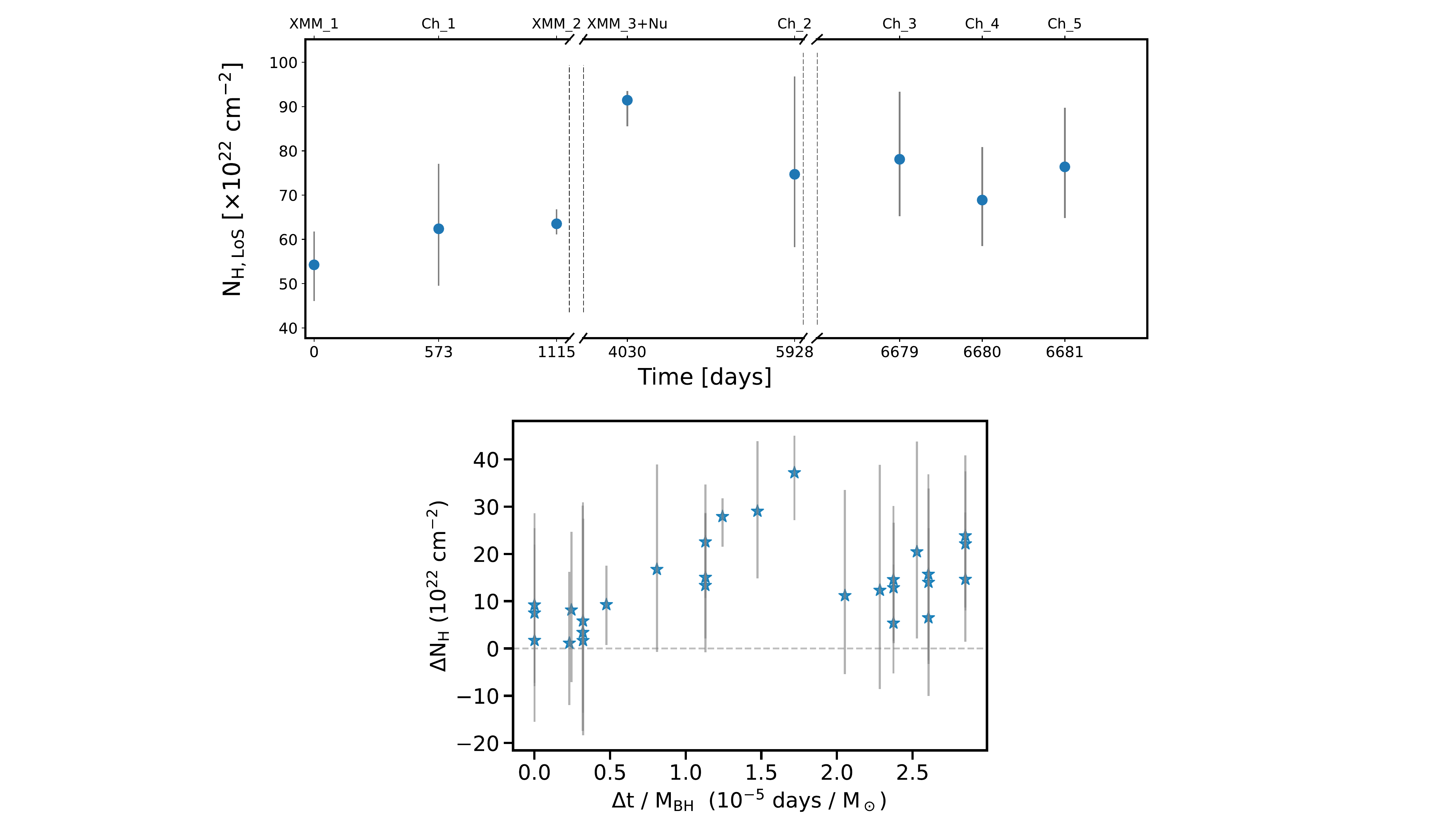}
    \caption[]
    {{MCG-03-34-064. {\it Upper panel:} evolution of the \nhlos. The x-axis is in linear scale with visual breaks to better visualize the observations.
    {\it Lower panel:} variations of \nhlos\ as a function of the time separation between observations, normalized by the black-hole mass, considering \skirt.}}
    \label{fig:mcg_plots}
\end{figure}

\clearpage
\newpage


\bibliography{biblio}{}
\bibliographystyle{aasjournalv7}



\end{document}